\documentclass[manuscript,nonacm]{acmart}

\usepackage[raggedrightboxes]{ragged2e}
\usepackage{tabulary}
\usepackage{tabularx}
\usepackage{longtable}
\usepackage{graphicx}
\usepackage{multirow}
\usepackage{subcaption}
\AtBeginDocument{%
  }

\setcopyright{acmlicensed}
\copyrightyear{2025}
\acmYear{2025}
\acmDOI{XXXXXXX.XXXXXXX}
\begin{document}

% \title{Bridging the Clinical and Technical Aspects of Micronutrient Assessment: A Comprehensive Review}

\title[Towards an Accessible, Noninvasive Micronutrient Status Assessment Method]{Towards an Accessible, Noninvasive Micronutrient Status Assessment Method: A Comprehensive Review of Existing Techniques}

% \author{Anonymous Author(s)}
\author{Andrew Balch}
\email{xxv2zh@virginia.edu}
\orcid{0009-0005-1282-7480}
\author{Maria A. Cardei}
\email{cbr8ru@virginia.edu}
\orcid{0009-0009-0047-6766}
\author{Sibylle Kranz}
\email{sk4gd@virginia.edu}
\orcid{0000-0001-5637-1706}
\author{Afsaneh Doryab}
\email{ad4ks@virginia.edu}
\orcid{0000-0003-1575-385X}
\affiliation{
  \institution{University of Virginia}
  \city{Charlottesville}
  \state{Virginia}
  \country{USA}
  % \postcode{22903}
}

\renewcommand{\shortauthors}{Balch, Cardei, Kranz, and Doryab}

%%
%% The abstract is a short summary of the work to be presented in the article.
\begin{abstract}
Nutrients are critical to the functioning of the human body and their imbalance can result in detrimental health concerns. The majority of nutritional literature focuses on macronutrients, often ignoring the more critical nuances of micronutrient balance, which require more precise regulation. Currently, micronutrient status is routinely assessed via complex methods that are arduous for both the patient and the clinician. To address the global burden of micronutrient imbalance, innovations in assessment must be accessible and noninvasive. In support of this task, this article synthesizes useful background information on micronutrients themselves, reviews the state of biofluid and physiological analyses for their assessment, and presents actionable opportunities to push the field forward. By taking a unique, clinical perspective that is absent from technological research on the topic, we find that the state of the art suffers from limited clinical relevance, a lack of overlap between biofluid and physiological approaches, and highly invasive and inaccessible solutions.
We present opportunities for future work to maximize the impact of a novel assessment method by incorporating clinical relevance, the holistic nature of micronutrition, and prioritizing accessible and noninvasive systems.

% Nutrients are critical to the functioning of the human body and their imbalance can result in detrimental health concerns. The majority of literature focuses on macronutrients, often ignoring the more critical nuances of micronutrient balance, which require more precise regulation. Currently, micronutrient status is routinely assessed via flawed methods, while technological methods are being developed in parallel to address the same scope. There is a lack of an interdisciplinary bridge between the two disciplines to design novel, noninvasive, and accessible methods that would revolutionize micronutrient assessment. 
% This article reviews current clinical and technological methods that assess micronutrient status, the budding field of precision nutrition, and emerging technological opportunities that pave the way for future work. 
% We find that technological methods are needed to fulfill the lack of data observed from clinical methods, and provide eight suggestions for future work to pursue novel, noninvasive, and accessible micronutrient status assessment methods.
% We also see that most novel methods do not compare micro measurements to a clinical gold standard.
\end{abstract}


\received{9 August 2024}
\received[revised]{12 April 2025}
\received[accepted]{18 May 2025}

% \begin{center}
%   \includegraphics[width=\textwidth]{Figures/teaser}
% \end{center}

\maketitle

\section{Introduction} \label{intro}
Balanced nutrition is important for the development and functioning of the human body and can have many downstream health effects. The World Health Organization (WHO, Table \ref{tab:abbreviations}) reports that individuals with proper nutrition have increased lifespans, are more likely to break cycles of poverty, and have lower risks of disease, which impacts productivity and mortality \cite{world_health_organization_nutrition_2024}. Malnutrition is the most critical issue related to nutritional intake, with undernutrition associated with about 45\% of deaths in children younger than five years old. Anemia is a major resulting disease, affecting 37\% of pregnant people and 40\% of children under 5. The WHO defines malnutrition as nutrient imbalance, which consists of deficiencies or excesses in essential nutrients \cite{WHO_malnutrition}. It is a condition that includes several interconnected factors: 1) being overweight, 2) obesity, 3) diet-related noncommunicable diseases like anemia, heart disease, and diabetes, and 4) undernutrition, which encompasses being underweight, wasting, and stunting, as well as a lack of essential micronutrients. Micronutrients are the vitamins (e.g. vitamins A, C, B12) and minerals (e.g. iron, calcium, iodine) which are critical to everyday human bodily function and can only be provided via dietary intake (i.e. they are not produced by the body).

According to \citet{hummell_role_2022}, malnutrition can be caused by insufficient intake, malabsorption, acute and chronic diseases, increased nutrient need, and weight loss surgery. The impact of acute or chronic diseases and weight loss surgery is self-explanatory, whereas other factors are less visible. Insufficient intake in particular can be influenced by a wide range of variables, making it a challenging problem. These factors include income, access to nutritious foods, culture, and upbringing (malnutrition patterns can be passed down to generations). Insufficient intake is a particularly high risk for low-income families. Furthermore, there is often a lack of awareness about insufficient intake, making behavior change more difficult. Malabsorption can also be a cause, as the absorption of nutrients found in food and supplements is highly individual and can be affected by diseases, genetic issues, and one's stage of life. In addition, individual nutrient requirements change over time, and several factors can increase nutrient demand, such as growth from infancy to adulthood, pregnancy, lactation, and recovery from illness or other trauma. 

Nutrients can be classified into two main categories: \textit{macronutrients} and \textit{micronutrients}. Both are essential in precise balance to prevent \textit{malnutrition}, also referred to as \textit{nutritional imbalance}. Macronutrients include carbohydrates, fats, and proteins, which are major energy sources and are required in larger amounts than micronutrients \cite{heitman_micronutrients_2022}. Although required in smaller amounts, optimal intake of micronutrients is crucial. Micronutrients include vitamins and minerals, and they play a critical role in chemical reactions that produce energy from macronutrients acquired from food, as well as other essential bodily functions \cite{world_health_organization_micronutrients_2025}.
Micronutrient intake through diet or supplements is crucial because our bodies cannot synthesize micronutrients, nor can they be substituted for one another \cite{biesalski_hans_micronutrients_2018}. A precise intake amount is required and slight deviations can result in either deficiency or excess, both with significant health impacts \cite{heitman_micronutrients_2022}. 
% Intervention programs aim to address deficiency and excess, but they rely on accurate status assessments to do so effectively.
For these reasons, micronutrient imbalance is a problem that is unique and deserving of attention.

\paragraph{Micronutrient Deficiency} The estimated number of people with micronutrient deficiency is 2 billion worldwide \cite{von_grebmer_synopsis_2014}.  It is estimated that micronutrient deficiencies are the cause of between 425,000 and 745,000 deaths in children under five years old \cite{brown_increasing_2021}. An estimated 56\% of preschool-aged children and 69\% of non-pregnant females of reproductive age worldwide suffer from at least one micronutrient deficiency \cite{Stevens2022Micronutrient}. Globally, the most widespread micronutrient deficiencies occur in iron, iodine, folate, vitamin A, and zinc \cite{bailey_epidemiology_2015}. The population of the United States (US) has significant risk of micronutrient deficiency due to the prevalence of a high-energy, low-nutrient diet \cite{the_institute_for_functional_medicine_micronutrient_2020, drake_micronutrient_2018}. The US' National Health and Nutrition Examination Survey (NHANES) estimates that 31\% of the US population is at risk for micronutrient deficiency, with calcium, potassium, iron, and vitamins A, D, C, and E being of particular concern \cite{the_institute_for_functional_medicine_micronutrient_2020, drake_micronutrient_2018}. Since NHANES is a survey of self-reported dietary intake with minimal biochemical testing, it is likely that this number is actually underestimating the true burden in the US \cite{drake_micronutrient_2018}. 

Deficiencies can cause developmental issues, metabolic disorders, impaired immune system, altered endocrine and cognitive functioning, chronic disease, and more. For example, high magnesium intake and circulating status (via urine) is directly associated with a greater risk of cardiovascular disease (CVD) \cite{rosique-esteban_dietary_2018}. While developed countries are certainly at a risk for micronutrient deficiencies, those most at risk are people living in underdeveloped countries. Populations that also bear a significant burden include developing countries as well as children less than five years old, pregnant people, and victims of chronic disease \cite{drake_micronutrient_2018, bruins_maximizing_2016, Stevens2022Micronutrient, world_health_organization_micronutrients_2025}. Risks of inadequate intake are exacerbated since different micronutrients are needed in different amounts at different stages in life \cite{biesalski_hans_micronutrients_2018}.  Vitamin A, iron, and zinc are among the most significant micronutrient deficiencies observed in children \cite{Stevens2022Micronutrient}. In the first 1000 days of life, iron, iodine, folate, and vitamin D have high dietary requirements, and a failure to meet them could result in poor physical and cognitive development. During adolescence, iron, calcium, folate, and vitamin D intake is critical, especially for those who menstruate. Pregnancy sees an increase in iron, folate, vitamin B12, and vitamin D requirements. The elderly are more at risk for vitamin D, B12, and B6 deficiency, and medications that influence micronutrient status or absorption must be heavily considered during this stage. Micronutrient deficiencies rarely manifest alone, and it is common to see multiple deficiencies arise simultaneously \cite{bailey_epidemiology_2015}. 

\paragraph{Micronutrient Excess} Like deficiency, micronutrient excess can also lead to health detriments, with the risk increasing as intake levels surpass an individual's upper limit \cite{gernand_upper_2019}. Toxicity from micronutrient excess can occur in any person who exceeds this limit. However, there is an exceptional risk for vulnerable populations, namely infants, young children, and pregnant people \cite{ishaq_effect_2024, pike_excess_2019, gernand_upper_2019, espinosa-salas_nutrition_2024}. For example, an excess of iron increases the risk of diarrhea, sepsis, meningitis, and gut inflammation for infants and young children, with a lethal dose of 150 mg/kg \cite{pike_excess_2019}. Similarly, surplus iron can lead to an increased risk of gestational and type 2 diabetes for pregnant people \cite{pike_excess_2019}. Elevated health risks are also a concern with an excess of vitamin D in infants, calcium in pregnant people, vitamin A in infants, young children, and pregnant people, and iodine in all three of these groups in addition to breastfeeding parents \cite{pike_excess_2019, espinosa-salas_nutrition_2024}.
Individuals who are more likely to be taking supplements (especially multivitamins) to address an initial deficit could also be at a greater risk for micronutrient toxicity \cite{bruins_addressing_2015, gernand_upper_2019}. 
% This is a concern with population-level nutrition interventions, such as food fortification, that aim to increase micronutrient intake without surpassing the recommended value. 
Research suggests that ingesting an excess of micronutrients through diet alone is unlikely and may only occur for those who take supplements \cite{espinosa-salas_nutrition_2024, bruins_addressing_2015, pike_excess_2019}. 
An excess of certain micronutrients can interfere with the absorption of others, potentially leading to secondary deficiencies. For instance, elevated calcium levels may reduce the absorption of magnesium. Additional information on micronutrient interactions are provided in Tables \ref{tab: char-ws}, \ref{tab: char-fs}, and \ref{tab: char-min}. There is minimal research on micronutrient excess because it is usually limited to the above population groups. 
% Thus, most methods to assess micronutrient status have highlighted nutrient deficiency. 
As a result, the work covered in this review largely focuses on micronutrient deficiency. For more information on excessive micronutrient intake and its specific effects for each micronutrient, we recommend referring to an up-to-date nutrition overview (e.g. \citet{espinosa-salas_nutrition_2024}).

\paragraph{Micronutrition Assessment}
When a micronutrient imbalance is identified, it is addressed through intervention. This involves modifying diet, providing micronutrient supplements, or on a population scale, fortifying food products such as iodized salt and fortified flour \cite{bailey_epidemiology_2015}. Interventions need to be carefully planned and monitored to avoid providing too little or too much of a particular micronutrient. There is a strong need for tools that can help guide intervention programs to effectively reach at-risk populations while minimizing the impact on those who have sufficient nutrient levels. This requires employing "different risk assessment methods to make the monitoring process more efficient, reliable, and cost-effective" \cite{bailey_epidemiology_2015}. 
% Thus, micronutrient assessment is an important research target, as it drives novel methods of intervention to maintain overall nutrient balance.
% Although the use of biomarkers is important for assessing micronutrient balance, intervention programs suffer from a lack of biomarker data \cite{brown_increasing_2021}. Accessing population-level biomarker data is difficult due to political, economic, logistic, and expertise-related issues. 
% Some proposed solutions include adding more biomarker analysis to existing surveys that can collect and test biosamples for diseases like malaria. Another suggestion is to research less invasive methods of sample collection and more efficient sample analysis, including point-of-care (PoC) methods. Additionally, the implementation of a Micronutrient Data Generation Initiative has been proposed to improve micronutrient-related data through advocacy, technical, and financial support \cite{brown_increasing_2021}.
%Nutrition plays a significant role in people's health, and micronutrients in particular deserve the attention of forthcoming research initiatives. 
% The health consequences of micronutrient imbalances, as well as the pursuit for effective micronutrient intervention programs needed as a result, require good micronutrient status assessment methods.
% Nutritional status assessment is the evaluation of an individual's dietary intake, physical condition, and health markers to determine their overall nutritional health.
\textit{Nutritional status assessment} is a way to evaluate an individual's overall nutritional health and is necessary to identify, prevent, and address any imbalances.
% is necessary to evaluate an individual's balance of nutrients and confront the burden of malnutrition.
This review focuses on micronutrient status assessment methods in particular because we find that existing methods are insufficient and new techniques are under-researched. 
% The visibility of micronutrient status suffers as a result of this lack of attention, further perpetuating associated health issues.
% Current methods, such as dietary logging, are inherently flawed and not well-established. 
% On the other hand, Indirect Calorimetry and formulas for estimating energy expenditure (macronutrient status) have been proven at the individual level \cite{reber_nutritional_2019}.

We propose that an ideal assessment method is both \textit{accessible} and \textit{noninvasive}. An \textit{accessible} method is one that is available to more individuals, while maintaining clinical effectiveness. This can be achieved through improvements in cost, efficiency, or mobility. Such a method should ultimately alleviate the need for laboratory tests, decrease reliance on high-effort and subjective surveys (such as dietary intake logging), and increase efficiency and accuracy for clinicians. A \textit{noninvasive} method would either eliminate the need for a biofluid sample or obtain it in an unobtrusive manner. Noninvasiveness is therefore interconnected with accessibility, as a method that is invasive (e.g. requiring a blood draw) is less accessible (e.g. places burden on the patient and on the clinician). An ideal method that is both accessible and noninvasive could obtain micronutrient status regularly, without much effort for the patient or clinician. To maintain effective and clinically-relevant assessment methods, it is also essential to balance the accessibility and noninvasiveness of a method with its ability to be a \textit{sensitive} (high true positive rate) and \textit{specific} (high true negative rate) assessment of micronutrient status. 
Because every country in the world is impacted by malnutrition \cite{WHO_malnutrition}, it is clear that an accessible and noninvasive assessment method is needed to enable routine assessment globally. This is especially true in countries with high-fat, low-nutrient diets such as the US, in low- and middle-income nations, and in areas of the world where complex, blood-based analyses are generally infeasible.
% The problem of micronutrient imbalance is clinical. However, several technical advancements have also been made in an attempt to resolve it. As technology improves and permeates through healthcare, we see it increasingly becoming a valuable tool in micronutrition. 
In our review, we aim to consider both clinical and technical aspects of micronutrition assessment and explore the potential for novel, accessible, and noninvasive methods. We argue that such an interdisciplinary approach is key to finding optimal solutions. 
% There are several technological works that attempt to design this type of ideal method, however we find that they still fall short. We present this review to highlight current and potential work that aims towards an accessible and noninvasive micronutrient status assessment method.

In the subsequent sections, we discuss \textit{biofluid} and \textit{physiological} analysis methods for micronutrient status assessment. Within the context of this paper, biofluid analysis refers to techniques that measure the amount of a micronutrient biomarker present in a biofluid as a means to infer the underlying nutritional status of an individual, whereas physiological analysis focuses on the bodily signs and signals which are symptomatic a micronutritional imbalance. For the latter, we focus on providing a brief overview of clinical physical assessment in nutrition and describe relevant applications of optical sensing.

\subsection{Paper Scope}
Few existing papers that review novel micronutrient assessment methods are comprehensive and clinically relevant \cite{udhani_comprehensive_2023, shi_development_2022, nimbkar_microfluidic_2024, kalita_point--care_2022, huey_review_2022, campuzano_pursuing_2024}. \citet{campuzano_pursuing_2024} and \citet{kalita_point--care_2022} focus particularly on electrochemical sensors, with the latter considering few vitamins and mostly out-of-body status. \citet{huey_review_2022} focus on vitamin A assessment and emphasizes the limitations of relying on centralized laboratories and specialized equipment, underscoring the need for more portable and accessible diagnostic methods. \citet{nimbkar_microfluidic_2024} focus on microfluidic assessment and \citet{shi_development_2022} concentrate specifically on wearable sensors. \citet{udhani_comprehensive_2023} provide a detailed review on biomarker-based micronutrient detection, focusing on the analytical chemistry aspect.
% \citet{udhani_comprehensive_2023} focus on various micronutrient biomarkers, with a discussion on point of care testing and commercial products. 
% \citet{shi_development_2022} focus specifically on wearable sensors. \citet{nimbkar_microfluidic_2024} review overall nutritional biomarkers, specifically with microfluidic assessment. 
% \citet{kalita_point--care_2022} review nanobiosensors for determining vitamin deficiency, mainly focusing on electrochemical sensors for few vitamins, and slight insights into in-body status. They note difficulties detecting multiple vitamins at once and integrating sensors into accessible technology. 
% \citet{huey_review_2022} review portable assessment methods for vitamin A status in biological samples. 
% \citet{campuzano_pursuing_2024} reviews electrochemical sensors for medicine and nutrition. Shows that a lot of people think electrochemical sensors are the key to precision nutrition, but we need to consider a lot more.

The novelty of our paper lies in its comprehensive review of accessible and noninvasive micronutrient assessment methods, uniquely emphasizing clinical relevance—an aspect often overlooked in prior work. It includes a background on micronutrients, tailored for a non-clinical perspective. We also explore a wide range of techniques for assessing biofluids and physiology, including assay-based technologies, electrochemistry-based methods, spectroscopy-based approaches, and analytic methods utilizing machine learning (ML). This review offers a high-level summary, with detailed methods available in dedicated sources.
% We note that while hair and nail samples can be valid subjects of micronutrient analysis, we focus on biofluids in our review of assessment techniques since nutritional analyses of hair and nails measure the metabolism of a nutrient in cells rather than the in-body status of the nutrient \cite{han_hair_2016, bali_quantitative_2023}. 
We contribute several relevant tables throughout the paper and provide a comprehensive reference about micronutrients and their assessment in the Appendix. Ultimately, this review calls for actionable potential opportunities to advance these methods.

% With the aim of pushing innovative research towards an accessible, noninvasive method of assessing micronutrient status, 
This review consists mostly of methods that are both accessible and noninvasive, or at least one of these. We additionally include few methods that are neither accessible nor noninvasive, yet are interesting and valuable for future work within this scope. 
% This review provides a high-level overview of each method, and we encourage readers to refer to dedicated sources for a more in-depth understanding of specific methodologies.
This review does not include methods for assessing micronutrient levels in entities outside of humans, such as food or pharmaceuticals. We note that while hair and nail samples can be valid subjects of micronutrient analysis, we focus on biofluids in our review of assessment techniques since nutritional analyses of hair and nails measure the metabolism of a nutrient in cells rather than the in-body status of the nutrient \cite{han_hair_2016, bali_quantitative_2023}. Our focus is on vitamins and trace minerals (micronutrients), thus we do not discuss nutrients such as proteins, fats, and carbohydrates (macronutrients) or major minerals (e.g. potassium, calcium, etc.). These are sometimes considered separate from micronutrients because of their higher intake requirements and larger quantities within the body.
% nutrients which are found in small amounts in the body (micronutrients), and does not include others found in larger concentrations, such as major minerals or macronutrients.
Lastly, this review does not include alternative medicine techniques or methods, as this is out of the scope. 
% Additionally, while the review discusses limitations of existing methods and actionable suggestions for improvement, future work should investigate further details on clinical implementation and inclusion into public health initiatives.

This review is structured as follows. 
% We first briefly mention the review methodology in Section \ref{methodology}.
% clarify some key ideas the paper relies upon and 
We first present background information in Section \ref{background section}. Next, we discuss micronutrient status assessment methods based on biofluid analysis in Section \ref{biofluids section}, and briefly touch on physiology-based methods in Section \ref{phys section}, focusing on optical sensing techniques. We summarize the current gaps in the literature and present suggestions for future work in Section \ref{Discussion}. 
% Most importantly, areas for interdisciplinary future work are presented that bridge the clinical and technological aspects. 
Finally, we end with a conclusion in 
% conclude by asserting how progress in these areas would improve global nutrition and fundamentally transform how it is addressed in 
Section \ref{conc}, followed by reference tables
% on micronutrient characteristics, physiological symptoms related to imbalance, and gold-standards for assessments 
in the Appendix (\ref{appendix}).

\section{Background} \label{background section}
In this section we describe micronutrient characteristics, their presence in biofluids, and their impacts on physiological functioning.
% We begin with a summary of micronutrients, their characteristics, their presence in biofluids, and their impacts on our physiological functioning.
Such a background is necessary to contextualize solutions to emerging micronutrient status assessment methods because of the variety of possible approaches. 
\textit{In-body} assessment methods refer to the levels of a micronutrient present within the body, reflecting the current internal state of an individual's micronutrition. On the other hand, \textit{out-of-body} assessment methods involve external indicators used to infer internal status, such as physiological symptoms measurable through wearable sensors or physical exams. We focus on both approaches, as each provides relevant insights into micronutrient status and offers the potential for non-invasive and accessible assessment methods.

\subsection{Micronutrient Characteristics} 
Micronutrients are divided into three categories: water-soluble vitamins, fat-soluble vitamins, and minerals \cite{callahan_classification_2020}. Water-soluble vitamins, such as vitamin C and the B vitamins, are absorbed directly into the bloodstream and are quickly excreted in urine. They are not stored for long periods in the body, so regular intake is necessary to prevent deficiencies, and there is less concern about toxicity from excess intake. On the other hand, fat-soluble vitamins like A, D, E, and K are absorbed into lymph vessels along with dietary fats. Fat-soluble vitamins are stored in larger quantities in fatty tissues and the liver, so deficiencies take longer to develop, and daily intake is less critical. However, due to their efficient storage and the lack of a rapid excretion mechanism, toxicity is more of a concern.

Minerals can be categorized as major or trace minerals based on the daily requirement \cite{callahan_classification_2020}. Major minerals, such as sodium, potassium, chloride, phosphorus, and magnesium, are required in amounts greater than 100 mg per day, while trace minerals like iron, copper, zinc, selenium, iodine, chromium, fluoride, and manganese are needed in amounts of 100 mg or less per day. Minerals are water-soluble and are absorbed directly into the bloodstream, sometimes with the help of transport proteins. It is important to note that minerals have an electric charge, and their function and storage can be influenced by various factors. For example, certain minerals carry a positive charge, such as sodium (Na+) and potassium (K+), while others are negatively charged, like chloride (Cl-). Their electric charge plays a crucial role in mineral homeostasis, influencing absorption efficiency, competition for transport and storage, bioavailability, and electrolyte balance. For a more in-depth discussion of these interactions, we refer the reader to \cite{Solomons1986, Varvara2023}. A comprehensive summary of the characteristics of micronutrients is presented in the Appendix in Tables \ref{tab:char-vitamins-water-soluble}, \ref{tab:char-vitamins-fat-soluble}, and \ref{tab:char-minerals}.

Dietary guidelines for intake vary across organizations and countries, and are based on factors such as age and sex. These guidelines are not standardized due to the individualized nature of micronutrient metabolism, the diversity of micronutrients, their interactions, and the specific values at which they are needed. The US, for example, utilizes dietary reference intervals such as the Recommended Dietary Allowance (RDA) \cite{national_institutes_of_health_nutrient_nodate}. The RDA represents the average daily level of intake sufficient to meet the nutrient requirements of 97-98\% of healthy individuals. The US National Institutes of Health (NIH) also defines terms to describe intake levels, such as an Upper Limit (UL), highlighting the potential for excess intake. In cases where there is insufficient evidence to establish these guidelines, there exists the more general term, Adequate Intake, which establishes a lower bound of nutritional intake necessary to meet a healthy nutritional state within a population \cite{institute_of_medicine_us_subcommittee_on_interpretation_and_uses_of_dietary_reference_intakes_using_2000}. For example, Adequate Intake for the nutrient biotin (vitamin B7) in human milk-fed infants is defined by the biotin content of human milk itself, because there is a lack of data available to scientifically determine an RDA for this population. Adequate Intake further emphasizes the uncertainty surrounding the appropriate levels of micronutrients in diet.

Micronutrient status assessment usually relies on the combination of the analysis of micronutrient biomarkers in biofluids as well as the physiological symptoms presented by an imbalance. 
Micronutrient biomarkers (predominantly measured by laboratory analyses of biofluids) are still debated, but can generally serve as a reliable \textit{optimal reference standard} for assessment methods \cite{elmadfa_developing_2014}. An optimal reference standard is the clinically agreed-upon technique for determining a patient's in-body status of a micronutrient (e.g. vitamin B9), and describes the target biomarker (e.g. folate), biofluid matrix to be analyzed (e.g. blood serum), and analysis method (e.g. LC-MS/MS) to use for accurate assessment. These standards are continuously debated by the clinical community, and are typically defined by current knowledge of ``the chemistry, absorption, distribution in the body, and metabolism" of a nutrient \cite{elmadfa_developing_2014}. Because these dynamics can be highly individualized, validating a standard requires well-controlled studies that involve specific dietary interventions and the subsequent evaluation of the standard's ``specificity, sensitivity, and suitability for various population subgroups" \cite{elmadfa_developing_2014}.

Certain micronutrient imbalances are associated with observable physical or physiological symptoms, such as those in the skin, eyes, autonomic functioning, etc., which may present opportunities for developing noninvasive assessment methods tailored to specific nutrient deficiencies
% Micronutrient imbalances have well-documented physical symptoms, which provides opportunities to apply noninvasive methods to determine micronutrient status 
\cite{hummell_role_2022, mauldin_performing_2021, esper_utilization_2015, dibaise_hair_2019, radler_nutrient_2013}. These symptoms provide noninvasive insights into the bodily storage of micronutrients, internal processes, and can help identify appropriate biomarkers for further testing. Section \ref{subsec_phys symptoms} discusses micronutrient effects on physiological processes in more depth.
It should be noted that some physical symptoms of micronutrient imbalances only arise when the imbalance becomes quite severe. This can happen slowly, in some cases over the course of several months (i.e. iron), and can result in irreversible symptoms (e.g. night blindness as a result of severe vitamin A deficiency; Tables \ref{tab: char-ws} - \ref{tab: char-min}). While physical symptoms still provide critical insights into the overall burden of micronutrient imbalance in a community, sole reliance on them for assessment can complicate preventative treatment. 
Quantitative, biofluid-based assessments for biomarkers are not immune to similar issues faced by physiological assessments (e.g. status of the biomarker plasma retinol decreases only after vitamin A stores in liver and eyes have nearly depleted), emphasizing the importance of considering both physiological and biofluid-based assessments in the practice of clinical nutrition.

\subsection{Micronutrients in Biofluids} \label{subsec_biofluids}
Biofluid analysis for micronutrient status remains challenging since it is often unclear how other biofluids reflect the optimal reference standard matrix for that micronutrient. Some biofluids include blood, saliva, sweat, tears, urine. Besides blood, each of these can be collected and analyzed noninvasively. Blood and urine are clinically relevant for representing in-body micronutrient status, as will be demonstrated in Section \ref{clinical/biochemical analysis}. However, evidence is less clear for saliva, sweat, and tears \cite{holler_micronutrient_2018}. Although it will not be covered in depth, it is also worth mentioning human milk as a potential biofluid. Breast milk may have implications on the developmental outcomes of infants whose primary source of nutrition is human milk, but this is actively debated \cite{lockyer_breast_2021}. As mentioned earlier, iodine excess in breastfeeding parents can also express itself in human milk, placing infants at an increased risk \cite{pike_excess_2019}. 

In the clinical literature, the equivalence of saliva, sweat, and tears to blood and urine in micronutrient status assessment remains ambiguous. 
% The challenge lies in determining how accurately biofluid composition reflects micronutrient status compared to the gold-standard matrix.
% The central issue of biofluid analysis for micronutrient status is that it remains unclear how well biofluid composition reflects that of the gold-standard matrix for that micronutrient. 
Some evidence for correlation of micronutrient levels in saliva with blood levels was found on a by-micronutrient basis \cite{holler_micronutrient_2018}. For example, serum and saliva levels of vitamin D were found to have a correlation of 0.56, measured using a total vitamin D (25-hydroxy vitamin D) kit with
the electrochemiluminescence technique \cite{bahramian_comparing_2018}. Additionally, some correlation of iron levels in saliva and serum (a blood derivative) have been found. However, validity remains inconclusive as some sources have found a high positive correlation between salivary and serum levels \cite{Canatan2012, Haji-Sattari2019}, while others report a high negative correlation \cite{Flora2021,Lokesh2021, Gawaly2020}. The techniques used to study this correlation involve enzyme-linked immunosorbent assays (ELISAs), laboratory assays, spectrophotometry, and chemiluminescence methods.
% While not micronutrient-specific, one paper claims saliva is a comparatively less complicated biofluid while still containing “proteins, nucleic acids, mucins, amino acids, enzymes, and primary metabolites, which are highly informative biomarkers for various physiological conditions of the body” \cite{bec_near-infrared_2020}.  

Additional research aims to quantify micronutrient status with saliva as well as tears \cite{ruiz-valdepenas_montiel_decentralized_2021, sempionatto_eyeglasses-based_2019}. The three types of human tears, basal, reflex, and emotional, differ in their chemical composition, which can influence which type is most appropriate to collect for specific analyses \cite{Liang2023}. \citet{sempionatto_eyeglasses-based_2019} in particular argue that tears are a good biofluid for analysis since they are noninvasive, less complex than blood yet still contain a "variety of biomarkers", and they "reflect concurrent blood levels" because of passive leakage of compounds from blood plasma.
It is important to note, however, that neither of these works (\cite{ruiz-valdepenas_montiel_decentralized_2021, sempionatto_eyeglasses-based_2019}) provide comparisons to the in-body status of their target micronutrients as measured by optimal reference standard, urine or blood-based assays. Additionally, it is important to consider the potential ethical implications of tear inducement and collection. One study used Schirmer strips, a common method for collecting tears in healthcare settings, and found that most participants considered the process acceptable \cite{Quah2014Patient}. Specifically, 70\% did not mind the procedure, and 74\% preferred tear collection over venous blood sampling or other forms of biofluid collection such as urine.

Sweat receives significant attention in emerging micronutrient detection methods, especially in wearables. The composition of collected sweat varies depending on whether it is produced actively (through exercise or heat exposure) or passively (through methods that encourage sweat production without physical activity, such as sweat patches or chemical stimulation) \cite{Klous2020}. Active sweating in particular usually produces higher concentrations of sodium, chloride, and metabolites.
\textit{Eccrine sweat} is a clear, odorless fluid secreted by eccrine sweat glands, which are essential for thermoregulation \cite{Hodge2022}. It is the primary sweat type used for micronutrient analysis, as \textit{apocrine sweat}—the other major type—is thicker and predominantly composed of lipids and proteins, making it less suitable for this purpose \cite{Heikenfeld2023}.
Therefore, when we discuss sweat, we refer to eccrine sweat, which is mostly sodium and chloride, with smaller amounts of micronutrients and metabolites (at the micro and nanomolar scale), similar to or smaller than their concentrations in blood plasma \cite{baker_physiological_2020}. Micronutrients found in sweat include potassium, calcium, magnesium, iron, copper, zinc, vitamin C, and vitamin B1. Out of the compounds in sweat, mostly sodium and chlorine ions are well-studied. Some research has explored water-soluble vitamins in sweat, such as vitamin C and vitamin B1, but there is no such attention on fat-soluble vitamins. 
% Some metabolites include glucose,
% % (concentration in blood is 100x lower)
% lactate, ammonia, urea, bicarbonate, amino acids, ethanol, and cortisol. 
Many confounding factors can impact sweat composition, such as
% whether the sweating is passive or active (see above). Active sweating usually produces higher concentrations of sodium, chloride, and metabolites. 
the contamination of sweat by skin-derived substances, notably iron \cite{baker_physiological_2020}. This is combated by pre-rinsed skin, removal of initial sweat (concentrations stabilize after 20 to 30 minutes of sweating), and the analysis of cell-free sweat. The region and method of collection can also have an impact, as many micronutrient concentrations can vary two to four times depending on the region. Finally, sweat can be reabsorbed into the body. Skin temperature and the flow rate of sweat both impact the rate of this reabsorption. 

The clinical literature notes a general "lack of association between dietary micronutrient intake and corresponding sweat micronutrient concentrations" \cite{baker_physiological_2020}. This lack of an association exists in comparison to blood as well. A review by \citet{baker_physiological_2020} finds that there is no established correlation between sweat and blood composition, and there is "little support for using sweat as a surrogate for blood". 
Concentrations of minerals in sweat are much more varied than in plasma, likely because minerals bind to carrier proteins in blood. The review also reports little to no correlation between sweat and blood concentrations of vitamin C and iron status. This finding for iron is echoed in another paper that found both iron and calcium have no correlation between sweat and blood concentrations \cite{baker_physiology_2019}. However, it was reported in the same paper that iron concentrations in sweat have been observed to be lower in anemic patients and higher in patients undergoing iron therapy. 
% We further discuss sweat as a tool for micronutrient detection in Section \ref{tech section}.
% We did not find similar work from the clinical perspective of the composition of micronutrients in saliva and tears.

Finally, it is important to consider the potential impact of time lag variability in biomarker expression across different body fluids. This variability arises from nutrient kinetics and bioaccessibility, particularly in relation to absorption, distribution, metabolism, and excretion (ADME) within the human body \cite{Mengucci2020, Gonzalez2023}. However, studies described in this section do not explicitly account for these temporal effects, which are crucial for accurately correlating blood biomarkers with those present in other biofluids. At present, the temporal dynamics governing these relationships remain insufficiently understood, and require further investigation.

% \subsection{Physiological Effects Associated with Micronutrient Imbalance}
 \subsection{Micronutrient Effects on Physiological Processes} \label{subsec_phys symptoms}
% Micronutrient imbalance is a critical issue. Imbalance can be caused by deficiency or an excess of micronutrient intake. Dietary guidelines for intake vary across organizations and countries, and are based on age and sex. There is no one established guideline and vary because of how individualized micronutrient metabolism is, as well as the diversity of micronutrients, their interactions, and specific values in which they are needed. In the US, dietary reference intakes are used, which most commonly consists of the Recommended Dietary Allowance (RDA). This represents the "average daily level of intake sufficient to meet nutrient requirements of 97-98\% of healthy individuals" \cite{NIH_Nutrient_Recommendations}. The National Institutes of Health defines more terms to describe intake levels such as an upper limit, highlighting the potential for excess intake. Some guidelines such as Adequate Intake exist when there is not enough evidence to establish stronger guidelines, further emphasizing the uncertainty of micronutritional guidelines.

%\subsubsection{Micronutrient deficiency}
Most of the physiological effects of micronutrients are related to the autonomic functions of the body. Autonomic functioning refers to bodily functions controlled by the autonomic nervous system, which regulates involuntary processes and rhythms such as breathing, heart rate, and digestion \cite{waxenbaum_anatomy_2023}.
The specific importance of micronutrients to the proper functioning of the nervous system has been documented by several papers \cite{georgieff_nutrition_2007, beitzke_autonomic_2002, chiu_impact_2021}.  For example, a study of vitamin B12 deficiency compared responses to a 60-degree passive head up tilt test between a control group, a vitamin B12 deficient group, and a group with diabetes mellitus \cite{beitzke_autonomic_2002}. They found that the deficient group had comparable autonomic neuropathy to the diabetic group. Our exploration of observable impacts of deficiencies on autonomic functioning found three main affected areas relating to biofluid composition and physiological effects: general symptoms, cardiac function, and sleep.

\subsubsection{General Symptoms}
%\subsubsection{Physiology of Micronutrient Imbalances}
Deficiencies can be classified as either clinical or subclinical based on their severity \cite{tardy_vitamins_2020}. Most deficiencies result in symptoms of general fatigue, lethargy, irritability, muscle pain, weakness, and headaches. Clinical deficiencies often have more distinguishable symptoms, while subclinical deficiencies are limited to the above non-specific ones. Deficiencies of vitamin C, B vitamins, iron, magnesium, and zinc have been linked to fatigue more so than others \cite{tardy_vitamins_2020, azzolino_nutritional_2020}. While energy and fatigue are more subjective and can can rely on subject-reporting, there are some established and validated assessment methods such as the Multidimensional Fatigue Inventory \cite{smets1995multidimensional} and the SF-36 Vitality Scale \cite{ware2001sf}. A comprehensive reference for physiological symptoms associated with deficiency is lacking in literature, so we provide one in Tables \ref{tab:phys-vitamins-water-soluble}, \ref{tab:phys-vitamins-fat-soluble}, and \ref{tab:phys-minerals} within the Appendix.

\subsubsection{Cardiac Function}\label{Nervous-discussion}
One main aspect of autonomic functioning affected by nutrient imbalance is cardiac functioning. Several studies explore the interaction between heart rate variability (HRV) and micronutrient deficiencies. Components of HRV are associated with parasympathetic (PNS) and sympathetic nervous system (SNS) activity \cite{shaffer_overview_2017}. High-frequency (HF) bands reflect PNS activity and correspond to the respiratory cycle, while low-frequency (LF) bands reflect PNS, SNS, and baroreceptor activity. Vitamin B12 deficiency is one of the most documented in terms of impact to HRV, with evidence that it lowers HRV overall, impacting sympathetic indices the most \cite{aytemir_assessment_2000, sozen_autonomic_1998, lopresti_association_2020, beitzke_autonomic_2002}. Supplementation of B12 was also demonstrated to return HRV indices to a comparably normal state \cite{aytemir_assessment_2000}. Deficiency of vitamin D was found to lower HRV as well \cite{lopresti_association_2020}. Calcidiol (25(OH)D) levels, a form of vitamin D, was shown to be associated with the ratio of LF to HF HRV power \cite{mann_vitamin_2013}. This metric is sometimes called sympathovagal balance, and is intended to be a measure of 'balance' between SNS and PNS activity, but there has been debate over this interpretation \cite{shaffer_overview_2017}. Iron-deficiency anemia (IDA), an advanced form of iron deficiency, has more conflicting evidence of HRV impacts, with some studies finding no difference in HRV indices versus the control \cite{tuncer_variabilidade_2009} while others were able to find a difference in the IDA group \cite{jibhkate_assessing_2019, yokusoglu_altered_2007}. 

Impacts of micronutrients on blood pressure have also been studied \cite{chiu_impact_2021, beitzke_autonomic_2002, toru_autonomic_1999}. The supplementation of potassium, magnesium, zinc, vitamins C, D, B6, and a decreased intake of sodium and selenium can "positively modulate blood pressure levels" \cite{chiu_impact_2021}. The aforementioned study involving responses to a head up tilt test in vitamin B12 deficient people found a drop in systolic blood pressure 60 beats after the test \cite{beitzke_autonomic_2002}. This finding aligns with previous work suggesting that a dip in blood pressure when standing up from sitting or lying down is a symptom of vitamin B12 deficiency \cite{toru_autonomic_1999}. 
% While not explicitly cardiac-related, this same study found an increase in sympathetic skin response, a measure of the change in voltage of the skin when electrical stimulation is applied, after treatment for vitamin B12 deficiency. 

As an aside, pulse-oximiters, smartwatches, and other health sensors or even smartphones can readily measure continuous cardiac function through photoplethysmography (PPG). Pulse rate (PR), HRV, and blood pressure can be derived from PPG \cite{mousavi_blood_2019, liu_assessment_2021}.

\subsubsection{Sleep}
Another area of research is the role of micronutrient status in sleep. Sleep duration is associated positively with iron, zinc, and magnesium and negatively with copper, potassium, vitamin A and vitamin B12 levels \cite{ji_relationship_2017, beydoun_serum_2014}. Sleep quality increases with zinc, magnesium, and vitamin B9 status and is negatively associated with vitamin B12 status \cite{cherasse_dietary_2017, jahrami_association_2021, beydoun_serum_2014}. There are conflicting findings for iron. One study reports that iron status is not proven to be correlated with sleep quality \cite{ji_serum_2017}, while another claims that supplementation had positive effects on sleep disorders \cite{leung_iron_2020}. Sleep deprivation is also connected to micronutrition through its influence on hormones that regulate stress and the immune system. Sleep deprivation can decrease levels of cortisol while increasing ghrelin, a hormone linked to hunger \cite{kim_impact_2015}. This shift can lead to increased hunger, and a continuous lack of sleep was even found to have a positive correlation with obesity (which is interconnected with malnutrition \cite{WHO_malnutrition}). Studies also found that after temporary sleep deprivation there were decreased magnesium levels measured via red blood cell testing and reduced zinc levels in plasma tests \cite{lopresti_effects_2020}.

\section{Biofluid Analysis Methods for Micronutrient Status Assessment}\label{biofluids section}

This section describes how biofluid analysis has been leveraged to assess micronutrient status. These assessment methods target particular biomarkers within a biofluid that are indicators of micronutrient status. Reliable biomarkers are a research challenge themselves (which AI techniques may address \cite{cote_artificial_2022}), but clinical literature suggests that micronutrient biomarkers are more established and specific than macronutrient biomarkers \cite{elmadfa_developing_2014}. 
% Several different biomarkers exist for each macronutrient in the body, thus making general macronutrient assessment from biomarkers challenging. 
Most micronutrients have one or two specific biomarkers associated with their circulating status that are considered to be the \textit{optimal reference standard} for status assessment (Table \ref{tab: assessment}). Although some optimal reference standards are still debated, their existence makes the evaluation of novel assessment methods more straightforward. 

In this section we will discuss clinical biochemical analysis, followed by other innovative technologies.
When reviewing the non-clinical methods (Sections \ref{subsec_assays} to \ref{subsec_analytic}), we particularly note what each approach claims to assess, the method by which they conduct this assessment, how their method was evaluated, and the clinical relevance of their implementation. 
Critical to accessibility is each method's platform. Point-of-care (PoC) devices are compact and portable enough to be deployed for use where needed, 
% (e.g. at the bedside, in the field, or in the patient’s home),
as opposed to traditional `benchtop’ technologies that are restricted to a laboratory setting. The term PoC is widely applicable, so 
% to a wide array of devices and platforms, so
% that can be varying degrees of accessible, 
we label methods more specifically to compare 
% opt for more specific descriptors 
% with the aim of differentiating 
the advantages and disadvantages of each, while still acknowledging that they are considered PoC. Specifically, the terms `portable', `smartphone-based', and `wearable' all imply PoC devices but also suggest varying levels of accessibility and ubiquity.

% the terms `portable’, ‘smartphone-based’, and ‘wearable’ are all PoC, but carry with them additional connotations for how accessible, and pervasive, these platforms are. 
% Additionally, in the methods we discuss, some are described as ``low-cost" based on how they are characterized in their original reference. We do not apply a standardized cost metric, and use the term ``low-cost" only when it is explicitly stated by the original authors.

We further expand on clinical relevance by noting that this includes the target biomarker, the assessment method, and the concentrations of that biomarker that are evaluated. A clinically relevant method should closely align with the optimal reference standard on these factors (Table \ref{tab: assessment}). To aid future work that may wish to integrate or innovate on a particular method, we explicitly mention when a study does not demonstrate this agreement (e.g. assessing RBP for vitamin A status) and/or what assessment methods were used during evaluation (e.g. ELISA). 
For micronutrients with rapid turnover (i.e. water-soluble vitamins) or without risks of excess, a lack of sensitivity in the upper spectrum of the clinically relevant concentrations (Table \ref{tab: assessment}) are not a major limitation for methods which primarily aim to identify deficiency. Regardless, such a limitation is still noted for completeness.
Last, we provide tables that group together similar works and summarize the pertinent details and quantitative results of their evaluation, if available. For conciseness, this information is omitted from the body of the text, and we encourage readers to instead refer to the relevant tables.

% \subsection{Anthropometric Measurements}
% These measurements include weight, BMI, skinfold measurements, body composition, bioelectrical impedance analysis (BIA), creatinine height index, and more \cite{reber_nutritional_2019}. These measurements, however, vary based on demographics, and BIA in particular is highly affected by hydration status \cite{elmadfa_developing_2014}. Generally, this aspect rarely provides direct insights into micronutrient status.
% \subsection{Direct Assessments}
%\subsection{Biofluids Analysis Methods}

\subsection{Clinical Biochemical Analysis} \label{clinical/biochemical analysis}
Clinical biochemical analysis involves laboratory testing of biomarkers found in urine, blood, or other biosamples \cite{reber_nutritional_2019}. Results can be influenced by several factors and need to be interpreted in the context of other aspects of the patient's health. Additionally, biochemical testing is often time- and resource-intensive. Despite this, biochemical analysis describes clinical optimal reference standard methods for quantifying the circulating micronutrient status in the body \cite{zhang_review_2018, holler_micronutrient_2018}. These methods can be roughly separated into various types of assays, and liquid chromatography (LC)-coupled spectroscopy.

One type of assay which is popular for clinical micronutrient assessment is a microbiological inhibition assay. These assays work on the principle that specific micronutrients are needed for the growth of certain bacteria, and this growth can be measured to indicate the amount of a micronutrient present in a sample \cite{zhang_review_2018}. Microbiological inhibition assays were previously the widely-accepted optimal reference standard, but improvements in LC and spectroscopy highlight their relatively poor precision and accuracy. These flaws have relegated microbiological inhibition assay methods to be used mostly during screening or in resource-constrained testing, except for some micronutrients, where they remain the standard. Another common type of assay applied in studies is the enzyme-linked immunosorbent assay (ELISA). An ELISA test is used for measuring antibodies in blood, and is a useful clinical screening tool for further testing \cite{kinman_elisa_2012}. While not the optimal reference standard for the assessment of micronutrient status, ELISA tests can yield valuable data for studies that are time or resource limited. 
% For this reason, the below methods that include comparison to an ELISA test are explicitly noted, as this reflects the authors’ effort to benchmark their approach against a widely accepted and clinically relevant standard. 
Lastly, we find that antibody or immunoassays as well as colorimetric and fluorometric assays are popular in emerging accessible and non-invasive assessment technologies.

Modern optimal reference standards overwhelmingly apply LC-coupled spectroscopy \cite{zhang_review_2018, holler_micronutrient_2018} (Table \ref{tab: assessment}). LC is defined as "a separation process used to isolate the individual components in a mixture" \cite{chemyx_basic_2017}. High-performance LC (HPLC) uses pressure to facilitate the separation process, reducing the time required. It is commonly coupled with mass spectroscopy (MS) in optimal reference standard approaches \cite{zhang_review_2018, holler_micronutrient_2018}. Spectroscopy is the "investigation and measurement of spectra produced by matter interacting with or emitting electromagnetic radiation" \cite{noauthor_spectroscopy_2019}. Every molecule reacts to the applied radiation in a unique way that allows us to "detect, determine, or quantify the molecular and/or structural composition of a sample" \cite{noauthor_spectroscopy_2019}. MS is the most important subfield of spectroscopy to understand for biochemical analysis. It measures the mass-to-charge ratio of the molecules in a sample as a way to determine and quantify the composition of molecules in the sample \cite{noauthor_spectroscopy_2019}. This is done by vaporizing the molecules in a sample into gas-phase ions, which are then sorted by their mass-to-charge ratios. We will discuss other forms of spectroscopy and their utility for micronutrient status assessment further in Section \ref{subsec_spec}.

Matrices are the biosamples that are the subjects of the aforementioned methods of analysis. Most are blood based, but in a few cases urine is used in the optimal reference standard (mostly for water-soluble vitamins) \cite{holler_micronutrient_2018}. Blood matrices are whole blood, washed red blood cells, plasma, and serum. Whole blood is blood as it is from the vein (venous blood). Washed red blood cells are red blood cells that have been separated from the other components of blood such as plasma, platelets, and white blood cells \cite{keir_washed_2016}. Plasma is obtained by adding an anticoagulant to whole blood and placing it in a centrifuge \cite{cell_guidance_systems_serum_2021}. Serum is obtained similarly to plasma, except the blood is allowed to clot before centrifuging. However, blood can also be obtained from the capillaries (capillary blood) as opposed to the vein, often via a finger prick. Capillary blood is often easier and cheaper to obtain, especially via untrained personnel, but contains a mix of venous and arterial blood, together with interstitial fluid which surrounds cells in the body \cite{royal_validity_2022}. Because of this, test results on capillary vs venous blood can differ (e.g. hemoglobin concentrations are higher in capillary blood) and so the two should not be considered interchangeably.

%Please add the following packages if necessary:
%\usepackage{booktabs, multirow} % for borders and merged ranges
%\usepackage{soul}% for underlines
%\usepackage[table]{xcolor} % for cell colors
%\usepackage{changepage,threeparttable} % for wide tables
%If the table is too wide, replace \begin{table}[!htp]...\end{table} with
%\begin{adjustwidth}{-2.5 cm}{-2.5 cm}\centering\begin{threeparttable}[!htb]...\end{threeparttable}\end{adjustwidth}
% \renewcommand{\arraystretch}{1.5}
\setlength{\extrarowheight}{5pt}
% \begin{table}[!htp]\centering
\tiny
\begin{longtable}{p{0.4in}p{0.4in}p{1.2in}p{0.6in}p{1in}p{0.8in}p{0.5in}}
\caption{Optimal Reference Standard Methods of Assessing Micronutrient Imbalance. Information from \cite{holler_micronutrient_2018, zhang_review_2018, mueller_aspen_2017, berger_espen_2022, national_institutes_of_health_dietary_nodate}.}\label{tab: assessment}
\\ \toprule
Micronutrient &Method &Biomarker &Matrix &Intervals &Impact of Inflammation & Approx. Cost (US; Walk-In Lab)\\\midrule
\endfirsthead 
\\ \toprule
Micronutrient &Method &Biomarker &Matrix &Intervals &Impact of Inflammation & Approx. Cost (US; Walk-In Lab)\\\midrule
\endhead
\hline \multicolumn{7}{r}{{Continued on next page}} \\ \hline
\endfoot
\bottomrule
\endlastfoot
Vitamin B1 &Erythrocyte transketolase activity coefficient assay &Increase in erythrocyte transketolase activity &Washed red blood cells &Deficient: >25\%; Insufficient: 15-25\%; Sufficient: <15\% &None on direct plasma levels &\$65\\
Vitamin B2 &Erythrocyte glutathione reductase activity coefficient assay &Increase in erythrocyte glutathione reductase activity &Washed red blood cells &Deficient: >40\%; Insufficient: 20-40\%; Sufficient: <20\% &Decrease in plasma levels (erythrocyte assays are more stable) &\$120\\
Vitamin B3 &LC-MS/MS &Niacin metabolites (NMN and 2-pyr, limited representation of stores and recent intake) &Urine &Deficient: <5.8; Insufficient: 17.5-5.8; Sufficient: >17.5 mcmol/day &Lack of evidence &\$179\\
Vitamin B5 &LC-MS/MS &Pantothenic acid (requires enzyme pretreatment) &Whole blood &Deficient: <0.22; Sufficient: 0.35-0.59 mg/L &Lack of evidence &\$149\\
Vitamin B6 &HPLC or LC-MS/MS &pyridoxal phosphate (PLP) &Plasma or serum &Sufficient: >4.94 or >7.41 $\mu$g/L plasma PLP (varies by source) &Decrease in plasma PLP, no effect on RBC concentration &\$69\\
Vitamin B7 \cite{eng_identification_2013} &LC-MS/MS for biotin; gel densitometry for MCC/PCC \cite{mall_biotin_2010} &Biotin (less sensitive); holo-MCC and holo-PCC (only reliable markers) &Urine for biotin; WBCs for MCC/PCC &
Sufficient: 4.4-31 $\mu$g/day urinary biotin, 8.2 arbitrary units holo-MCC, 9.1 arbitrary units holo-PCC &None on biomarkers &\$199\\
Vitamin B9 &LC-MS/MS &Folate &Serum for altered exposure and recent intake; Red blood cell for long term/3 month status and storage levels &Sufficient: >3 ng/mL serum, >140 ng/mL RBC &Lack of evidence &\$29\\
Vitamin B12 \cite{adas_medical_knowledge_team_vitamin_2023}&GC-MS &B12, confirmed with methylmalonic acid (MMA, also related to B2, B6, folate); no single 'optimal reference standard' &Plasma or serum &Deficient: <200-250 pg/mL B12, >0.03 mg/L MMA (some debate over this) &Association with increased B12 levels &\$35\\
Vitamin C &HPLC &Ascorbate &Plasma (some claim serum should be avoided) &Deficient: 1.94; Insufficient: 2.11-4.05; Sufficient: 4.05 mg/L &Decrease in plasma ascorbic acid (rapid, decrease when CRP >10 mg/L, normal values not detected if CRP >40 mg/L) &\$49\\
\cmidrule{1-7}
Vitamin A &LC-MS/MS &Retinol (only sensitive to deficiency or excess in storage, affected by infection and protein/zinc deficiency). Best method is to indirectly measure reserves in liver over several days of administration &Plasma or serum &Severely deficient: <0.1; Deficient: 0.1-0.2; Sufficient: 0.3-1; Toxic: >1 mg/L retinol &Decrease in serum retinol (adjustment equations exist but are not universally applicable, e.g. BRINDA \cite{luo_using_2016}) &\$58\\
Vitamin D \cite{ross_dietary_2011}&LC-MS/MS &25(OH)D (calcidiol) &Plasma or serum &Deficient: <12; Insufficient: 12-<20; Sufficient: 20-50; Toxic: >50 ng/mL (not definitively established/linked to clinical outcomes, varies based on assay and lab) &Decrease in plasma levels (all values below reference ranges with CBP >40 mg/L) &\$59\\
Vitamin E &LC-UV &Ratio of Vit E to total blood lipids &Plasma or serum &Insufficient: <5.17 mg/L Vit E, <0.8 mg Vit E/g total lipid; Sufficient: 8.6-13 mg/L Vit E (adults have higher levels) &Some effects (blood concentrations less interpretable at CRP >80 mg/L) &\$48\\
Vitamin K &Immuno-based assays &Plasma phylloquinone (usually for short term intake, no single `optimal reference standard'); prothrombin time (time to blood clot, only clinically relevant measure); variety of other 'functional' biomarkers &Plasma &Deficient: <0.15; Sufficient (fasting): 0.15-1 $\mu$g/L &Status associated with lower inflammatory marker concentration &\$96\\
\cmidrule{1-7}
Iron \cite{gloria_iron_nodate,gerber_iron_2024}&Electrochemi-luminescence immunoassay (ECLIA) &Ferritin for deficiency (first phase, evaluates storage, inflated by infection); Iron increase after supplementation for malabsorption; Hemoglobin used to confirm IDA &Serum &Iron-deficiency anemia: <10, Deficiency: 10-30 $\mu$g/L ferritin &Ferritin may be inflated, falsely normal/misleading (adjustment equations exist but are not universally applicable, e.g. BRINDA \cite{luo_using_2016}) &\$29\\
Copper &ICP-MS &Copper or ceruloplasmin (CP), neither reliable &Serum &Depletion: <50.8 (copper); Deficient: 50.8-76.2 (copper, high CRP); Sufficient: 63.5-158.9 $\mu$g/dL (copper), 180-400 mg/L (CP) &Increase in plasma concentrations &\$33\\
Zinc \cite{pat_bass_symptoms_2023}&Atomic Absorption Spectroscopy (AAS) &Zinc (levels are halved by Systemic Inflammatory Response Syndrome, can be normal with clinical symptoms present, levels vary with time of day so it is recommended that albumin and CRP changes are taken into account) &Plasma or serum &Deficient: 70 women, 74 men; Insufficient: 70/74-80; Sufficient: 80-120 $\mu$g/dL &Decrease in plasma levels (significant when CRP exceeds 20 mg/L, adjustment equations exist but are not universally applicable, e.g. BRINDA \cite{luo_using_2016}) &\$38\\
Iodine &ICP-MS &Iodine &Urine (24h or random), serum less recommended &Depletion: <20, NA; Deficient: 20-100, <40; Sufficient: 100-300 $\mu$g/24hr urine, 40-100 $\mu$g/L serum (levels should be higher in those who are pregnant or lactating) &Lack of evidence &\$89\\
Selenium &AAS, ICP-MS &Selenium (recent intake) or Selenoprotein P &Plasma or serum &Deficient: <60; Sufficient: >60; Toxicity: >474 to 948 $\mu$g/L Se (intervals vary by source and population: women and black people have naturally lower concentrations) &Decrease in plasma levels proportional to inflammation, can be adjusted for  &\$99\\
Magnesium &AAS &Magnesium &Serum (little correlation with overall status or tissue stores) and urine (after supplementation) &Deficient: <18.23; Sufficient: 18.23-23.1 mg/L serum Mg & &\$28\\
\end{longtable}
\normalsize

\subsection{Assay-Based Technology} \label{subsec_assays}
% A summary of assay-based methods is found in Table \ref{tab: emerging-assay}. 

% \subsubsection{Background}
%\paragraph{Background} 
% \subsubsection{Lateral Flow Assays} 
One of the largest areas of work applicable to micronutrient status explored by this review is quantitative assays. Although we cover many different types of quantitative assays, a comprehensive review of lateral flow quantitative assays by \citet{urusov_towards_2019} provides a robust background for how these devices work. Lateral flow (immunochromatographic) assays indicate that a target compound is either present in the sample or present in excess of a particular threshold, usually via staining on the test membrane. These types of assays are useful for tests that benefit from quick conclusions (e.g. pregnancy tests). 
% Lateral flow quantitative assays simply extract quantitative information about the target analyte from the lateral flow assay itself.
%\paragraph{Extracting Quantitative Information} 
The most prevalent approach for extracting quantitative information from lateral flow assays is via optical signal registration \cite{urusov_towards_2019}. This method involves the analysis of absorbed and reflected light from the test surface and the staining upon it, similar to the practice of spectrophotometry. The test/control ratio is a common metric used to quantify the magnitude of this staining relative to a control or reference area. Some commercially-available devices are limited to automatically confirming the presence of the test line, while others can use line intensity to calculate analyte content. \citet{urusov_towards_2019} note that portability has become a recent focus in this market, and this is not just limited to specialized devices in a portable platform. Smartphones have been successfully used for optical signal registration, even with fluorescent labels that decrease detection limits. Some manufacturers provide their own smartphone apps for quantitative analysis, but controlling for lighting and positioning is a challenge. To address this issue, another approach is the use of a standardized scanning device to collect image data and the off-device analysis of the image by specialized software. More experimental approaches, such as magnetic and electrically conductive labels, have also surfaced but have yet to mature. Tables \ref{tab: emerging-assay-sweat}, \ref{tab: emerging-assay-smartphone}, and \ref{tab: emerging-assay-commercial} present a summary of assay-based methods developed for micronutrients.

\begin{figure}
    \centering
    \includegraphics[width=0.95\linewidth]{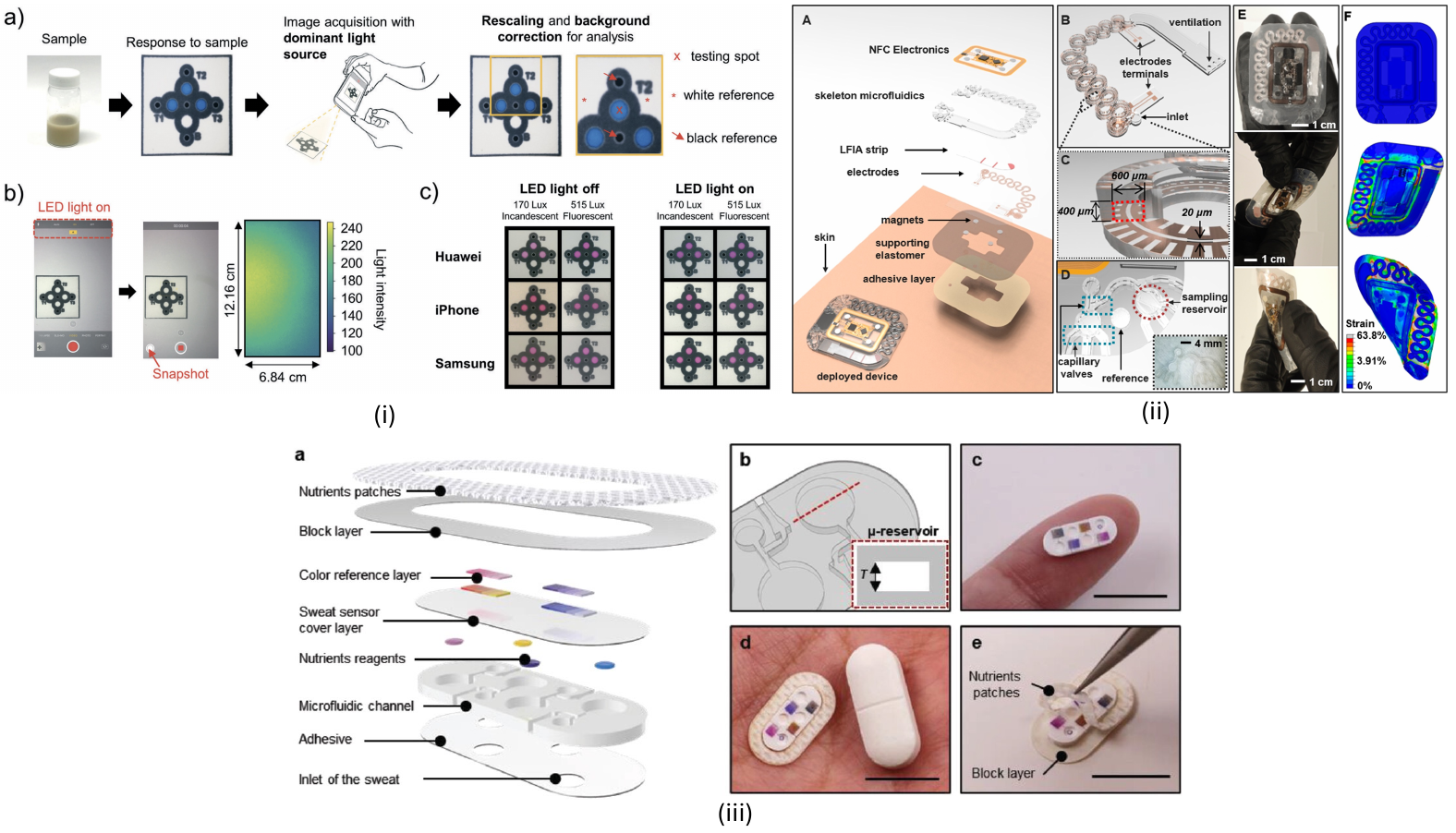}
    \caption{(i) A smartphone camera and LED flash can be used to analyze results from a colorimetric assay. Used with permission from \citet{kong_accessory-free_2019}. (ii) The design of a multiplexed sweat sensor using microfluidics. Used with permission from \citet{kim_soft_2020}. (iii) The design of a colorimetric sweat sensor using microfluidics. Used with permission from \citet{kim_skininterfaced_2022}.}
    \label{fig:sweat_figures}
    \Description{(i)(a) Graphic describing the process of taking a sample, adding it to the test area of the assay, capturing an image of the assay's response to the sample with a smartphone, then analyzing the image using the test spots and black/white reference areas embedded into the test.
    % [from left to right] a vial labeled "sample", a right arrow; the testing surface with visible changes labeled "response to sample"; a right arrow; a smartphone pointed at the test surface to take an image labeled "Image acquisition with *dominant light source*"; a right arrow; a close up of the test area, indicating where the testing spots, white reference, and black reference exist on the area labeled "*Rescaling* and *background correction* for analysis" 
    (b) A smartphone screen is shown taking an image of the test area with the flash on. Beside, a heatmap of light intensity from the captured image.
    % [from left to right] screen capture of a smartphone camera app with the test area in view, a red box indicates that the smartphone's LED flash is on; a right arrow; screen capture of 
    (c) A comparison of images of the test area, taken with Huawei, iPhone, and Samsung brand smartphones, with the phone's LED flash one or off, under two external lighting conditions.
    (ii)(a) Exploded-view of the skin-interfaced, multiplexed assay showing its electronic componentry.
    (b) Labeled diagram of the skeleton microfluidics and electrodes. The sample enters through an inlet, travels through the microfluidic channels to interact with the electrodes, then exits out a ventilation port.
    (c) Labeled diagram showing the tolerances of the skeleton microfluidics and electrodes.
    (d) Labeled diagram of the device's integrated LFIA strip, showing the capillary valves, sampling reservoir, and reference area.
    (e) Three images of the device being handled and bent.
    (f) Corresponding strain maps of the device being handled and bent.
    (iii)(a) Exploded-view of a skin-interfaced colorimetric sweat sensor, highlighting the channels, chemical reagents, reference areas, and nutrient supplementation patches.
    (b) Labeled diagram showing the shape and size of a microfluidic reservoir.
    (c) The sweat sensor is shown fitting easily on someone's fingertip.
    (d) The sweat sensor held in someone's palm, next to a pill for scale.
    (e) The nutrient patch is pulled back to reveal the block layer between the assays and the supplementation patch.
    For more information on each figure, please refer to their respective sources.
    }
\end{figure}

\subsubsection{Sweat-Based Colorimetric Sensors} Colorimetrics uses reagents to react with an analyte and change color to indicate concentrations of target substances. This color can be analyzed by a camera, such as in a smartphone as demonstrated in Figure \ref{fig:sweat_figures}(i) by \citet{kong_accessory-free_2019}. 
Most sweat-based sensors use microfluidic devices for the collection of sweat \cite{jo_review_2021}. These devices use very tiny valves and channels to "capture and store sweat on the surface of the subject’s skin" via the natural pressure of sweat glands \cite{jo_review_2021}. \citet{sekine_fluorometric_2018} demonstrated a microfluidic skin patch with fluorometric probes that aimed to analyze chloride, sodium, and zinc in sweat. 
% We are most interested in the implications of their sensor for zinc assessment, since this is a trace mineral. 
A smartphone attachment was designed to take light from the camera’s flash and pass it through an excitation filter, allowing only a particular wavelength through. 
% This excites the fluorometric probes, which emit light that can be captured by the smartphone camera. An emission filter is used to only let fluorescence emission wavelengths through. 
Similar to other quantitative assays, normalized intensity can be calculated against a reference and used to determine biomarker concentration and calibration curves for each nutrient were determined by known concentrations in spiked synthetic sweat.
% (6.54-196.2 $\mu$g/dL for zinc, which covers the physiological range)
The authors claimed a strong correlation and accuracy, but do not report exact results.
Field studies were conducted where sweat was induced in human volunteers (undisclosed sample size). Measured concentrations were compared to ion chromatography (chloride), atomic absorption spectrometry (AAS; sodium), and inductively coupled plasma mass spectrometry  (ICP-MS; zinc). Again, no statistical analysis of results is presented and it appears zinc measurement was the most inaccurate and subject to the most variance.

A paper by \citet{kim_skininterfaced_2022} described on-body colorimetric measurement of vitamin C, calcium, zinc, and iron using sweat as a biofluid (Fig. \ref{fig:sweat_figures}(ii)).
% Again, the micronutrients vitamin C and zinc are the main focus of our analysis.
A bespoke colorimetric assay was used for each micronutrient, assessed with known concentrations in a buffer solution. Since temperature and pH can affect colorimetric results, the authors conducted tests within the normal range of body temperature and pH in sweat, finding only a slight shift in results. Uniquely, micronutrients could be supplemented transdermally through the patch itself. Multiple on-body tests were conducted with 7 volunteers (4 male). Sweat was induced by a sauna before and after supplementation (either orally or transdermally). Patch-based measurements were found to be correlated with ICP-MS results. Although the paper claimed that "sweat chemistry correlates, at least semiquantitatively, to plasma chemistry" for these micronutrients \cite{kim_skininterfaced_2022}, this claim is based on the time dynamics of concentrations after supplementation rather than a comparison to an optimal reference standard status assessment (Table \ref{tab: assessment}). Furthermore, the clinical literature points out that that vitamin C and iron concentrations in sweat have been shown to have little to no correlation with levels in blood \cite{baker_physiological_2020, baker_physiology_2019}. Therefore, we argue that this device provides insights into the rate of excretion of these micronutrients rather than their status. Lastly, there is no analysis of possible measurement bias induced by transdermal supplementation at the point of sweat collection. 

\subsubsection{Sweat-Based Multiplexed Sensors} Multiplexed analyses combine and analyze data from assays with multiple sensors such as electrocardiogram (ECG), temperature, electrodermal activity (EDA), HRV data and more. Thus, these multiplexed sensors can enable more complex physiologic monitoring and diagnosis. For example, EDA measures the change in skin conductance caused by sweat, an indicator of nervous system arousal \cite{posada-quintero_innovations_2020}, which can be associated with micronutrient status (Section \ref{subsec_phys symptoms}). 

\citet{kim_soft_2020} developed an on-body biosensing platform that could collect and analyze cortisol, glucose, and vitamin C in sweat using microfluidics (Fig. \ref{fig:sweat_figures}(iii)). The device included a lateral-flow assay for cortisol and fluorometric assays for glucose and vitamin C. Assay results were imaged with a smartphone (with special lenses in the case of fluorometry) and were analyzed to yield quantitative results. There were also electrodes for sweat rate and EDA.
% with Near-Field Communication (NFC) and Radio Frequency (RF) to power them and communicate results. 
The focus in the paper was on stress indicators. Field-testing of the device for cortisol assessment involved subjecting 2 participants to “intensive work periods” (interrupted sleep schedule and caffeine intake) for 7 days then rest (regular schedule) for 14 days \cite{kim_soft_2020}. Additional tests with 2 different participants for all target biomarkers involved subjecting participants to intensive work followed by a regular schedule and vitamin C supplementation for 14 days.
Spikes in vitamin C associated with intake could be observed under these conditions.

% The authors also claim they were able to assess changes in circadian rhythms through cortisol, but no trends in glucose were found. 

For a more in-depth review of the nuances of creating wearable multimodal sensors with sweat collection and analysis capabilities, we direct the reader to \cite{yokus_integrated_2021}.

\begin{table}[!htp]\centering
\caption{Assay-Based Methods Using Sweat}\label{tab: emerging-assay-sweat}
\tiny
\begin{tabular}{p{0.5in}p{0.6in}p{0.4in}p{0.5in}p{0.5in}p{0.6in}p{0.75in}p{0.6in}p{0.25in}}\toprule
% \multicolumn{9}{c} {Assay-Based Methods} \\
Method &Platform&Targets &Analytes &Evaluation &Control &Results &Notes &Source
\\\midrule
% Sweat-based colorimetric sensors
Fluorometry &Wearable patch &Chloride, sodium, and zinc &Sweat &Human subjects&Ion chromatography (chloride), AAS (sodium), ICP-MS (zinc) &Zinc measurement was the most inaccurate and subject to the most variance &Statistical analysis of results is not presented &\cite{sekine_fluorometric_2018} \\
Colorimetry &Wearable &Vitamin C, calcium, zinc, and iron &Sweat &Human subjects &ICP-MS on diluted sweat samples and supplementation &Correlations with ICP-MS of 0.926 for Vit C, 0.743 for calcium, 0.895 for zinc, and 0.963 for iron, time dynamics of measurements after supplementation were in line with those of blood &Sensor can also supplement micronutrients transdermally &\cite{kim_skininterfaced_2022} \\
% Sweat-based multiplexed sensors
Fluorometric and lateral flow assays &Wearable with smartphone analysis &Cortisol, glucose, and vitamin C &Sweat &Human subjects &ELISA for cortisol, controlled stress and diet &Cortisol aligned with circadian rhythm changes and had $R^2$ of 0.7974 with ELISA, observed spikes in Vit C with intake, no trends in glucose &Measurement of sweat rate and EDA, with NFC and RF to power them and communicate results &\cite{kim_soft_2020} \\
\bottomrule
\end{tabular}
\end{table} 

\subsubsection{Smartphone-based Quantitative Assays}
%For remaining applications in this section, we focus on smartphone-based quantitative assays for micronutrient status assessment because these most closely align with our goal of developing an accessible method for measuring status. 
Smartphones are increasingly being used as analytical platforms for quantitative assays. Often, the phone is either used to photograph the assay results for quantitative analysis \cite{lee_smartphone_2014, lee_nutriphone_2016, srinivasan_ironphone_2018, serhan_total_2020, serhan_2021, ferreira_new_2021, dortez_integrated_2023, prakobdi_non-invasive_2024}, or it communicates with a more specialized and standardized sensing device \cite{lu_rapid_2017, lee_flexible_2016, vemulapati_quantitative_2017}. 
%\textbf{Vitamin D Measurement} 
\citet{lee_smartphone_2014} demonstrated the use of smartphones to image and quantify vitamin D (calcidiol) levels from an immunoassay. 
% When a sample is introduced to the assay, the detection area develops a more 'intense' color at lower concentrations of vitamin D. 
After the sample is deposited on the test and incubated for a few hours, the assay was imaged using a custom smartphone accessory.
% that positions the strip in front of the camera and illuminates the back with a diffused LED light. 
% Concentrations can then be quantified from the image by comparing the brightness of the detection region to the reference area. 
The device was evaluated using three levels of known concentrations
% (selected from deficiency thresholds): 15, 40, and 110 nM/L. However, these concentrations are 
that span from deficiency to sufficiency, but not excess, although this range is debated (Table \ref{tab: assessment}). 
% Although this range is debated. 
Results were compared to an ELISA test.
%\textbf{Vitamin B12 Measurement} 
% Motivated by the asymptomatic nature of Vitamin B12 deficiency, 
The same researchers developed a smartphone-based assay method (dubbed ``the Nutriphone") for B12 quantification \cite{lee_nutriphone_2016}. 
% Although lateral flow assays generally have a LoD that is too high for B12 status, the authors implemented a "'spacer pad' for increasing the duration of the key competitive binding reaction" that decreases the LoD into a viable range\cite{lee_nutriphone_2016}. 
% The ratio of test to control line intensities extracted from the smartphone image are used as an input to a 4-parameter logistic curve to output an estimated vitamin B12 concentration.
% (from 0-$\sim$441 pg/mL). 
Twelve human subjects provided capillary blood samples from a finger prick, and assay results were compared to an Immulite 2000 immunoassay system. On these samples, the Nutriphone failed to accurately determine B12 levels above 441 pg/mL. Unlike the Immulite, this solution appears to be insensitive to the upper spectrum of serum vitamin B12 concentrations (Table \ref{tab: assessment}). In addition, while the authors do not specify the form of vitamin B12 their device aimed to investigate, the reported molecular weight is closest to the non-optimal reference standard cyanocobalamin.
As reported in Table \ref{tab: assessment}, vitamin B12 has no single optimal reference standard, but it is often confirmed with methylmalonic acid.
The authors suggest that future work should aim to be more effective at lower limits of detection and better account for interferents in whole blood. 
An evolution of the Nutriphone assesses iron, as ferritin
% Iron, as ferritin, has likewise been targeted by this group 
\cite{srinivasan_ironphone_2018}. 
% It directly addresses the hemoglobin-based methods of iron deficiency detection we discussed, stating that “screening and diagnosis for ID [Iron Deficiency] is often limited to proxy hemoglobin measurements alone” \cite{srinivasan_ironphone_2018}. 
This assay was evaluated in-lab with known concentrations of ferritin in spiked buffer (n=27) and serum samples (n=12) to optimize performance. 
% Test concentrations were in the range of 3-556 $\mu$g/L for spiked buffer and 5.78-888 $\mu$g/L for serum, which covers the physiologically-relevant range. 
Human trials were also conducted (n=20) and results were compared to the Immulite 2000. 

\begin{figure} [t]
    \centering
    \includegraphics[width=0.7\linewidth]{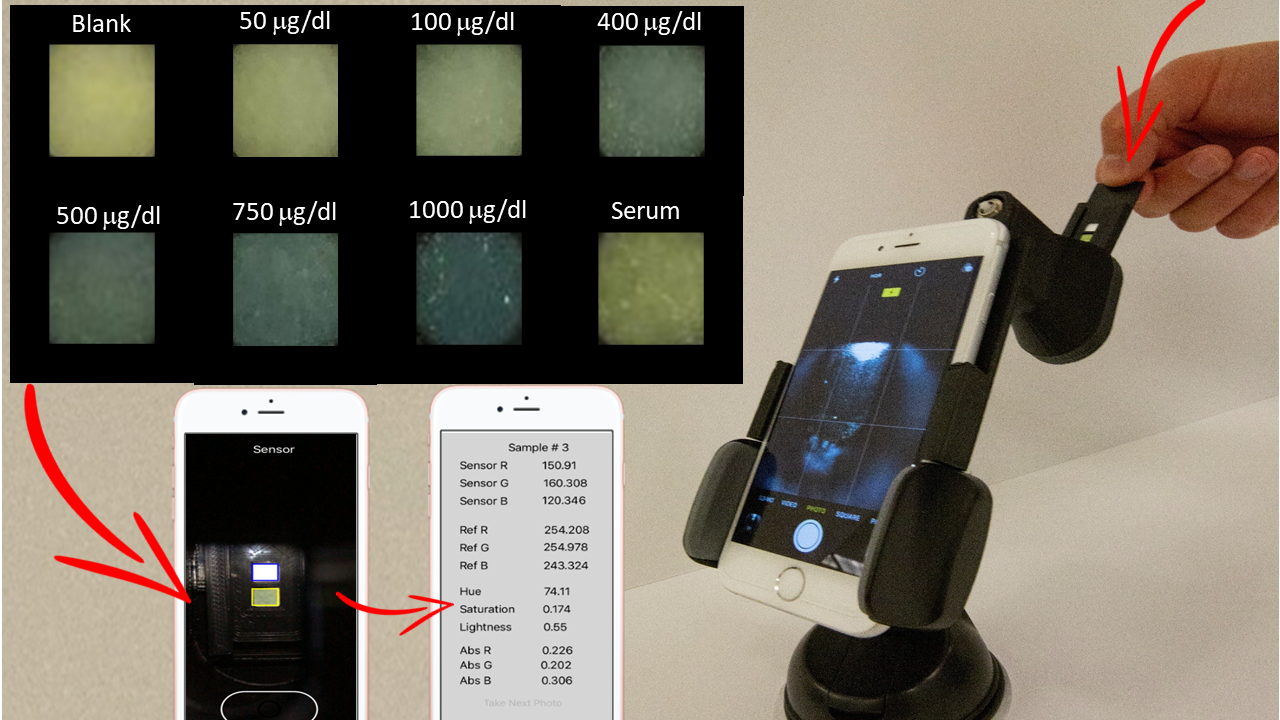}
    \caption{The use of a smartphone-based assay for the quantitative assessment of iron in serum. Used with permission from \citet{serhan_total_2020}.}
    \label{fig:smartphone-iron}
    \Description{Top left: a table comparing images of the test area for a blank sample, a serum sample, and iron concentrations ranging from 50 $\mu$g/dL to 1000 $\mu$g/dL. The test area clearly darkens compared to the blank and serum samples as iron concentration increases. Bottom left: a smartphone is shown capturing an image of the test area and displaying the results of the image analysis to the user. Right: the full test setup consists of a smartphone in holder, with the test strip placed into the attachment described in \citet{serhan_total_2020}}
\end{figure}

\citet{serhan_total_2020} had a similar goal of using a smartphone-based assay to measure total iron in serum (Fig. \ref{fig:smartphone-iron}). This paper focused on total iron instead of ferritin (the clinically-accepted biomarker for imbalance; Table \ref{tab: assessment}) because it is "the most direct metabolite in the [iron] panel" \cite{serhan_total_2020}. Total iron can provide valuable insights into iron status, thus it is a worthwhile target even if it is not the optimal reference standard \cite{national_institutes_of_health_iron_2023}. This is the first work to consider the risks of excess iron, purposefully designing the assay to be sensitive to both deficiency and excess with a “dynamic range of 50-300 $\mu$g/dL” \cite{serhan_total_2020}. 
% Like the prior studies we have seen, the image is captured by placing the test strip in a holder in front of a smartphone camera and illuminating it with diffused LEDs. 
Twenty capillary blood samples were collected via a finger prick, and assay results were compared to optimized multi-well plate spectrophotometry. Additional analysis found that their approach had a coefficient of variance of 10.5\%, compared to 2.2\% for the lab tests. One unique form of validation here that did not appear in other work was specificity testing using interferent analytes. 
The preceding method was approved upon with a new system to measure total iron levels from whole blood, consisting of an iOS smartphone application, a 3D printed sensing chamber, and a vertical flow membrane-based sensor strip \cite{serhan_2021}. 
% Their new approach filters out the cellular components of whole blood and carries out a colorimetric chelation reaction. A resulting color change is detected by the smartphone device within 5 minutes. 
% Their sensor strip has four unique membranes that work to absorb and separate a drop of blood, as well as promote the movement of transferrin-bound iron to mix with reagents existing in the strip where the colorimetric reaction is carried out. 
The smartphone application's accuracy and precision were tested against a reference imaging software (ImageJ) for the same colorimetric sensing strip.
% and resulted in a signal to noise ratio >40 and a mean bias of 2\% with an upper limit of $\sim$210 $\mu$g/dL. 
They compared their iron detection technique to a spectrophotometry-based laboratory test for iron detection on 14 venous blood samples from 9 volunteers (7 male) and found greater limit of detection (LoD) than the laboratory method (2.2 $\mu$g/dL)
% and achieved a calibration slope of 0.0004 au $\mu$g$^{-1}$ dL$^{-1}$, a correlation plot with slope of 1.09, a coefficient of determination (R$^2$) of 0.96, and a mean bias of 5.3\% \cite{serhan_2021}. 
The authors discuss future steps of expanding this tool toward measuring total iron binding capacity and saturation levels. 
% The main limitation mentioned is that they manually diluted some samples to represent a low iron concentration.

\citet{ferreira_new_2021} developed an assay for urine, a biofluid that is less invasive than blood and more representative of status than sweat. Their paper presented a paper-based colorimetric assay for urinary iron and quantification of results from images.
The assay itself contained four columns of five sample units each. This design allows for replicate results and outlier exclusion. 
% After 20 minutes, the assay was imaged and the intensities of each sample unit was analyzed in imageJ to yield absorbance. 
Blank (water) samples were used to obtain a baseline signal intensity for absorbance calculations and calibrate for urine color interference.
Calibration curves were determined using iron standards in water and synthetic urine. It was found that the phosphate and citric acid in the synthetic urine significantly interfered with the slope of the calibration curve.
The sensor was evaluated using volunteer urine samples (n=26) pre-treated with nitric acid. The results were compared to AAS and found to be similar. \citet{dortez_integrated_2023} showcase another colorimetric assay for serum iron, quantified by smartphone images. 
% Most interesting part is the use of anti-transferrin to strip the iron ions from transferrin before quantification.
% Samples are preprocessed with anti-transferrin to strip iron (Fe3+) ions from the protein transferrin, which are then reduced to Fe2+ before the sample is deposited on the assay. A smartphone and light box were used to capture images of the test area, which are analyzed to yield quantified results. 
The assay was developed using iron standard solutions and evaluated using diluted serum spiked with Fe3+. It was found that the assay could analyze multiple samples simultaneously, allowing for auto calibration of test samples against a certified reference control. \citet{prakobdi_non-invasive_2024} claimed to present a noninvasive saliva-based screening test for IDA using a nitrocellulose lateral flow system to measure iron (Fe3+) levels in spiked saliva. 
% The method uses a capillary flow-driven microfluidic device and tests iron's reaction with bathophenanthroline (Bphen) and ferrous (Fe2+) ions to measure iron (Fe3+) levels in spiked saliva. 
A reaction's color change was analyzed, and the results indicate a linear response in the 100-2000 $\mu$g/dL range (falling far above the lower limit for `normal' serum iron results \cite{chad_haldeman-englert_iron_nodate}).
The study used pooled commercial saliva, and it did not compare measured iron levels with a clinical optimal reference standard for iron status assessment. In addition, we note the general ambiguity of whether saliva levels of iron are an accurate indicator of circulating blood levels (Section \ref{subsec_biofluids}).

%\paragraph{Vitamin A Measurement} 
The smartphone moved into a supporting role in \citet{lee_flexible_2016}. The authors implemented a paper-based microfluidic immunoassay for vitamin A (as retinol binding protein or RBP), iron (as ferritin), and C-reactive protein (CRP) in a unique platform that resembles a flat, portable index-card design. Most notably, it was able to analyze whole blood with minimal pre-treatment because of built-in plasma separation (similar to \cite{serhan_2021}). The device included competitive assays for CRP and RBP and a sandwich assay for ferritin. Sandwich-based assays are more sensitive, while competitive assays are more effective for analytes of larger concentrations.
CRP is a useful inclusion since both iron and vitamin A assessment are impacted by inflammation (Table \ref{tab: assessment}). 
On-device light emitting diodes (LEDs) and photodetectors analyzed the assay, sending results and various test metrics to a smartphone app over NFC. The added ability to store this data in a remote server made the device a powerful tool for population-level screening. 
The device was evaluated on whole blood samples (n=95), each run 3 times, and compared to ELISA. A documented 84.4\% of sample was male, and 6 samples were spiked with ferritin, CRP, and RBP to assess a wider range of concentrations. 
For evaluation, a physiologically-relevant cutoff of 15 $\mu$g/L was set for ferritin deficiency (although this is insensitive to the deficiency upper limit of 30 $\mu$g/L; Table \ref{tab: assessment}).
% Median CVs vs ELISA: 2.5\% vs 6.9\%, 10.8\% vs 3.9\%, and 3.9\% vs 2.67\% for ferritin, RBP, and CRP, respectively. Deficiency AUC of 0.907 for iron (< 15 ng/mL), 0.7139 for VA (< 16.3 $\mu$g/mL), 0.9756 for acute inflammation (> 10 ng/mL).
The decision to make this device single-use is perplexing, especially since a reusable version was used for testing. We would also like to note that RBP is not the clinically-accepted biomarker for vitamin A in serum, which should be measured directly instead (Table \ref{tab: assessment}).  

% In \citet{lu_rapid_2017}, the Erickson lab deviates from smartphone image analysis and moves to a bespoke device.
% Their 

The approach in \citet{lu_rapid_2017} achieved simultaneous quantification of vitamin A (as RBP), iron (as ferritin), and CRP on a single test strip using multiple fluorescent markers and immunoassays.
% With multiple tests on a single strip, precautions must be taken to ensure no one test impacts the others. To achieve this, each target has a different colored fluorescence tag at a fixed distance from a single control line. 
The paper proposed a reusable, standalone reader (the TIDBIT). 
% The reader has a camera-based design with focusing lens and long-pass (535 nm) optical filter. 
% Six blue LEDs with band-pass optical filters (458 nm) are used for fluorescent excitation of the assay tags. 
Interestingly, quantitative results were only presented to the user if they are within a "physiologically relevant dynamic range" of 2.2-20 $\mu$g/mL for RBP, 12-200 $\mu$g/L for ferritin, and 0.5-10 $\mu$g/mL for CRP \cite{lu_rapid_2017}. We note that the ferritin detection range sufficiently covers iron deficient status, but does not extend into the threshold which indicates iron-deficiency anemia (IDA; <10 $\mu$g/L; Table \ref{tab: assessment}). Although a complete blood count (CBC) test that measures hemoglobin is the standard method for diagnosing anemia more generally, an additional ferritin assessment is still useful to determine if the patent suffers from IDA (the most common type of anemia) \cite{national_heart_lung_and_blood_institute_iron-deficiency_2022}. Furthermore, ferritin status can inform the clinician whether iron supplementation is an appropriate treatment for the anemia, as over-supplementation and excess can lead to adverse effects \cite{national_heart_lung_and_blood_institute_anemia_2022}. To assess the TIDBIT device, forty-three human serum samples were purchased from a commercial vendor for testing and compared against ELISA \cite{lu_rapid_2017}. While the R² for the RBP assay was lower than for the other biomarkers, the authors explained that the assay was optimized for high sensitivity and specificity near the diagnostic cutoff for vitamin A deficiency, rather than for precise quantification across the entire physiological range.
% Evaluation results for RBP 
% were poor, with no explanation of why this may be. 

This TIDBIT device was applied further in \citet{vemulapati_quantitative_2017}, which examined vitamin D (as 25(OH)D3)
% A “novel elution buffer that separates 25(OH)D3 from its binding protein in situ” enables 
via the assessment of capillary blood from a finger prick with no pre-treatment.
% Uses their TIDBIT device, a “smartphone-assisted portable imaging device that can autonomously perform the necessary image processing”.
In testing, the test/control ratio was highly correlated to vitamin D concentrations in standard solutions.
% (R2=0.95) with a coefficient of variance of 1.7\% at 40.1 ng/mL and 14.7\% at 50.1 ng/mL. 
Commercial serum standards highly correlated with assay results (4-parameter logistic curve), with coefficient of variance (CoV) of 2.63\% at 34 ng/mL and 11.2\% at 0 ng/mL.
Human trials with serum (n=21) and capillary blood (n=6) samples were conducted, and results were compared to LC-MS/MS measurements. 
% Serum tests showed R2 of 0.91. 
The accuracy of deficiency detection was assessed for serum but not whole blood with an area under the curve (AUC) of 0.836 for deficiency cutoffs of 20 ng/mL and 1 for 12 ng/mL. Only the latter aligns with the general clinical threshold for deficiency (Table \ref{tab: assessment}). 
% RMSE of assay serum sample results with LC-MS/MS was 2.16 ng/mL. For finger prick tests the R2 with LC-MS/MS was 0.94, but the accuracy of deficiency detection was not assessed.

% Smartphone-based
\begin{table}[!htp]\centering
\caption{Assay-Based Methods Using Smartphones}\label{tab: emerging-assay-smartphone}
\tiny
\begin{tabular}{p{0.5in}p{0.6in}p{0.4in}p{0.5in}p{0.5in}p{0.6in}p{0.75in}p{0.6in}p{0.25in}}\toprule
% \multicolumn{9}{c} {Assay-Based Methods} \\
Method &Platform&Targets &Analytes &Evaluation &Control &Results &Notes &Source
\\\midrule
Quantitative immunoassay &Smartphone-based &Vitamin D (calcidiol)&Serum &Known sample solutions and human subjects &ELISA &Errors “at same order” as ELISA & No statistical analysis of results &\cite{lee_smartphone_2014} \\
Quantitative immunoassay &Smartphone-based &Vitamin B12 (cyanocobalamin) &Capillary blood sample &Human Subjects &Immulite 2000 Immunoassay &Correlation of 0.93 with control, ~85\% specificity and ~60\% sensitivity for deficiency detection &Used synthetic form of B12; poor accuracy outside of deficiency range &\cite{lee_nutriphone_2016} \\
Quantitative immunoassay &Smartphone-based &Iron (total) &Capillary serum sample &Human Subjects &Multi-Well Plate Spectrophotometry &$R^2$ of 0.98 with control, CoV of 10.5\% &Conducted specificity testing using interferent analytes; considered toxicity &\cite{serhan_total_2020} \\
Quantitative immunoassay &Smartphone-based &Iron (total) &Capillary blood sample &Human Subjects &Laboratory developed test: spectrophotometry-based technique & Correlation plot with slope of 1.09, R$^2$ of 0.96, and a mean bias of 5.3\%& Improved on control LoD; manually diluted some samples to represent a low iron concentration &\cite{serhan_2021} \\
Quantitative immunoassay &Smartphone-based or portable device (TIDBIT) &Iron (ferritin)&Capillary blood sample &Human Subjects &Immulite 2000 Immunoassay &Correlation of 0.92 with control, sensitivity of 0.9 for deficiency detection & &\cite{srinivasan_ironphone_2018} \\
Quantitative immunoassay &Portable device (TIDBIT) &Vitamin D3 (calcidiol) &Serum and whole blood&Commercial standards; human subjects&Known solutions; LC-MS/MS& CoV with standards of 2.63\% at 34 ng/mL and 11.2\% at 0 ng/mL; $R^2$ of 0.91 for serum tests, 0.94 for capillary blood tests&&\cite{vemulapati_quantitative_2017} \\
Colorimetric assay &Paper-based &Iron (total) &Urine &Human subjects &AAS &RSD of 9.5\% &Urine citric acid was found to interfere with results &\cite{ferreira_new_2021} \\
Colorimetric assay &Paper-based &Iron (Fe3+) &Serum &Spiked, diluted from human subjects &Spiked known concentration &Error of 3.7\% and RSD of 1\% &LoD of 0.3 $\mu$g/mLl &\cite{dortez_integrated_2023} \\
Colorimetric assay &Paper-based & Iron (Fe3+) &Saliva &Spiked pooled commercial saliva &Spiked known concentration &$R^2$ of 0.99 &It is debated whether salivary iron reflects circulating status &\cite{prakobdi_non-invasive_2024} \\
Opto-electronic immunoassay &Card &Vitamin A (RBP), Iron (ferritin), and CRP &Whole blood &Human subjects &ELISA &CVs of 2.5\% for ferritin, 10.8\% for RBP, and 3.9\% for CRP &Cutoff for iron deficiency was insensitive to upper limit; RBP is not the clinically-accepted biomarker for vitamin A status; Device transmits results to a smartphone over NFC &\cite{lee_flexible_2016} \\
Multiplexed quantitative assay &Portable device (TIDBIT) &Vitamin A (RBP), Iron (ferritin), and CRP &Serum &Human subjects &ELISA &$R^2$ of 0.56 for RBP, 0.92 for ferritin, 0.88 for CRP &Ferritin range does not cover anemia; RBP is not the clinically-accepted biomarker for vitamin A status; RMSE for RBP was 21 $\mu$g/mL &\cite{lu_rapid_2017} \\
\bottomrule
\end{tabular}
\end{table} 

\subsubsection{Commercial Products} 
Commercially-available devices have emerged in recent years to provide point of care (PoC) testing for some micronutrients \cite{abebe_low_2019, bloom_diagnostics_overview_2021, Fiddler2023, albrecht_rapid_2021}. 
One study has explored the utility of a commercial iCheck FLUORO device to assess vitamin A concentrations in human milk (human milk vitamin A or HMVA) \cite{abebe_low_2019}.  HMVA is critical since it is the primary source of vitamin A for breastfeeding children. If there is a vitamin A deficiency in human milk, it is likely to cause developmental issues for a child. The authors collected human milk samples and socio-demographic and anthropomorphic data from lactating mothers in the Mecha district, Ethiopia (n=104). This region was selected because prior studies applying this device for HMVA assessment recommended further investigation of populations at greater risk of vitamin A deficiency.
Concentrations of vitamin A in human milk were measured by iCheck FLUORO and compared to HPLC. 
The commercial device was found to overestimate low HMVA concentrations and had a weak overall correlation with HPLC results.
% Mean HMVA was 41\% of the estimated average HMVA concentrations of developing countries. 
Therefore, the paper concluded that studies which assess vitamin A intake among breast-feeding children in developing countries should not assume average HMVA.
It was argued that devices like the FLUORO are needed to monitor HMVA status, especially for intervention programs that typically assume average HMVA. Still, they must be “reliable across a range of HMVA concentrations” \cite{abebe_low_2019}.

\citet{albrecht_rapid_2021} likewise studied the efficacy of the Quidel Inc Sofia fluorescent immunoassay for serum vitamin D (as calciol). It should be noted that calciol is an inactive form of vitamin D, distinct from the optimal reference standard of calcidiol for vitamin D status. The assay was analyzed by the Sofia Analyzer, a PoC device for immunoassay analysis. 
% Sample is centrifuged, combined with reagent, incubated for 5 minutes, then deposited onto assay and placed in the Sofia device. After 5 minutes, the device quantifies VD3 from fluorescence using an undisclosed algorithm.
A total of 324 samples were collected and 296 were used (229 female). Additionally, 433 tests were run using both frozen (208) and fresh samples (88). 
Notably, the researchers also assessed random error and inter-operator reliability for the device.
Because only one sample had a concentration above 100 ng/mL, the authors recommend additional testing for concentrations above 80 ng/mL (well into the range of toxicity; Table \ref{tab: assessment}). 

Bloom Diagnostics is a home use `lab' device that analyzes single-use qualitative test strips to quantitatively assess the status of in-vitro (in-body) biomarkers, similar to the TIDBIT \cite{vemulapati_quantitative_2017, lu_rapid_2017, bloom_diagnostics_overview_2021}. Tests available for Bloom include thyroid-stimulating hormone (TSH), ferritin, CRP, and estimated glomerular filtration rate with cystatin C. While Bloom approaches the goal of accessible nutrition assessment, its assays still require the user to collect a sample themselves. Depending on the test, this could involve a finger prick, coaxing the blood into a collection tube, and depositing it properly onto the assay.
% In our experience using the device, we found this to be non-trivial and intimidating for a first-time user.
VitaScan is another commercial PoC device that tests for iron deficiency, and they validate results against the clinical optimal reference standard for in-body iron measurement \cite{Fiddler2023}. The device is not yet released, but it is planned to assess vitamins B12, D, and A and CRP in the future. The method is still invasive as it utilizes capillary blood obtained from a finger prick and also requires the user to obtain the sample themselves. 
% There are areas for improvement with respect to invasiveness. 

% Commercial
\begin{table}[!htp]\centering
\caption{Commercial Assay-Based Methods}\label{tab: emerging-assay-commercial}
\tiny
\begin{tabular}{p{0.5in}p{0.5in}p{0.5in}p{0.4in}p{0.4in}p{0.6in}p{0.75in}p{0.8in}p{0.25in}}\toprule
% \multicolumn{9}{c} {Assay-Based Methods} \\
Method &Platform&Targets &Analytes &Evaluation &Control &Results &Notes &Source
\\\midrule
Fluorometry &Portable device &Vitamin A (retinol and retinyl esters) &Breast milk &Human subjects &HPLC &Weak correlation ($R^2$=0.59, p<0.001), but mean difference was “not statistically different from zero” &Of major concern was the ability for the breast milk to satisfy the vitamin A requirements of children &\cite{abebe_low_2019} \\
Immuno- fluorescence &Portable device &Vitamin D (cholecalciferol) &Serum &Human subjects &Abbott Alinity i immunoassay & $R^2$ of 0.89; SE of 0.16 at 10 ng/mL, 0.19 at 12 ng/mL, and 0.35 at 30 ng/mL & Standard error lower than control; recommends additional testing for excess status; target is not the clinically-accepted biomarker for Vitamin D &\cite{albrecht_rapid_2021} \\
Quantitative immunoassay &Home use device &Iron (ferritin), other tests &Capillary blood sample &Not published &Not published &Not published &Commercially available &\cite{bloom_diagnostics_overview_2021}\\
\bottomrule
\end{tabular}
\end{table} 

% \subsubsection{Limitations}
% Although there are several wearable, smartphone, and point of care assay devices, their accessibility in some cases is limited by specialized assay chips and most require invasive biofluids for analysis. Saliva and sweat have been proposed as alternatives to a finger prick, but their ability to reflect in-body status of a given micronutrient is often unknown, debated, or disproved. This is especially true for sweat-based sensors, as we will expand on in the following section.

\subsubsection{Comparison}
Although they are the most accessible and noninvasive, sweat-based assays are generally not a viable alternative to clinical methods of micronutrient status assessment given the current clinical knowledge about the correlation between sweat and blood concentrations of micronutrients. A critical tradeoff exists between accessibility and accuracy, yet until the clinical literature has reached an agreement on which alternative biosamples (i.e. not blood or urine) are most appropriate for the assessment of in-body status, it may be more productive to pursue accessibility through device design rather than through the biosample analyzed so as to not sacrifice clinical-relevance.

Microfluidics and smartphone-based quantification systems may deliver on accessibility and noninvasiveness. The former enables analysis on a small volume of a biosample, meaning that less of the sample needs to be collected (via potentially invasive means in the case of blood). The latter decreases reliance on commercial assay analyzers, which may only operate on proprietary assays, and enables accessibility through portability and cost-effectiveness.

As opposed to other types of assays (e.g. immunoassays), colorimetric assays have been overwhelmingly applied to the assessment of iron (with one expanding to sweat vitamin C, calcium, and zinc). This may indicate the poor versatility of colorimetric assays, whereas other assay techniques are able to assess a wider variety of biomarkers, largely through adjustments to the antibodies used in the assay. 
The trend of multiplexed assays in both sweat-based and non-sweat-based sensors is therefore a promising means to balance the pros and cons of different assay types. While multiplexed solutions may increase cost, we argue that the potential to assess multiple biomarkers at once significantly decreases the burden on the patient and the clinician caused by the need to use multiple assays/devices. 

\subsection{Electrochemistry-Based Methods}
%\paragraph{Background}
Electrochemical analysis is a method that has been applied in literature to quantify levels of micronutrients in biofluids such as saliva, sweat, tears, urine, and blood. \citet{huang_electrochemical_2021} summarizes the basis of electrochemical sensors with respect to vitamins, but we see work applying these ideas to minerals as well. The concentration of vitamins in an electrolyte (water or fat/organic solution that allows for the transfer of electrons) can be quantified by measuring electrical properties at a working electrode. The most common measurement techniques are \textit{voltammetry} and \textit{amperometry}. Voltammetry applies a varying voltage to the electrolyte and measures the resulting current, while amperometry applies a constant voltage and measures the resulting current over time \cite{harvey_114_2024}. 
Below, we dive into novel methods that were evaluated in biofluids, roughly divided into voltammetry and amperometry. For more detailed discussions, we direct the reader to recent reviews focusing on electrochemistry-based methods (e.g. \cite{huang_electrochemical_2021, Pakeeza2024, Kirazolu2023, sardar2023}). 
A summary of electrochemistry-based methods reviewed herein is found in Tables \ref{tab: emerging-electro-volt}, \ref{tab: emerging-electro-amp}, and \ref{tab: emerging-electro-oth}. 

\subsubsection{Voltammetry}
Voltammetry has been a popular method for the assessment of micronutrients in biofluids. 
\citet{revin_simultaneous_2012} proposed a novel electrode for the simultaneous measurement of vitamins B2 (riboflavin), B9 (folic acid), and C (ascorbic acid). 
% The authors claim the sensor was developed with a physiological pH in mind, but the buffer used was pH 7.2 instead of the 7.35-7.45 of blood. 
Peak currents for each vitamin were well separated at mixtures of various concentrations. 
% Linearity in buffer was demonstrated between 3.76-33.88, 8.83-79.5, and 5.28-47.6 mg/L (r2>0.99 for all) with LoDs of 0.017, 0.11, and 0.126 mg/L, respectively. 
For vitamin C in particular, the tested linear range of the sensor was insensitive to the lower limit of sufficiency and below (Table \ref{tab: assessment}). Additionally, the analyzed biomarkers for vitamins B2 and B9 differed from their clinical optimal reference standard.
Selectivity analysis showed that linearity in each vitamin was maintained even in the presence of elevated concentrations of all other vitamins. No interference from other common physiological interferents was found.
Two plasma samples from a clinical laboratory were diluted and tested before and after spiking with vitamin standards, with good recovery. 
% We note that direct riboflavin assessment is not the most clinically relevant indicator of B2 status (Table \ref{tab: assessment-ws}).
% Peak currents for all three vitamins were still distinct, and >99\% recoveries of the spiked concentrations was found.
Another electrode was developed by \citet{jothimuthu_zinc_2013}, examining zinc. 
% with a linearity that covers the physiologically-relevant range. 
Interestingly, a sample pH of 6 was optimal for zinc assessment, which is more acidic than most biofluids (e.g. blood has a pH of 7.35-7.45).
The zinc content in both acetate buffer and spiked, HCl-diluted serum was evaluated. 
% The linear range for the sensor was 32.7 to 327 $\mu$g/dL (physologically relevant) with an r2 of 0.994 and LoD of 6 mcM.
A square wave voltammetric sensor was developed for the simultaneous measurement of glutathione (GSH), nicotinamide adenine dinucleotide (NADH) and folic acid (vitamin B9) \cite{raoof_high_2015}. A single urine and serum sample was collected for evaluation and spiked with known amounts of the targets. The authors did not present serum results. 
% Linearity of 0.006-161, 1-650, and 3-700 mcM and LoDs of 0.002, 0.3 and 1.0 mcM, respectively for each target in PBS. R2 for all calibration equations was >0.99. 

\citet{kim_wearable_2015} designed a square wave anodic-stripping voltammetric sensor to monitor zinc in sweat during physical activity. The sensor itself was wearable, printed on tattoo transfer paper.
% The authors thoroughly evaluated the sensor, testing its repeatability and ability to hold up against bending and stretching. 
The assessment was conducted with standard zinc solutions in a buffer medium.
% covering the physiological concentrations of zinc in sweat. 
On-body experiments (7 participants, 5 male) demonstrated the sensor's ability to assess zinc in cycling-induced sweat, which was found to be close to the physiological range.
% A linear calibration equation was used to quantify zinc concentrations, which were found to be close to the physiological range of zinc in sweat. 
Again, no clinical reference standard assessment methods for individual zinc status were used for comparison. However, this is less of a concern since this work explicitly focuses on providing insights into zinc excretion through sweat rather than determining in-body status. 
% The authors note the potential of applying this approach to other minerals found in sweat. 

Gao et al. examined zinc as well as copper using voltammetry \citet{gao_wearable_2016}. A wearable electrochemical sensor was created to assess Zn, Cd, Pb, Cu, and Hg ions in sweat and urine. Uniquely, the sensor incorporated skin temperature measurement for calibration, and to compensate for the influence of temperature on electrochemical signals. This was important, as peak current was shown to increase with temperature.
The device was developed with spiked synthetic sweat samples at concentrations an order of magnitude lower than is observed in blood (Table \ref{tab: assessment}). Calibration curves demonstrate a linear relationship between peak current and concentration for the ions, but quantitative metrics were not reported.
The authors conducted a human study with a single participant for on and off-body measurements with ICP-MS as a control. The measured and controlled concentrations for Zn and Cu were similar, but statistical analysis of the results was not conducted.
In \citet{stankovic_electroanalytical_2016}, a novel “boron-doped diamond electrode” for vitamin B12 (as cyanocobalamin) quantification was studied.
% B12 concentrations were linear between 2-11 and 11-35 mcM/L (r2>0.99) with LoD of 0.7 mcM/L in buffer with a pH of 2. 
Interference analysis showed a 10\% signal change in the presence of a ``10-fold excess of vitamin B6" \cite{stankovic_electroanalytical_2016}.
The sensor was evaluated in four spiked urine samples, diluted, and pH adjusted to 2. In this case, the electrode analyzed cyanocobalamin, which is the synthetic form of vitamin B12.

\citet{sempionatto_eyeglasses-based_2019} developed an eyeglasses-based platform to conduct electrochemical analysis of tears, assessing the concentration of glucose, alcohol, vitamins B2, B6, and C. Tears were induced with menthol sticks before being collected and analyzed by the glasses. The sensor itself used square wave voltammetry (SWV) for vitamin measurement, demonstrated only as a proof of concept.
% and chronoamperometry for glucose and alcohol. Alcohol, glucose, and vitamin measurements were all evaluated separately. Alcohol levels were compared to blood alcohol content (BAC) from a breathalyzer and glucose status was compared to a commercial glucometer. Finally, 
% Vitamin assessment was included as a proof of concept to determine if detection was possible via the proposed method. 
After a baseline was acquired with the sensor, tears were induced from 3 participants and analysis was conducted every 30 minutes for 2 hours after taking a multivitamin. 
% The sensor showed distinct peaks at potentials of -0.55V, 0.2V and 0.6V for B2, C and B6 respectively. 
Peak potentials emerged for each vitamin and were verified with known concentrations of vitamins added to baseline tear samples. 
Because this was a proof of concept, no comparison to optimal reference standards with blood or efforts to quantify in-body vitamin levels were made. 
% We would like to note that this time range for detection after supplementation is only possible for water-soluble vitamins. It is unclear whether tear samples could reflect levels of fat-soluble vitamins, given their inability to be dissolved into water-based solutions. 

A microfluidic, graphine-oxide-based sensor chip has been applied for the quantification of ferritin in serum \cite{garg_microfluidic-based_2020}. Notably, the sample must be pumped through the sensor, where cyclic voltammetry was performed continuously with an external potentiostat.
% In PBS, the linearity of the sensor fell between 7.81 to 500 ng/mL (compared to 0-1500 ng/mL for ELISA; physiologically-relevant) with an r2 of 0.996 and an LoD of 0.413 ng/mL. 
An evaluation was conducted with spiked serum samples 
% (31.25, 62.5, and 125 $\mu$g/L) 
(which were not sensitive to deficiency)
and compared to ELISA. The sensor overestimated concentrations <100 $\mu$g/L by $\sim$10\% and underestimated the larger concentration by $\sim$4\%. 
% R2 with ELISA was 0.966.

\citet{sun_bendable_2021} focused on reusability in vitamin C assessment. 
% Solution is designed to be incredibly low-cost, with the analyzer being reusable.
Their device used cyclic voltammetry to determine whether the vitamin C content present in the sample is normal or deficient (<4.93 mg/L). This deficiency threshold was greater than what is reported by the clinical literature, exceeding even the limit for sufficiency (Table \ref{tab: assessment}).
Interestingly, their device was also self-powered, using vitamin C as a biofuel.
% A linear relationship between voltage and vit C content was found for 0-17.6 and 17.612-52.8 mg/L, with a detection limit of 9.9 $\mu$g/L (physiologically-relevant). On raw human samples, an r2 of 0.984 (p<0.001) was found with HPLC results. 
In a trial for scurvy detection (n=22), the device correctly determined the 4 deficient individuals (ground truth by HPLC). The authors also demonstrated its potential to screen for patients exhibiting a medical condition, identifying 30 patients (total sample size unclear) who suffered from vitamin C deficiency during routine checkups. Of these, >85\% had a medical condition associated with inflammation and oxidative stress. 

On-body electrochemical sensing of vitamins B6 (pyridoxine), C (ascorbic acid), D3 (calciol), and E (alpha-Tocopherol), as well as 9 amino acids and several macros in sweat, was enabled by \citet{wang_wearable_2022}. Of the biomarkers examined for each vitamin, only the vitamin C biomarker aligned with the clinical optimal reference standard biomarker. Sweat was passively sampled using iontophoresis in a watch-based platform.
Voltammetry was used to detect vitamins indirectly. 
% Target molecules adhere to a polymetric layer, which changes the peak height current density of redox-active nanoreporters. 
Quantitative results were not reported, but concentrations of vitamins appeared to linearly correlate with peak height current density.
The authors noted the flexibility of this approach to measure numerous other biomarkers.
Human trials were conducted with healthy volunteers and patients but only examined amino acids. This exemplifies the tendency of mainstream research to ignore micronutrients.

\citet{kumar2023} developed a manganese dioxide nanoparticle–bimetallic metal-organic framework composite to detect vitamin D3 in spiked human plasma. 
% The authors applied cyclic voltammetry with a potential window of 0–1.5 V at a scan rate of 50 mVs$^{-1}$. 
% Linearity of 0-75 ng/mL (r=0.965)
Voltammetry measurements were compared to a optimal reference standard for vitamin D detection, HPLC with ultraviolet detection (HPLC-UV), and obtained similar values.
\citet{Seker2024} designed a touch-based sensor that simultaneously monitored zinc and ascorbic acid (vitamin C) levels after supplementation. The technique measured fingertip sweat and uses SWV for zinc detection and potentiometric measurement for ascorbic acid detection. 
% TESTED range of 35.2-176.1 mg/L for AA, 50-250 $\mu$g/dL for Zinc. Linearity not reported
Lastly, \citet{shi_wireless_2024} assessed an NFC-powered sensor for riboflavin (B2) in sweat. 
% B2 standards were prepared at different pH (4-7) and concentrations (3.76-376.4 $\mu$g/L). Demonstrates a limit of detection at 0.45 $\mu$g/L and an r2 of 0.9935. 
Selectivity testing was conducted by the addition of common sweat molecules into the standard, which did not significantly influence results. Uniquely, a pH sensor was incorporated into the device to account or the influence of pH on measurements. Human trials involved subjecting participants to exercise (n=1) or heat stress (n=2) to induce sweat after supplementation. Sweat samples were analyzed by the device and compared to HPLC sweat measurements for the exercise trials and fluorescence spectroscopy for the heat-stress trials. The time-dynamics of the device results followed the general trend of the control analysis methods, exhibiting more variance compared to urine results.

\begin{table}[!htp]\centering
\caption{Electrochemistry-Based Methods Using Voltammetry}\label{tab: emerging-electro-volt}
\tiny
\begin{tabular}{p{0.6in}p{0.5in}p{0.45in}p{0.3in}p{0.5in}p{0.6in}p{0.7in}p{0.85in}p{0.2in}}\toprule
% \multicolumn{9}{c}{Electrochemical Methods} \\
% \cmidrule{1-9}
Method &Platform&Targets &Analytes &Evaluation &Control &Results &Notes &Source
\\\midrule
% Voltammetry
Voltammetry &Benchtop &Vitamins B2 (riboflavin), B9 (folic acid), and C (ascorbic acid) &Plasma &Human subjects &Spiked known concentrations &>99\% recovery of spiked concentrations &Tested in 7.2 pH buffer, which is slightly lower than pH of blood ($\sim$7.4); direct riboflavin is not the gold-standard biomarker for vitamin B2; folic acid is the synthetic form of folate; Linear range for vitamin C did not reach below the lower limit for sufficiency &\cite{revin_simultaneous_2012} \\
Anodic stripping voltammetry &Benchtop &Zinc &Serum &Human subjects &Spiked known concentrations &Peak current decreased with concentration, but were lower in magnitude than buffer &A sample pH of 6 was necessary for optimal performance &\cite{jothimuthu_zinc_2013} \\
SWV &Benchtop &Vitamin B9 (folic acid), GSH, and NADH &Urine and serum &Human subjects (urine only) &Spiked known concentrations &Accurate recovery of spiked concentration & Simultaneous determination of targets; unclear whether pre-existing urine composition biased recovery; claims serum evaluation but this is not presented; folic acid is the synthetic form of folate &\cite{raoof_high_2015} \\
Square wave anodic stripping voltammetry &Wearable tattoo&Zinc &Sweat &Zinc stock solutions &Known zinc solutions &$R^2$ of 0.999 for measured current vs stock solutions, LoD of 0.05 $\mu$g/mL &Zinc content in actual sweat from single participant was close to physiological range &\cite{kim_wearable_2015} \\
Square wave anodic stripping voltammetry &Wearable patch &Zn, Cd, Pb, Cu, and Hg ions &Sweat and urine &Human subjects &ICP-MS &Similar to control, but provided no statistical analysis &Included temperature sensor to account for the influence of skin temperature on peak current &\cite{gao_wearable_2016} \\
SWV &Benchtop &Vitamin B12 (cyanocobalamin) &Urine &Diluted from human subjects &Spiked known concentrations &98-105\% recovery &Cyanocobalamin is the synthetic form of B12 &\cite{stankovic_electroanalytical_2016} \\
SWV and chronoamperometry &Eyeglasses&Vitamins B2, B6, and C, alcohol, and glucose &Tears &Human subjects &Breathalyzer BAC, commercial glucometer, vitamin supplementation &Correlations of 0.852 with BAC, ~0.7 with glucometer, distinct voltage peaks for each vitamin &Glucose and alcohol was main focus, vitamin assessment included as a proof of concept; exact form of each vitamin is not known &\cite{sempionatto_eyeglasses-based_2019} \\
Cyclic voltammetry &Benchtop &Iron (ferritin) &Serum &Spiked from human subjects &ELISA &$R^2$ of 0.966 and lower linearity than control; tended to overestimate concentrations <100 $\mu$g/L &Range of spiked concentrations was not sensitive to deficiency; studied the impact of pH and interferent compounds &\cite{garg_microfluidic-based_2020} \\
Cyclic voltammetry &Portable &Vitamin C &Serum &Human subjects &HPLC & $R^2$ of 0.984 (p<0.001) and 100\% accuracy in deficiency detection &Targets scurvy (extreme deficiency) but the threshold was set above sufficiency &\cite{sun_bendable_2021} \\
Voltammetry &Wearable patch &Vitamins B6 (pyridoxine), C (ascorbic acid), D3 (calciol), E (alpha-Tocopherol), and other macronutrients and amino acids &Sweat &Not reported for vitamins &Not reported for vitamins &Observable linear relationship between vitamin concentration and peak height current density &Only vitamin C aligns with clinical standard, quantitative results for vitamins were not reported &\cite{wang_wearable_2022} \\
Voltammetry & Benchtop & Vitamin D3 & Plasma & Spiked from human subjects & Spiked known concentrations and HPLC-UV & LoD of 1.9 ng/mL; RSD of 0.3-2.6\% and recovery of 96-102\% &Exact form of D3 (gold-standard 25(OH)D3/calcidiol or calciol) not reported &\cite{kumar2023} \\
SWV and potentiometric measurement &Portable &Zinc and vitamin C (ascorbic acid) & Fingertip sweat & Human subjects & Supplementation & Both micros could be analyzed over time simultaneously & No comparison to clinical assessment of status or statistical analysis of results; Vitamin C range far exceeded physiological concentrations in plasma & \cite{Seker2024} \\
Differential pulse voltammetry &Wearable patch &Vitamin B2 (riboflavin) &Sweat &Human subjects &HPLC &$R^2$ of 0.9783 with sweat HPLC and 0.87 with urine fluorescence spectroscopy &Urine included as comparison to sweat status; incorporates pH sensor to control for influence of pH on measurements; direct riboflavin is not the gold-standard biomarker for Vitamin B2 status &\cite{shi_wireless_2024} \\
\bottomrule
\end{tabular}
\end{table}

\subsubsection{Amperometry}
Vitamins B2, B9, C, and D, as well as the mineral iron, have been assessed in biofluids using amperometry \cite{maiyalagan_nanostructured_2013, sempionatto_epidermal_2020, zhao_wearable_2021, ruiz-valdepenas_montiel_decentralized_2021, zhu_nonenzymatic_2023, Ning2024}.
\citet{maiyalagan_nanostructured_2013} confronted a major limitation of glassy carbon electrode-based sensors for vitamin B9 (folic acid): interference from vitamin C. Their nanofiber-modified electrode successfully avoided this interference.
% The performance of the sensor peaked at pH 7.2, which is slightly lower than the pH of blood.
% In spiked buffer, current response was linear within 26.484-264.8 ng/mL, with an r2 of 0.9901 and a LoD of 26.484 ng/mL. 
In evaluation, two serum samples were collected and evaluated before and after spiking with 4.41 $\mu$g/L of folic acid. The peak current increased accordingly, allowing for greater than 99\% recovery.
Vitamin C is examined by \citet{sempionatto_epidermal_2020}, who deployed amperometry and immobilized ascorbate oxidase in a tattoo-based platform. Tests on human subjects (n=4) focused on the temporal characteristics of the current response, finding peak response in sweat 90 minutes after supplementation and a return to baseline 180 minutes after, in line with the plasma response of vitamin C. Tears and saliva were noted as other possible biofluids, with tears yielding a similar temporal profile on a single subject (albeit with different peak currents). The authors also experimented with supplementing vitamin C through orange juice, claiming that the response from sweat samples increased in line with increasing vitamin C content (n=2). Crucially, no statistical analysis was conducted on the results of the experiments, such as correlations between intake and measured current. There were also no comparisons made to clinical reference methods of vitamin C assessment, though this was stated as a subject for future work. 
% \citet{oshin_graphene-based_2020} describes the measurement of ferritin via a "graphene-based field effect transistor functionalized with anti-ferritin antibodies". Their sensor uses an electrolyte buffer solution as an analyte, and the change in resistance across the sensor is measured. While the goal of the device is to non-invasively detect iron deficiency in children through saliva samples, no tests on biosamples are reported. The authors claim a detection range of 5.3 ng/L to 0.5 $\mu$g/L with the sensor, which is far lower than the physiologically-relevant range for ferritin in serum (Table \ref{tab: assessment-min}).

A wearable, electrochemical device to measure vitamin C levels in sweat, urine, and blood was proposed by \citet{zhao_wearable_2021}. The device was used in a study where 6 male participants, aged 20-30, were given vitamin C as emergen-C brand supplements. The sensor was wearable, but no on-body measurements were made. Instead, urine and induced sweat were collected three hours post-intake and measured with the device. Blood samples were collected from a single participant in a separate study, analyzed with the device, and compared to results from urine and sweat. This research did not compare device results to any optimal reference standard for vitamin C assessment; instead, it relied on intake, which (as we will see in Section \ref{phys section}) is a poor equivalent for in-body status due to individual differences in micronutrient absorption. In addition, we note that the vitamin C supplement, emergen-C, contains several other nutrients that could influence the results of the analysis. 

One paper proposed simultaneous measurement of vitamins C and D from a single saliva sample \cite{ruiz-valdepenas_montiel_decentralized_2021}. Their sensor combined an electrocatalytic vitamin C (ascorbic acid) amperometric assay and competitive vitamin D (25(OH)D3) immunoassay. Vitamin D was the clear focus of the paper, and the vitamin C sensor received little to no attention aside from some analysis of potential cross-talk between the two sensors. 
% LoDs for 12 ng/mL of vitamin D in whole saliva and 29 ng/mL in saliva diluted with phosphate buffer are reported. These former meets the threshold for deficiency in blood stated in Table \ref{tab: assessment-fs}. 
The sensor was applied in a study that supplemented vitamins to 3 participants and used the device to analyze saliva samples at increasing time intervals from intake. No optimal reference standard assessment method was used to evaluate the sensors, although the authors advocated for this evaluation and the development of truly quantitative sensors in future work. 
Another flexible, electrochemical sweat biosensor for vitamin C used polyaniline film modified with phytic acid \cite{zhu_nonenzymatic_2023}. The biosensor was validated with synthesized vitamin C samples of known concentrations. 
% Two linear ranges: 0.5-10 and 10-500 mcM/L with LoD of 0.17 mcM/L
Four human subjects were given supplements and had their sweat collected 3 times over 90 minutes. The sensor detected a general increase in current from the sweat samples over time, with variation across subjects. Saliva tests were also conducted. Peak current occurred 60 minutes after supplementation, in line with results from \cite{sempionatto_epidermal_2020}.
Another team of researchers designed a finger-actuated wirelessly-charging wearable that measures vitamin C and levadopa (a central nervous system agent) levels from sweat \cite{Ning2024}. The system had a microfluidic chip with a self-driven pump and anti-reflux valve, a flexible wireless circuit board, and a companion smartphone app. They ran a study with five healthy participants whose sweat was collected and measured after exercising and ingesting vitamin C tablets as well as fava beans \cite{Ning2024}. The results were not compared to clinical results as a optimal reference standard baseline.

% Amperometry
\begin{table}[!htp]\centering
\caption{Electrochemistry-Based Methods Using Amperometry}\label{tab: emerging-electro-amp}
\tiny
\begin{tabular}{p{0.55in}p{0.5in}p{0.45in}p{0.5in}p{0.5in}p{0.6in}p{0.7in}p{0.7in}p{0.2in}}\toprule
% \multicolumn{9}{c}{Electrochemical Methods} \\
% \cmidrule{1-9}
Method &Platform&Targets &Analytes &Evaluation &Control &Results &Notes &Source
\\\midrule
Amperometry &Benchtop &Vitamin B9 (folic acid) &Serum &Human subjects &Spiked known concentrations &>99\% recovery of spiked concentration &Robust against interference from ascorbic acid; sensor performance peaked at pH 7.2, slightly lower than pH of serum; folic acid is the synthetic form of folate &\cite{maiyalagan_nanostructured_2013} \\
Amperometry &Wearable tattoo&Vitamin C (ascorbic acid) &Sweat, tears, and saliva &Human subjects &Supplementation &Similar time dynamics to plasma for tears and sweat & &\cite{sempionatto_epidermal_2020} \\
Amperometry&Wearable sensor chip&Vitamin C (ascorbic acid) &Sweat, urine, and blood &Human subjects &Supplementation/ intake, blood measurements with the same sensor &Urine and sweat measurements increased after intake, with inter-trial variation. Correlations with blood sensor measurements were 0.81 for sweat and 0.72 for urine &Used emergen-C as a vitamin C supplement, which contains several other micronutrients &\cite{zhao_wearable_2021} \\
Amperometry&Wearable sensor chip &Vitamin C (ascorbic acid) &Sweat and saliva &Stock solutions and human subjects &Known solutions and supplementation &$R^2$ of 0.99 and LoD of 0.0299 $\mu$g/mL for stock solutions, was able to detect general increase in current after supplementation & &\cite{zhu_nonenzymatic_2023} \\
Electrocatalytic vitamin C amperometric assay and competitive vitamin D immunoassay &Portable &Vitamins C (ascorbic acid) and D (calcidiol) &Saliva &Human subjects &Vitamin supplementation &Observable rise and drop in levels over time &No quantitative results; saliva sample must be pre-treated; Vitamin C has little evaluation &\cite{ruiz-valdepenas_montiel_decentralized_2021} \\
% Amperometry with a graphene-based field effect transistor&Point of care&Ferritin &Electrolyte buffer solution &Ferritin stock solutions &Known ferritin concentrations &Resistance measurements decreased with increasing ferritin concentrations, detection range of 5.3 ng/L to 0.5 $\mu$g/L &Goal was to non-invasively detect iron deficiency in children through saliva samples &\cite{oshin_graphene-based_2020} \\
Chronoamperometry & Wearable sensing system & Vitamin C (ascorbic acid) and Levodopa & Sweat & Human subjects & Vitamin supplementation &
% The sensitivities of the levodopa sensor and the vitamin C sensor are 0.0073 and 0.0018 $\mu$A$\cdot$$\mu$M$^{-1}$, respectively, and the 
Detection correlation coefficients of both exceed 0.99; both sensors have a wide linear detection range of 0-17.6 mg/L and 0-1000 $\mu$M, respectively, and low detection limits of 0.05 mg/L and 17.9 $\mu$M, respectively. & The system is wireless, battery-free, flexible, finger-actuated, and self-pumping &\cite{Ning2024} \\
\bottomrule
\end{tabular}
\end{table}

\subsubsection{Impedance Analysis}
% Two techniques do not fit neatly into the above categories: body impedance measurement scanning \cite{heo_novel_2021} and electroacupuncture \cite{noauthor_support_2023}.

\citet{heo_novel_2021} focused on analyzing solely vitamin D (calcidiol) status using body impedance. They explored the correlation between vitamin D levels in blood, body composition, blood parameters from checkup, and arm impedance (from wrist to elbow) in 26 patients (14 male) to calibrate an impedance measurement frequency for vitamin D \cite{heo_novel_2021}. The motivation was that "body fat accumulates vitamin D," and body fat can be measured by impedance measurement \cite{heo_novel_2021}.

% Vitastiq is the most immediately relevant device for micronutrition insights, with the company claiming that the device can non-invasively measure 26 different vitamins and minerals \cite{noauthor_support_2023}. The basis of the device's technology is electroacupuncture. It measures the electrical resistance at an 'acupuncture point' and compares it to a 'calibration point'. These points are very specific, and the user is taken through a tutorial to locate them properly \cite{kritsonis_hands-_2017}. Minimal details about how the device functions and electroacupuncture provide insight into micronutritional status are available. The information the user receives is only qualitative, and consumer reviewers noted a lack of confidence in the results \cite{kritsonis_hands-_2017}. As such, the manufacturer notes that Vitastiq is not a medical device and should only be used as an indicator of "general vitamin trend" \cite{noauthor_support_2023}.

\begin{table}[!htp]\centering
\caption{Other Electrochemistry-Based Methods}\label{tab: emerging-electro-oth}
\tiny
\begin{tabular}{p{0.55in}p{0.5in}p{0.45in}p{0.5in}p{0.5in}p{0.6in}p{0.7in}p{0.7in}p{0.2in}}\toprule
% \multicolumn{9}{c}{Electrochemical Methods} \\
% \cmidrule{1-9}
Method &Platform&Targets &Analytes &Evaluation &Control &Results &Notes &Source
\\\midrule
Arm impedance measurement scan &Non-mobile, clinical device &Vitamin D (calcidiol) &Impedance measurement at 21.1 Hz &Human subjects &Vitamin D status by blood test (method unspecified) &$R^2$ of 0.75 (regression with Vitamin D level) &Also evaluated regression models with medical checkup and body composition analysis data &\cite{heo_novel_2021} \\
% Electroacupuncture &PoC &26 different vitamins and minerals &Electrical resistance on the surface of the skin &Not published &Not published &Not published &Should only be used as an indicator of "general vitamin trend" &\cite{noauthor_support_2023} \\
\bottomrule
\end{tabular}
\end{table}

\subsubsection{Comparison}
Across voltammetry and amperometry, vitamin C, B vitamins, and zinc are the most common targets of electrochemistry-based methods. However, a few do target vitamin D, which is the only target of the impedance analysis technique. Voltametric devices were more often applied to blood or urine biosamples, and the associated studies relied less on supplementation as a control compared to amperometry-based methods. Amperometric devices appeared to focus more on a wearable form-factor. The single study using impedance analysis reported results which were less accurate than those reported by voltametric or amperometric approaches to vitamin D assessment \cite{heo_novel_2021}. This may indicate that body impedance analysis is not a promising method for future research, but we add that the results of this study may be more trustworthy than others due to its larger sample size and evaluation in the context of optimal reference standard vitamin D status assessments using blood. For these reasons, we cannot conclude whether impedance analysis is less promising than amperometry or voltammetry.

\subsection{Spectroscopy-Based Methods} \label{subsec_spec}
% \paragraph{Background}
Spectroscopy is the most common method of optimal reference standard micronutrient status assessment. However, there is still a lot of work to do to make spectroscopic approaches more accessible and less invasive. We begin with some definitions of common terms in the field. \textit{Spectroscopy} is "the investigation and measurement of spectra produced by matter interacting with or emitting electromagnetic radiation" \cite{noauthor_spectroscopy_2019}. \textit{Spectrometry} is the application of spectroscopy; the way in which quantitative measurements are obtained. When we speak of a \textit{spectra}, we mean any measurement that is a function of wavelength or frequency. A sample will absorb or emit these spectra when electromagnetic radiation of a known wavelength is applied. Spectra are measured by a detector, the \textit{spectrometer}. Because the level of radiation applied is known, analyzing the resultant spectra after it interacts with the sample provides information about the sample. 
We have previously described MS and LC-coupled spectroscopy, the types of spectroscopy used by most optimal reference standard clinical biochemical analyses (Section \ref{clinical/biochemical analysis}. As mentioned, these methods are expensive, non-specific, complex, and often require an invasively-collected biosample. While there have been strides to make MS more accessible \cite{crocombe_portable_2018}, we focus on alternative spectroscopic techniques that may yield micronutrient insights.
% Most spectroscopic methods for micronutrient assessment suffer from inaccessibility. Though less accessible, these give us insights into future novel technologies that can become more accessible.

\begin{figure}
    \centering
    \includegraphics[width=0.7\linewidth]{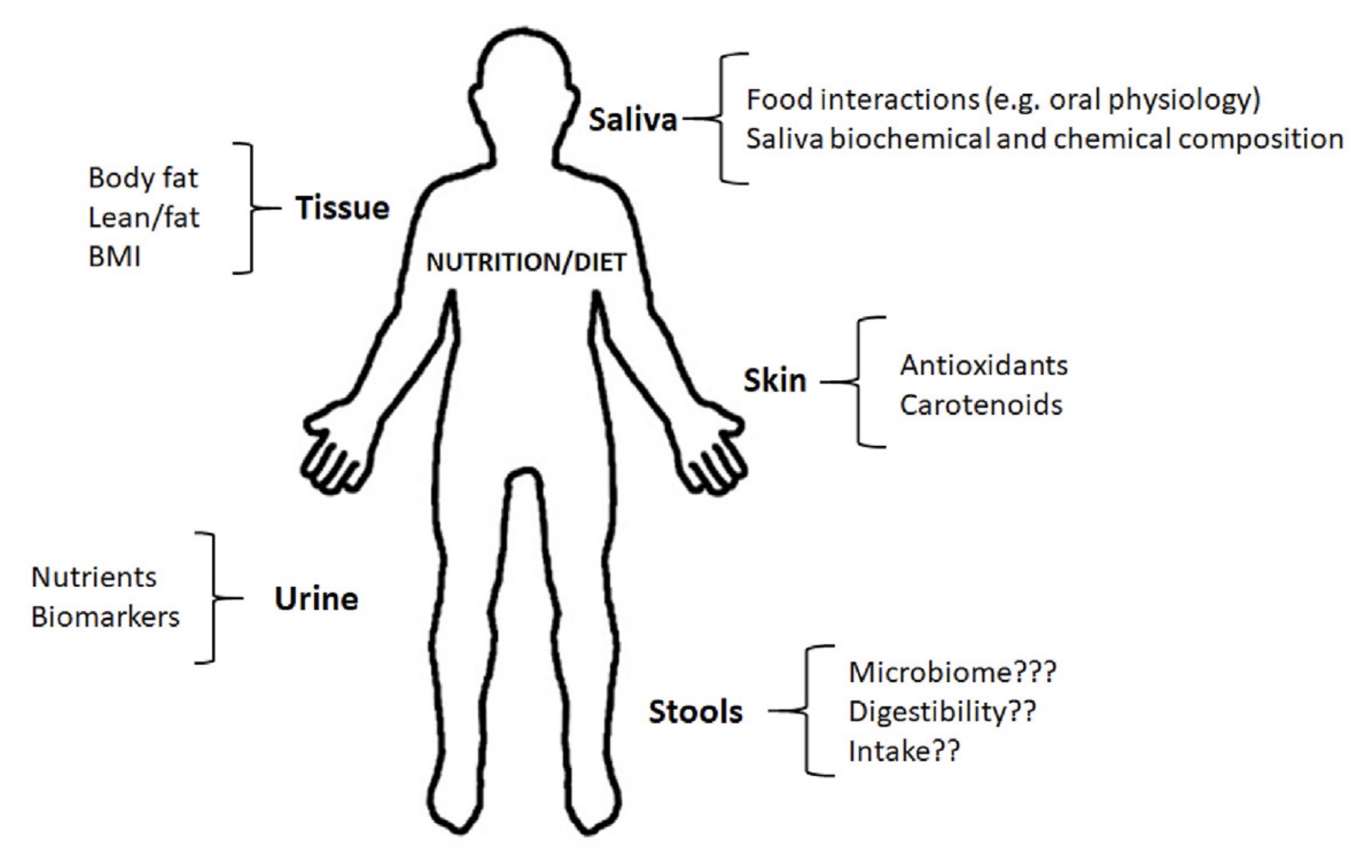}
    \caption{Vibrational spectroscopy targeted at saliva, tissue, skin, and urine have current applications in precision nutrition, while stools hold potential nutritional insights (indicated by question marks). Used with permission from \citet{dongdong_possibilities_2023}}
    \label{fig:vibrational-applications}
    \Description{The figure shows a clipart outline of the human body with five categories labeled: Saliva, Tissue, Skin, Urine, and Stools. Each of these categories has brackets detailing the specific nutritional insights that can be gained from each sample. Saliva provides information on food interactions and saliva biomechanical and chemical composition, tissue lends information about body fat, lean/fat, and BMI, skin provides information on antioxidants and carotenoids, urine shows nutrients and biomarkers, and stools provide insights on the microbiome, digestibility, and intake.}
\end{figure}

Categorized under emission spectroscopy, \textit{fluorescence spectroscopy} has been demonstrated for the measurement of primarily B vitamins in non-biological samples such as multivitamins and energy drinks, although vitamin B1 was measured in urine \cite{zhang_review_2018}. The B vitamins continue to get attention in \textit{near infrared (NIR) spectrophotometry} (750-2500 nm wavelength), where their measurement has been reported as well \cite{zhang_review_2018}. One review recognized the ability of \textit{vibrational spectroscopy}, which includes \textit{infrared (IR)} and \textit{raman spectroscopy}, to act as a tool for biofluid analysis in precision nutrition \cite{dongdong_possibilities_2023} (Fig. \ref{fig:vibrational-applications}). However, like other studies in nutrition, this review had few considerations for micronutrients and the approaches covered were largely concerned with general nutritional status or macronutrients. \citet{tsiminis_measuring_2017} noted the potential of raman spectroscopy for the measurement vitamin B12, though this has yet to be realized at in-body concentrations due to the low sensitivity of raman spectroscopy. In biosamples, spectrophotometry was applied to measure vitamin C \cite{holler_micronutrient_2018}. Spectroscopic skin tests and raman spectroscopy have also been noted as promising techniques in the assessment of provitamin A carotenoid status in the body \cite{holler_micronutrient_2018}. Measurement of several water-soluble vitamins in synthetic mixtures and dosage forms was achieved with \textit{derivative and multivariate spectrophotometry} \cite{zhang_review_2018}. Derivatives of UV spectrophotometry (185-400 nm wavelength) have also been particularly useful in the analysis of caffeine and B vitamins in energy drinks. The open-source Lumos platform also enables on-body spectroscopy \cite{watson_lumos_2022}.

% Another method involves developing a sensor that portrays spectroscopic capabilities when combined with micronutrients. \citet{Mughal2024} proposed a reduced graphene oxide to act as an optical sensing material of vitamins K1, K2, B6, and D3. The material was developed using the bismuth nanoparticle embedded polypyrrole nanocomposite (rGO/pPy/Bi NC). They collected blood from volunteers and obtained plasma and serum. Each vitamin was analyzed in a different concentration and ph level as it was combined with either an acid, base, or buffer. These were tested with  UV-vis spectrophotometry, and the proposed material was found to be a highly selective sensing probe that could be well differentiated with UV-vis spectrophotometry \cite{Mughal2024}. Though, we note that this work did not compare measured vitamin levels with clinical values.

\subsubsection{Accessible Spectroscopy}
Significant progress has been made to make spectrometry as a study more accessible and compact \cite{crocombe_portable_2018}. Major subfields of spectrometry (visible, Raman, mid-IR, NIR, MS, and hyperspectral imaging) have seen the development of portable or handheld devices. In some cases, such as visible, near-IR, and hyperspectral imaging, these can even be smartphone-based. This paves the way for more accessible, noninvasive techniques. A summary of spectroscopy-based methods is found in Table \ref{tab: emerging-spectro}. 

There was one application of spectrometry for vitamin D (calcidiol) measurement in our interest area of accessible approaches \cite{walter_all-optical_2020}. With human serum samples in mind, their sensor used surface plasmon resonance (SPR) together with smartphone-based spectrophotometry to assess vitamin D content. The general design of the sensor involved an optical waveguide to direct light from the smartphone flash through one or more SPR sensors, a diffraction grating, and finally into the smartphone camera where a spectra of pixel intensities can be extracted. The device was evaluated on spiked serum samples.
% 0, 25, 50, and 100 nM of vitamin D, covering the physiological range of 10 to 40 ng/mL. 
As the concentration decreased, the center of mass of the spectra shifted right, allowing for the detection of these concentrations. The paper claims a comparable LoD to optimal reference standard methods of LC-MS, but no quantitative estimates of vitamin D concentration, sample size, or statistical analysis of the results are provided. 

Some benchtop approaches to spectroscopic analysis of biofluids for micronutrient assessment have also been demonstrated \cite{peterson_photonic_2014, peterson_enhanced_2015, bi_performance_2016, Mughal2024}.
\citet{peterson_photonic_2014} presented a photonic crystal-based sensor for ferritin assessment. Photonic crystals were designed to accumulate a target biomolecule on their surface, which changes their reflected peak wavelength value (PWV) under a spectrophotometer.
The authors subjected their sensor to robust evaluation, utilizing human liver ferritin, commercial serum controls (Liquichek), and three different ELISA tests. The developed sensor held up against the ELISA tests, 
% with comparable linearity within 0-1200 $\mu$g/L (r2=0.98) at a wider dynamic range (26-2000 $\mu$g/L), 
but with a higher LoD (26 $\mu$g/L) that did not cover the lower end of iron deficiency. Bias by Bland-Altman analysis was similar to that of the BioVendor ELISA, and recovery from known serum controls was greater than 94\%.

Motivated by a specificity issue in the preceding sensor, \citet{peterson_enhanced_2015} employed iron-oxide nanoparticles to minimize non-specific signals. This time, the goal was soluble transferrin receptor (sTfR) quantification from serum, an indicator of iron supply to tissues \cite{erdman_jr_iron_2012}. 
It should be noted that sTfR is not influenced by inflammation to the same extent as serum ferritin, and is therefore a potentially more ideal biomarker. However, sTfR is more expensive to assay.
For this experiment, biomolecule interaction on the assay was quantified using the Biomolecular Interaction Detection system from SRU Biosystems Inc \cite{peterson_enhanced_2015}. The authors compared results from their assay on Liquichek control sera to ELISA and the previously developed photonic crystal assay. 
% Assessed using control sera from Liquichek 1, 2, and 3. Bias for these control samples was “0.18, 0.19, and -0.04 $\mu$g/mL, respectively”. LoD of 14 ng/mL was much lower than ELISA (min 100 ng/mL). 
The authors claimed that the bias of the assay was not “statistically different from the reference ELISA tests” \cite{peterson_enhanced_2015}. 

Moving from iron to vitamin C, \citet{bi_performance_2016} demonstrated the immobilization of ascorbate oxidase in a microfluidic channel, enabling the quantification of vitamin C with UV-visible spectroscopy. During analysis, the biosample was diluted in phosphate buffered saline (PBS) and “pumped through the microfluidic channel”. 
% Calibration was conducted at the absorbance peak for vitamin C of 266 nm using stock solutions (r2=0.9893).
A serum sample was obtained by a single healthy, female volunteer for evaluation. The sample was pretreated to remove proteins 
% with a molecular weight greater than 10 kDa
, and a few drops were added to the sensor. Even with extensive pretreatment, there was evidence of interference at 280 nm, close to the analysis peak of vitamin C at 266 nm. 
% The concentration of vitamin C in serum was determined to be within the physiologically relevant range. 
% The authors claim a limit of detection of 2.4+/-0.1 mcM (0.42 mg/L) and a limit of quantification of 7.9+/-0.5 mcM (1.4 mg/L)
No optimal reference standard measurement was provided for comparison. 
\citet{Mughal2024} mixed different electrolytic solutions with plasma and serum, and when paired with a novel, reduced graphene oxide, vitamins K1, K2, B6, and D3 could be individually identified using UV-visible spectrophotometry. This method did not compare the measured levels from 5 subjects with clinical values.

%Please add the following packages if necessary:
%\usepackage{booktabs, multirow} % for borders and merged ranges
%\usepackage{soul}% for underlines
%\usepackage[table]{xcolor} % for cell colors
%\usepackage{changepage,threeparttable} % for wide tables
%If the table is too wide, replace \begin{table}[!htp]...\end{table} with
%\begin{adjustwidth}{-2.5 cm}{-2.5 cm}\centering\begin{threeparttable}[!htb]...\end{threeparttable}\end{adjustwidth}
\begin{table}[!htp]\centering
\caption{Spectroscopic Methods}\label{tab: emerging-spectro}
\tiny
\begin{tabular}{p{0.6in}p{0.5in}p{0.4in}p{0.5in}p{0.55in}p{0.5in}p{0.75in}p{0.6in}p{0.2in}}
\toprule
Method & Platform & Targets & Analytes & Evaluation & Control & Results & Notes & Source \\
\midrule
SPR-coupled spectrophotometry & Smartphone-based & Vitamin D (calcidiol) & Serum & Spiked from human subjects & LC-MS & Claim comparable LoD to control, spectra shifted right with decreasing concentration & No quantitative vitamin D estimates or statistical analysis & \cite{walter_all-optical_2020} \\
Photonic crystal &Benchtop &Iron (ferritin) &Serum &Liquichek control sera & Known ferritin concentrations and multiple ELISA tests &Comparible recovery (>94\%) and bias (by Bland-Altman analysis) to best-performing ELISA &LOD higher than cutoff for anemia, lower end of deficiency &\cite{peterson_photonic_2014} \\
Sandwich iron-oxide nanoparticle immunoassay &Portable &Iron (STfR) &Serum &Liquichek control sera &ELISA &SD of 0.45 mcg/mL vs ELISA &&\cite{peterson_enhanced_2015} \\
UV-vis spectrophotometry &Benchtop &Vitamin C (ascorbic acid) &Serum &Human subjects &None &Within phsyiologically-relevant concentrations &Strong focus on the effectiveness of the immobilization technique, not vitamin C measurement &\cite{bi_performance_2016} \\
UV-vis spectrophotometry & Benchtop & Vitamins K1 (phylloquinone), K2 (menaquinone), B6, D3 (cholecalciferol) & Serum/plasma & Human subjects & Various sensing techniques (SWV, HPLC-MS/MS (ESI), SWAdSV, Thermal wave transport analysis, DP AdSV, SWASV, DPV, Electrochemical, Colorimetric aptasensor) & Limits of detection of vitamins K1, K2, B6, and D3 are 0.075, 0.1, 0.12, and 0.15 ng/mL, respectively. Limits of quantification are 0.29, 0.3, 0.38, and 0.48 ng/mL for vitamins K1, K2, B6, and D3, respectively. & Clinical gold-standard biomarkers are not used for all vitamins (except for K, where LoD is too high for deficiency); used bismuth nanoparticle embedded polypyrrole nanocomposite (rGO/pPy/Bi NC) as an optical sensing material &\cite{Mughal2024} \\
\bottomrule
\end{tabular}
\end{table}

% Optical spectroscopy &Wearable (watch and finger clamp) &Glucose &Optical analysis of the epidermis &Human subjects &Commercial glucometer &Correlations with control of 0.843 at 470 nm and -0.927 at 515 nm &Pilot study for platform &\cite{watson_lumos_2022} \\
% Spectrophotometry &Portable &Glucose &Glucose solution and blood analog &Known concentrations &Known concentrations &$R^2$ of 0.96 and RMSE of 11.1 mg/dL for glucose estimation, F1 of 0.99 for glycemia classification &Interesting experimental design, found different optimal peaks than \cite{watson_lumos_2022} &\cite{shokrekhodaei_non-invasive_2021} \\

\subsubsection{Comparison}
A wide range of spectroscopic methods have been applied for the quantification of different micronutrients, making it difficult to compare each approach. Iron assessment appears to be feasible in serum, but further development would be useful to increase the accessibility of a spectroscopic approach that targets this nutrient. Furthermore, clinical research could consider alternative biosamples for iron, to increase the noninvasiveness of spectroscopic techniques. Studies using spectrophotometric methods have yet to be thoroughly validated so as to indicate the utility of spectrophotometry over any alternative method.

\subsection{Biofluid Analytic Methods} \label{subsec_analytic}
% \subsubsection{Estimating Biochemical Results}
AI and ML can be used to detect micronutrient levels in individuals.
A common approach to estimating nutritional status is by making predictions from pre-existing biofluid analysis or demographic data using ML. Such methods have been applied to derive micronutrient-specific insights \cite{truijen_predicting_2021, kurstjens_automated_2022, khan_machine_2021, patino-alonso_predictive_2022, pullukhandam2024, luo_practical_2023}. A summary of biofluid analytic methods can be found in Table \ref{tab: emerging-ai}, where we observe a common trend of utilizing classical machine learning algorithms (logistic regression, gradient boost, naive bayes, random forest) for predictive modeling.

\subsubsection{Single Micronutrient Malnutrition Detection} 
Some studies focus on detecting deficiency of a specific micronutrient. Two such papers investigate iron status \cite{luo_using_2016, pullukhandam2024}, while a third examines vitamin D \cite{patino-alonso_predictive_2022}.
\citet{luo_using_2016} used hospital outpatient data collected over three months to predict whether a patient had normal or abnormal ferritin (iron) status using logistic regression. The collected data included age, sex, ferritin test results (used as markers), and other ‘predictor’ tests that were conducted in the main hospital lab only (n=5128, sex breakdown not reported). These authors considered the broader clinical usefulness of ML-powered insights, claiming that “predicted ferritin results may sometimes better reflect underlying iron status than measured ferritin” \cite{luo_using_2016}. This conclusion was based on dual independent review by 2 pathologists on 26 selected cases where measured and predicted ferritin were ‘highly discrepant’. They propose that predicted levels could be used to flag lab-measured ferritin for further review.
\citet{pullukhandam2024} use gradient boost on NHANES complete blood count (CBC) data to classify and explain IDA (n=19995, sex breakdown not reported). They found that the most critical features for IDA are low levels of hemoglobin, higher age, and a higher red blood cell distribution width.

For the prediction of vitamin D deficiency, \citet{patino-alonso_predictive_2022} applied ML (logistic regression, naive bayes,
and random forest) to anthropomorphic data of older Europeans (35-75 y/o; 50/50 males/females) given anthropomorphic features. A total of 501 participants contributed their “waist circumference (WC), body mass index (BMI), waist-to-height ratio (WHtR), body roundness index (BRI), visceral adiposity index (VAI), and the Clinical University of Navarra body adiposity estimator (CUN-BAE) for body fat percentage”. Vitamin D as 25(OH)D was measured by immunoassay and the threshold of deficiency was set to be 20 ng/mL (34.7\% prevalence). We note that this threshold more closely aligns with \textit{insufficiency} (Table \ref{tab: assessment}).
Logistic regression analysis found that the most significant features differed by sex. All but CUN-BAE were associated with vitamin D deficiency in males, while only CUN-BAE was associated in females. ML models for deficiency prediction were trained on each feature individually. The authors discovered that Naive Bayes was the top performer by AUC for WC, BMI, WHtR, and BRI but was bested by logistic regression for VAI and CUN-BAE.

\subsubsection{Multiple Micronutrient Detection} 
Since it is rare for micronutrient imbalance to occur in isolation, researchers have studied the ability to predict malnutrition of multiple micronutrients \cite{truijen_predicting_2021, kurstjens_automated_2022}.
\citet{truijen_predicting_2021} focus on micronutrient malnutrition in older populations, citing how malnutrition in older adults is often diagnosed too late despite the existence of screening methods. The goal of the study was to use logistic regression to classify each sample as having either no micronutrient deficiency or one or more deficiencies among vitamins C, B6, B12, selenium, and zinc, confirmed by blood tests. These particular micronutrients were selected because they interact less with each other (we add that B-vitamins do interact; Table \ref{tab:char-vitamins-water-soluble}), were among the most prevalent deficiencies, had clinically relevant cutoff points for deficiency. Logistic regression was applied to routine biochemical and diagnostic data from 9 years of United Kingdom NDNS for ages $\geq$50 (n=1518, 57.2\% female). This dataset suffered from ethnic disparities, with the authors noting that $\geq$95\% of NDNS participants were white. 

\citet{kurstjens_automated_2022} aimed to develop a random forest algorithm to assess risk of low body iron storage (ferritin plasma levels) in anemic primary care patients using CBC and CRP test results from 3 medical laboratories (n=2,935, $\sim$1,493 female. Two algorithms were developed, each based on laboratory ferritin results from different chemistry analyzers (from Siemens and Roche). Interestingly, the two most important features were both derived from CBC test results (Table \ref{tab: emerging-ai}). The authors took an important step to consider how such a model could assist a clinician by asking 4 professionals to indicate whether a patient had low ferritin based on CBC and CRP, with and without algorithm results. The found that the algorithm alone was more accurate than both scenarios. Detection of low vitamin B12 and B9 levels were also considered, but this yielded poor results with AUCs of 0.52 and 0.57 respectively. 

% Detecting multiple micronutrients in a complex sample is a critical challenge for spectroscopic methods, but computational approaches have aimed to solve this \cite{monakhova_chemometrics-assisted_2010}. Independent component analysis (ICA) is the traditional means of separating a multivariate signal into individual constituents, but ICA is often unreliable for spectra of a complex solution. Mutual Information Least Dependent Component Analysis (MILCA) is proposed as an alternative, searching for the least dependent components in a sample. The algorithm was applied to determine concentrations of vitamins C, B3, B6, A and E in various mixtures using spectra obtained with UV-visible spectrophotometry in the 200-450 nm range. In solutions made up of a combination of known concentrations of vitamin standards, MILCA outperformed other signal decomposition approaches. They also note that it is highly sensitive to factors that impact the state of a sample such as pH, hydrolysis, and redox reactions. This evidence that individual micronutrients can be quantified in a complex mixture with a method as relatively simple as spectrophotometry is promising for future accessible methods of status assessment.

\subsubsection{Adjusting Biomarkers for Inflammation}
When conducting biochemical analysis for micronutrient assessment, a common issue is the impact of inflammation on biomarker measurement. One R package aims to solve this problem and improve interpretability by adjusting biomarkers of micronutrients in the context of inflammation \cite{luo_practical_2023}. The package implements inflammation adjustment for retinol-binding protein, serum retinol, serum ferritin, sTfR, and serum zinc, using acid glycoprotein (AGP) and/or CRP as biomarkers for inflammation. The authors have also published a paper describing a procedure on when and how to apply their technique \cite{luo_practical_2023}.
%Please add the following packages if necessary:
%\usepackage{booktabs, multirow} % for borders and merged ranges
%\usepackage{soul}% for underlines
%\usepackage[table]{xcolor} % for cell colors
%\usepackage{changepage,threeparttable} % for wide tables
%If the table is too wide, replace \begin{table}[!htp]...\end{table} with
%\begin{adjustwidth}{-2.5 cm}{-2.5 cm}\centering\begin{threeparttable}[!htb]...\end{threeparttable}\end{adjustwidth}
\begin{table}[!htp]\centering
\caption{Analytic Methods}\label{tab: emerging-ai}
\tiny
\begin{tabular}{p{0.5in}p{0.5in}p{0.75in}p{0.75in}p{0.6in}p{0.75in}p{0.75in}p{0.2in}}\toprule
% \multicolumn{8}{c}{AI/ML Methods}\\\cmidrule{1-8}
Method &Targets &Data &Important Features &Ground Truth &Results &Notes &Source \\\midrule
Logistic regression &Normal or abnormal ferritin &Hospital outpatient data &"total iron-binding capacity, mean cell hemoglobin, and mean cell hemoglobin concentration" from \citet{luo_using_2016} &Ferritin test results &AUC of 0.97 &Predictions could be used to flag lab ferritin for review &\cite{luo_using_2016} \\
Gradient boost & IDA & US NHANES (CBC and serum ferritin) dataset and Kenyan nutrition dataset (for evaluation) & Low blood level of hemoglobin, higher age, and higher red blood cell distribution width & Serum ferritin & Precision AUC of 0.87 (training); recall of 0.98 (training) and
0.89 (evaluation) & Heavy class imbalance (4.9\% IDA vs. 95.1\% non-IDA) & \cite{pullukhandam2024} \\
Logistic regression, naive bayes, and random forest &Vitamin D deficiency &Anthropormorphic measurements of older Europeans &CUN-BAE for females, all others for males &Blood 25(OH)D by immunoassay &Max AUC of $\sim$0.53 for all features; LR best for VAI and CUN-BAE, NB for all others &Did not assess predictive ability of multiple features at once &\cite{patino-alonso_predictive_2022} \\
Logistic regression &Presence of micronutrient deficiency &UK NDNS, ages $\geq$50 &Low protein, energy intake, TC, hemoglobin, HbA1c, ferritin, vitamin D and high CRP &Blood test results for Vitamins C, B6, B12, selenium, and zinc &AUC of 0.79 &$\geq$95\% of NDNS participants were white &\cite{truijen_predicting_2021} \\
Random forest &Classify low body iron storage (plasma ferritin) &CBC and CRP tests in anemic primary care patients &Mean corpuscular hemoglobin and mean corpuscular volume &Ferritin results from two laboratory chemistry analyzers (two separate models) &AUC of 0.9 and 0.92 for each model, models were more accurate than professionals with and without access to results &Attempted Vitamin B12 and B9 deficiency detection with poor results (AUCs of 0.52 and 0.57) &\cite{kurstjens_automated_2022} \\

% Random forest &Childhood anemia &2011 Bangladesh demographic and health survey (BDHS) &Child morbidity regarding fever, household toilet facilities, child age &Reported anemia by BDHS &AUC of 0.6875 &Results limited by lack of clinical and dietary variables &\cite{khan_machine_2021} \\
% ANN &TC, LDL, HDL, and triglyceride estimation &Demographic features and indicators of obesity and heart disease &BMI–WHtR (waist to height ratio), BMI–SADHtR (sagittal abdominal diameter to height ratio) and WHtR–SADHtR &Standard lipid panel &Accuracies of 0.82 for TC, 0.793 for LDL, 0.81 for HDL, and 0.445 for triglycerides &Used established relationship between obesity and lipid panel results &\cite{vrbaski_lipid_2019} \\
% ANN &Predicting LDL-C &Dataset of electronic health records &NA &Direct LDL-C measurements &Correlation of 0.97 &Smaller confidence intervals than traditional formulas &\cite{oh_estimation_2022} \\
% Random forest regression &TC estimation &Clinical and anthropometric data collected by nutritionists during weight loss interventions &Basal metabolic rate, fat in the leg and trunk, total energy expenditure, waist-to-hip ratio, systolic BP, and age &Direct TC measurement &MAPE of 4.39\% &Glucose and triglycerides were also critical features, but must be assessed invasively &\cite{garcia-durso_non-invasive_2022} \\
\bottomrule
\end{tabular}
\end{table}

\subsubsection{Comparison}
\citet{luo_using_2016} and \citet{kurstjens_automated_2022} both utilize data obtained from patients during the course of their health care, whereas other sources of data came from surveys (e.g. the US NHANES). Collecting data during health care can result in a large amount of data to mine for micronutritonal insights without needing to rely on national surveys (which may not contain a wide variety of features) or conducting an independent assessment of a population. The prediction of iron (as ferritin) appears to be more successful than the prediction of other biomarkers. This is likely due to iron's large influence on hemoglobin, which was an important feature for all ML models that were trained to predict ferritin status \cite{luo_using_2016, kurstjens_automated_2022, pullukhandam2024}. 

\subsection{Biofluid Analysis Limitations} \label{subsec_biofluid limits}
The largest general limitation of existing biofluid-based assessments is that methods effectively indicating in-body status of a micronutrient often utilize an invasively-collected biosample (i.e. blood) for analysis. While noninvasive biosamples such as sweat and saliva were applied, the clinical research on micronutrients suggests that these samples may not accurately reflect micronutrient status compared to blood. As such, validation studies are required.
% Urine as a biofluid is under-utilized in methods outside of clinical biochemical analysis. It can be noninvasively collected, it is more researched than saliva or sweat, and, in some cases, it is a valid alternative to blood in clinical gold-standard biochemical assessments. Future methods should shift focus to biofluids that have proven clinical relevancy.
% It is for this reason that a proper clinical assessment of micronutrient status considers physical exams, dietary logs, clinical history, and biochemical analysis together \cite{reber_nutritional_2019}.
% Many of the novel micronutrient status assessment methods favor specialized, novel sensors and data analysis approaches, as opposed to being clinically-motivated and accurately assessing in-body status. 
We also observe that biofluid-based assessments are more specialized in nature. Frequently, a bespoke assay, device, or sensor is implemented for the assessment of only a single micronutrient. This is understandable considering the inherent difficulties in detecting and quantifying micronutrients that have unique metabolic pathways, are present in such limited quantities, and play different roles in bodily function. Furthermore, there is also a common need for a specialized reagent, buffer, or electrolytic solution to enable the analysis of each micronutrient (especially for assays and electrochemical methods). It is for this reason that multiplexed or simultaneous detection of multiple micronutrients is limited, despite their potential to enable more holistic status monitoring. However, we acknowledge that combining several specialized assays and unique sensor designs adds complexity to the measurement process, creating barriers to commercialization as well as widespread adoption. 

On the topic of commercialization, we posit that the lack of commercial, multiplexed, or PoC technologies for micronutrient assessment is a result of both the ingrained reliance on laboratory blood testing within health care institutions as well as the expensive and thorough human subjects evaluations necessary to bring an effective assessment technique to market.
We find that current methods do not consistently test the clinical biomarker for a given micronutrient (e.g. measuring vitamin D via calciol instead of the optimal reference standard calcidiol a.k.a. 25(OH)D). Even when the proper biomarker is examined, the device itself may be evaluated on concentration intervals that are not pertinent to the clinical spectrum of deficiency to excess (e.g. the linear range of a sensor may be able to indicate sufficient status, but the LoD is too high to infer deficiency). Again, we recognize that some methods are specifically designed to accurately detect a deficient status in a resource-constrained environment, or assess micronutrients where excess does not pose a significant health risk. In these cases, a lack of sensitivity in the upper spectrum is less of an issue.
Some research fails to include a clinically relevant biochemical test (e.g. ELISA or HPLC) for their target micronutrient as a scientific control. 
By contrast, a non-clinically relevant biochemical test either does not follow best practices in clinical nutrition, targets an alternative biomarker, or uses an alternative process from what is typically considered acceptable in modern clinical practice. Such a non-clinically relevant test can also simply rely upon supplementation rather than any substantive evaluation of micronutrient status (see below). Each of these oversights actively limits the clinical value, and therefore the real-world utility, of the proposed solution.

The lack of a clinically relevant biochemical test is confounded by the prevalence of supplementation-based experiments, where an assessment is conducted on a patient before, during, and/or after intake of the target micronutrient. The prescribed intake may be as specific as a supplement pill or as general as a food that is known to contain high amounts of a micronutrient. Because of individual variance in the absorption of different micronutrients, it is difficult to know the true impact of micronutrient intake, and therefore the practical accuracy of a status assessment, without also conducting a relevant biochemical test.
% Future technical methods of micronutrient assessment should shift focus, and prioritize meeting the clinical need.
Lastly, the study designs themselves often feature low sample sizes (1-5 participants) that make it difficult to assess the analytical validity of each method. We recognize that design-oriented research faces a larger hurdle when attempting to conduct human-subjects experiments that are clinical in nature, and several studies echo this point. 
Biofluid analytic methods have sample sizes that are (necessarily) much larger, but share issues related to the equitable representation of demographic groups within their sample.
Similar to the impact of omitting a relevant biochemical test, we argue that an emphasis on new technologies over more rigorous and larger-scale validation studies makes it difficult to judge the true effectiveness of a novel assessment method. 

Moving to the specific approaches, we find that clinical biochemical analysis is invasive, expensive to analyze, and the methods of analysis and thresholds for imbalance are debated \cite{reber_nutritional_2019}. Additionally, biomarkers are generally sensitive but not specific, and their analysis requires considering an extensive list of factors that alter the ability of a marker to indicate nutrient status (such as inflammation, disease, or medications) \cite{elmadfa_developing_2014}. 

Although there are several wearable, smartphone, and point of care assay devices, their accessibility in some cases is limited by specialized assay chips and most require invasive biofluids for analysis. Saliva and sweat have been proposed as alternatives to blood, but their ability to reflect in-body status of a given micronutrient is often unknown, debated, or disproved (Section \ref{subsec_biofluids}). 
% Electrochemical biosensors are most impacted by the ambiguous quality of sweat as a biosample for circulating micronutrient status. The physiology of sweat is discussed in depth in Section \ref{background section}, where we uncover that there is little to no evidence of a correlation between sweat and blood composition.
% , and there is "little support for using sweat as a surrogate for blood" \cite{baker_physiological_2020}.
% Interestingly, many works in electrochemical biofluids analysis attempt to position sweat as a valuable biomarker that can take the place of blood when assessing micronutritional status. 
One survey offers additional insight into the issues faced by wearable sweat sensors: low sweat rates, sample evaporation, skin contamination impacting sweat content, and the difficulty to access fresh sweat \cite{bariya_wearable_2018}. 
Assays and electrochemical devices also suffer from being more specialized in nature (complicating manufacture and integration) and not utilizing more ubiquitous methods of health monitoring such as smartphones and smartwatches. 

The field of spectroscopy shows great potential (especially IR and Raman), though we argue that there is not yet enough accessible, micronutrient-specific research that provides insights into in-body status. Most work in this area required benchtop analyzers instead of on-body approaches. With the latter however, one must take great care not to harm a user with the spectroscopic approach. UV spectroscopy in particular requires the application of UV light, which is broken into three types depending on its wavelength. Exposure to UVA (400 - 320 nm) and UVB (320 - 280 nm) radiation can cause damage to DNA and result in cancer, because of how this energy penetrates the skin \cite{dorazio_uv_2013}. UVC radiation from sunlight (280 - 100 nm) does not possess these same dangers, especially because it is absorbed by the skin, but direct exposure to artificial UVC radiation can still cause burns on the skin and eyes \cite{united_states_food__drug_administration_ultraviolet_2020}. The exact amount of UV radiation which is considered harmful varies with the area of exposure, amount of melanin content in the skin, and more, so we direct the reader to Table 2 in \citet{dorazio_uv_2013} for more information.

Prediction from clinical health data is able to combine and analyze a large breadth of features, but this data is often insufficient for micronutrition. The lack of micronutrition data availability poses a grand limitation for the ability to make strides in analytic techniques with AI/ML. As a result, \citet{brown_increasing_2021} and others urge for more micronutrition data.
% bringing us back to the overarching challenge of data availability. 
% Collecting data commonly requires a medical professional and/or lab access for analysis of relevant biomarkers. 
Additionally, the insights provided by these solutions are limited to a small set of micronutrients and/or indicate only a binary deficiency status, rather than a continuous one.

\section{Physiological Sensing for Micronutrient Assessment} \label{phys section}
Physiological assessment is as important as biofluid analysis, and some argue they should be considered together. 
% Notably, physiological sensing has primarily been applied to detect nutrient deficiencies and has not yet been explored for assessing nutrient excess. 
Many physiological symptoms of micronutrient deficiency (Tables \ref{tab: physio-ws} to \ref{tab: physio-min}) only manifest in severe cases, but the assessment of physiological symptoms remains a critical step in nutritional practice for two central reasons. First, it is useful when estimating the burden of micronutrient malnutrition in a population that is difficult to assess via biofluid methods. Second, physiological assessments provide insights into patient health that are unique and complementary to quantitative, biofluid-based assessments.
Though it is valuable, physiological analysis receives significantly less attention in emerging micronutrition research.
% Our discussion aims to enable further development of physiological sensing for micronutrition assessment by discussing the potential of optical sensors, such as those commonly found in ubiquitous devices like smartphones. 
A comprehensive overview of physical examination in clinical nutrition is out of scope for this review, so we direct the reader to \citet{reber_nutritional_2019} and \citet{hummell_role_2022} for more information.

Physiological sensing enables automatic detection of symptoms linked to micronutrient deficiency (Section \ref{subsec_phys symptoms} and Tables \ref{tab: char-ws} to \ref{tab: char-min}), supporting NFPEs. Incorporating metrics like heart rate, activity, sleep, and temperature can enhance personalization and reveal health trends over time.
% The importance of physiological sensing lies in its opportunity for automatic detection of physiological symptoms associated with micronutrient deficiency (Section \ref{subsec_phys symptoms} and Tables \ref{tab: char-ws} to \ref{tab: char-min}), aiding a Nutrition-Focused Physical Exam (NFPE). Furthermore, incorporating data on heart rate, physical activity, sleep quality, body temperature, etc. into micronutritional assessments has the potential to form a clearer picture of the individual patient, allowing for more personalized nutritional insights and the analysis of trends over time. 
\citet{king_application_2017} outline key criteria for effective wearable monitoring: noninvasive, user-friendly, reliable, and informative. \citet{witt_windows_2019} shows how raw data from sensors like PPG, ECG, accelerometers, EDA, and temperature can offer physiological insights about human physiology relevant to micronutrient imbalance. \citet{yokus_integrated_2021} also provide valuable design considerations for future wearable micronutrient assessment tools. The remainder of this section highlights innovative methods of optical sensors in assessing physiological signals.
% reviews techniques for yielding health insights from multiple sensors and their applications. Applications here focus on aspects of health that are not directly related to nutrition, however the authors suggest several criteria for practical wearable monitoring, such as being "noninvasive, intuitive to use, reliable, and provide relevant feedback to the wearer" \cite{king_application_2017}. \citet{witt_windows_2019} describes interesting ways that raw sensor data from wearables can be used to explore health. Different sensors such as PPG, ECG, accelerometer, EDA, and those collecting temperature and physical activity can give valuable insights into human physiology that can be associated with micronutrient imbalance. 
% Some examples are insights into HRV, sleep quality, stress levels, and behavioral patterns. % Some research has summarized the variety of common, external sensors available and their potential for integration into health and performance monitoring \cite{yokus_integrated_2021}.
While not explicitly wearables-focused, \citet{yokus_integrated_2021} provide some useful considerations for wearable devices that should be applied to future micronutrient assessing devices. 

\subsection{Applications of Optical Sensors in Assessing Physiological Signals} \label{camera_stuff}
\begin{figure}
    \centering
    \includegraphics[width=0.7\linewidth]{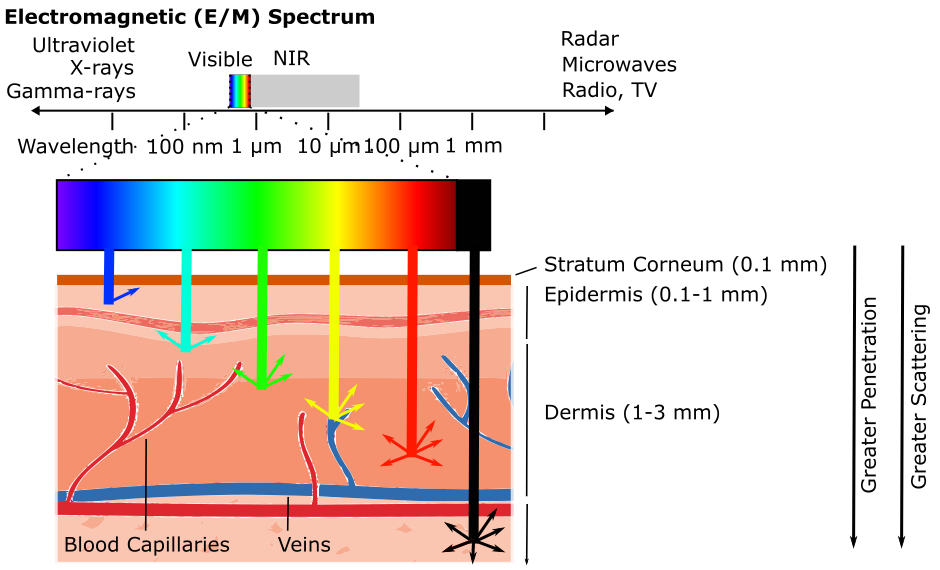}
    \caption{As light is applied to the skin, it is absorbed and reflected in different ways depending on its wavelength. Importantly, skin melanin content can impact the absorbance of visible light. For more information, please refer to \cite{austin_visible_2021}. Used with permission from \citet{mcduff_camera_2021}.}
    \label{fig:em-skin}
    \Description{The figure shows the distribution of different kinds of waves on the electromagnetic spectrum along a horizontal line. On the left are gamma rays, x rays, and ultraviolet light, in the middle of the spectrum is the visible light range (indicated by the classic rainbow), and on the right of the line are radar, microwaves, radio waves, etc. The wavelength of each is also shown along the line. Below this wavelength line, the figure provides a zoomed in image of the visible light spectrum with colors ranging from purple on the left to black on the right. Underneath the spread of these different colors is an image of skin with the stratum corneum at the top (.1 mm thick), the epidermis below it (.1-1 mm thick), followed by the dermis (1-3 mm thick) and including blood capillaries and veins. There are arrows that drop down from each of the colors on the visible light spectrum demonstrating how far they penetrate into the skin. Going from the left we have purple light with the smallest wavelength and the shallowest penetration into the skin. The figure also notes that with less penetration there is less scattering. There is a rather linear increase across the cyan, green, yellow, red and black colored wavelengths penetration and scattering. Black is located on the right and has the greatest penetration and scattering, extending all the way past the dermis.}
\end{figure}

% This subsection highlights innovative methods utilizing optical sensors for assessing physiological signals.
\citet{mcduff_camera_2021} reviews how camera sensors can be used for noninvasive physiological measurement, which can allow for nutritional insights. 
% There are two primary ways physiological signals can be extracted using optical sensors.
The analysis of motion artifacts can reveal minute subtleties in body motion over time that are caused by various physiological mechanisms (e.g. breathing). Also, camera sensors can measure the intensity and wavelength of light absorbed and reflected by our bodies, especially skin (Fig. \ref{fig:em-skin}). Differences in measured light over time can be observed and associated with physiological signals (e.g. heartbeat) and status (e.g. low blood oxygen saturation). However, it is important to note that skin melanin content can impact how light is absorbed and reflected \cite{austin_visible_2021}. A failure to properly account for these differences can (and has) resulted in racial biases (for example, pulse oximetry \cite{sjoding_racial_2020} and wrist-worn heart rate sensors \cite{koerber_accuracy_2023}). 

The type of optical sensor has a large impact on the signals that can be derived from it \cite{mcduff_camera_2021}. RGB cameras are found in most smartphones and, therefore, are the most prevalent. These operate largely in the visible spectrum of light (400 to 700 nm wavelengths), but they can often detect some light in the NIR range. NIR cameras are able to detect light in the 700 to 1000 nm range, and thermal cameras can go fully into the infrared spectrum of 2000 to 14000 nm. Thermal cameras, as their name suggest, can provide unique information over other sensors such as body temperature and sweat gland activation. However, this comes with a higher cost and lower resolution. Finally, multi and hyper-spectral cameras allow for the measurement of multiple wavelengths of light at once. This can also be achieved by combining signals from multiple sensors. 

Cameras have been extensively applied to the measurement of physiological vital signs \cite{mcduff_camera_2021, mcduff_iphys_2019, liu_rppg-toolbox_2023,nabian_open-source_2018, van_der_kooij_open-source_2019, guede-fernandez_photoplethysmography_2020}. One survey describes 5 physiological vital signs where optical sensor measurements been applied: pulmonary activity, EDA, blood oxygen saturation, glucose status, and cardiac activity \cite{mcduff_camera_2021}. PPG signal capture has been achieved with RGB cameras \cite{van_der_kooij_open-source_2019, guede-fernandez_photoplethysmography_2020}. NIR cameras are also useful for PPG in use cases like sleep monitoring that require environments with low visible light \cite{mcduff_camera_2021}. 
% Some notable works that apply PPG include an open-source method for remote PPG using a consumer-grade webcam \cite{van_der_kooij_open-source_2019}. Smartphone cameras have also been used to conduct PPG through the fingertip, with the smartphone's flash as a light source \cite{guede-fernandez_photoplethysmography_2020}. Measurements of pulmonary activity commonly utilize videos from an RGB camera to assess body motion in the torso, mouth, and nostrils. Multiple open-source toolkits have been published for physiological sensing using cameras \cite{liu_rppg-toolbox_2023,nabian_open-source_2018, mcduff_iphys_2019}.
% Key metrics are the breathing rate, breathing rate variability, and tidal volume (the amount of air moving through the lungs during the respiratory cycle). 
% Blood oxygen saturation represents the ratio of oxygenated to deoxygenated blood hemoglobin. 
When measuring blood oxygen saturation, multi or hyper-spectral cameras are preferred. Measuring multiple wavelengths of light at once benefits the simultaneous measurement of oxy and deoxy-hemoglobin, and therefore blood oxygen saturation. Recently, \citet{sharma_mobispectral_2023} proposed a smartphone-based hyperspectral imaging platform and used it to identify organic fruits. 
Optical sensor-based techniques could be use to detect and investigate other physiological symptoms in the fingertip \cite{hasan_smarthelp_2018, wang_hemaapp_2016, umar_real_2020}, mouth \cite{welikala_automated_2020, xue_oral_2022}, and eyes \cite{suner_prediction_2021, park_mhealth_2020, kumar_mobile_2021, kasiviswanathan_semantic_2020}. 
% This is best exemplified by approaches to hemoglobin estimation. 

% \paragraph{Hemoglobin Estimation} 
Optical sensors have been successfully applied to noninvasively assess a major protein, hemoglobin. Hemoglobin allows for oxygen transport in red blood cells and is produced by iron, vitamin B9, and vitamin B12 \cite{mayo_clinic_staff_anemia_2023}. A deficiency in these micronutrients can result in lowered hemoglobin, which manifests as anemia. A review of the state of the art emphasizes the need for an affordable and accessible method of hemoglobin measurement, which can be realized with commodity smartphone cameras \cite{hasan_noninvasive_2021}. Various works have used smartphone cameras as a way to estimate hemoglobin status noninvasively \cite{hasan_noninvasive_2021, wang_hemaapp_2016, hasan_smarthelp_2018, suner_prediction_2021}.  
% They argue that leveraging smartphone cameras would make hemoglobin measurement more accessible, especially in underserved communities, due to the popularity of smartphones. 
They report that PPG signals derived from the fingertip and conjunctiva (skin behind the lower eyelids) under NIR light in 1070 and 850 nm wavelengths contain the most critical features to hemoglobin estimation. \citet{wang_hemaapp_2016} had users place their fingertip directly onto a smartphone camera while it recorded video to determine hemoglobin status. 
% The fingertip was illuminated with incandescent, white, and two IR wavelengths (970 and 880 nm) of light. 
% RGB time series waveforms were extracted from the video and used to measure the water and protein content of plasma. 
Analysis found that the blue spectra of plasma was the most important for protein composition, and therefore hemoglobin estimation. 
Classifiers were built to identify hemoglobin status as ``normal" or ``anemic" in the context of demographic averages. Optimal reference standard blood hemoglobin measurements (Masimo Pronto optical device) were used as ground-truth. 
Because camera-based assessment can exhibit bias with differences in skin color (see above), the authors carefully considered and reported the demographics of their study participants in addition to controlling for light absorption by skin tissue during experiments. 
A later paper devised a similar system, applying artificial neural networks (ANNs) and using only the smartphone's flash as a light source \cite{hasan_smarthelp_2018}. 
% Once again, participants took videos of their fingertip and the RGB intensities were used as inputs into an ANN which was trained on clinically measured hemoglobin levels as the target. The ANN also took the participant's age and gender as inputs to make its prediction. 
% These authors took important steps to make their model interpretable. 
By combining frames within the videos, the researchers were able to identify regions with high variation in predicted levels. 

The use of the conjunctiva as an ROI is gaining popularity due to the fact that it, like the fingernail bed and palmar creases, has no melanin and is devoid of “epidermis, dermis or subcutaneous fat which could impede the transmission of light” to deeper vascular layers \cite{suner_prediction_2021}. 
This means the blood vessels are easier to analyze, and there may be less bias due to skin color. 
These properties have been leveraged by \citet{suner_prediction_2021} to estimate hemoglobin concentration and screen for anemia using smartphone images of the conjunctiva. Spectral super-resolution (SSR) has been introduced to measure blood hemoglobin levels \cite{park_mhealth_2020}. This method is based on a wealth of existing research reconstructing hyperspectral images from RGB signals. Statistical learning was applied to approximate a hyperspectral image of the conjunctiva from a simple smartphone camera image. Using the hyperspectral data, the hemoglobin content in blood can be computed more effectively than the RGB data alone. 
% Linear regression with laboratory blood hemoglobin and the hyperspectral and SSR images showed an $R^2$ of 0.954 for the hyperspectral images and 0.912 for the SSR images on the validation sets. 
However, no analysis of skin color differences was conducted.

Techniques that use optical sensors benefit from the accessibility provided by the use of smartphones but their nutritional applications thus far are limited to macronutrients. They are also more susceptible to demographic biases in hardware, software, and datasets \cite{mcduff_camera_2021}. 

 \section{Implications for Future Micronutrient Status Assessment Methods} \label{Discussion}

The state of the art reveals limitations that hinder the assessment of micronutrient status in individuals in the following ways: (1) the lack of \textit{clinical relevance} in innovative approaches, (2) the absence of \textit{comprehensive} assessment techniques, and (3) the deficiency of \textit{accessible} and \textit{noninvasive} methods. 
Future work could aim to address these issues while considering real-world integration into clinical and public health settings.
% For this, the REASSURED framework, which stands for Real-time connectivity, Ease of specimen collection, Affordable, Sensitive, Specific, User-friendly, 
% (including cultural acceptability)
% Rapid and robust, Equipment-free/Environmentally friendly, and Deliverable 
% (i.e. accessible)
% to end-users \cite{land_reassured_2019}, offers a valuable guide.
% that aims to develop upon these emerging techniques should also consider how an accessible and noninvasive assessment method could ultimately be practically integrated into clinical practice and public health. 
Here, the requirements for medical diagnostic tests set by frameworks like REASSURED \cite{land_reassured_2019} are useful as they intersect with and expand beyond accessibility and non-invasiveness. REASSURED stands for Real-time connectivity, Ease of specimen collection, Affordable, Sensitive, Specific, User-friendly (including cultural acceptability), Rapid and robust, Equipment-free/Environmentally friendly, and Deliverable (i.e. accessible) to end-users \cite{land_reassured_2019}. 
First we briefly discuss how emerging methods align with these criteria, summarized in Table \ref{tab: criteria}, then we discuss opportunities to address the aforementioned limitations, summarized in Table \ref{tab:future_opportunities}.
% evaluates how emerging methods align with these criteria, though 
We note that definitive conclusions remain challenging due to the early stage of many technologies.

% , we identify how each of the emerging micronutrient assessment methods discussed in this review conform to all of the above criteria. Below, we discuss our reasoning for these decisions, and should note that it is difficult to speak conclusively about emerging techniques.

%Please add the following packages if necessary:
%\usepackage{booktabs, multirow} % for borders and merged ranges
%\usepackage{soul}% for underlines
%\usepackage{xcolor,colortbl} % for cell colors
%\usepackage{changepage,threeparttable} % for wide tables
%If the table is too wide, replace \begin{table}[!htp]...\end{table} with
%\begin{adjustwidth}{-2.5 cm}{-2.5 cm}\centering\begin{threeparttable}[!htb]...\end{threeparttable}\end{adjustwidth}
\begin{table}[!htp]\centering
\caption{Whether Methods Address Proposed and REASSURED \cite{land_reassured_2019} Criteria.}\label{tab: criteria}
\scriptsize
\newcolumntype{C}[1]{>{\centering\arraybackslash}p{#1}}

\begin{tabular}{p{0.6in}p{0.1in}p{0.2in}p{0.2in}p{0.2in}p{0.2in}p{0.2in}p{0.2in}p{0.2in}p{0.2in}p{0.1in}p{0.2in}p{0.35in}}\toprule
&\multicolumn{1}{C{0.35in}}{Optimal Reference Standard} &\multicolumn{4}{c}{Assays} &\multicolumn{3}{c}{Electrochemical} &\multicolumn{1}{C{0.25in}}{Spectro-scopic} &\multicolumn{2}{C{0.5in}}{Biofluid\par Analysis} \\\cmidrule{1-12}
Criteria &\multicolumn{1}{c}{-} &\multicolumn{1}{C{0.35in}}{Sweat \par Colorimetry} &\multicolumn{1}{C{0.35in}}{Multiplexed \par Sweat \par Sensors} &\multicolumn{1}{C{0.35in}}{Smartphone-based} &\multicolumn{1}{C{0.35in}}{Commercial} &\multicolumn{1}{C{0.35in}}{Voltammetry} &\multicolumn{1}{C{0.4in}}{Amperometry} &\multicolumn{1}{C{0.3in}}{Impedance} &\multicolumn{1}{c}{-} &\multicolumn{1}{C{0.15in}}{Single} &\multicolumn{1}{C{0.2in}}{Multiple} \\\midrule
Clinical\par relevance &\multicolumn{1}{c}{X} &\multicolumn{1}{c}{-} &\multicolumn{1}{c}{-} &\multicolumn{1}{c}{X} &\multicolumn{1}{c}{X} &\multicolumn{1}{c}{-} &\multicolumn{1}{c}{-} &\multicolumn{1}{c}{-} &\multicolumn{1}{c}{X} &\multicolumn{1}{c}{X} &\multicolumn{1}{c}{X} \\
Comprehensive assessment &\multicolumn{1}{c}{-} &\multicolumn{1}{c}{-} &\multicolumn{1}{c}{X} &\multicolumn{1}{c}{-} &\multicolumn{1}{c}{-} &\multicolumn{1}{c}{-} &\multicolumn{1}{c}{-} &\multicolumn{1}{c}{-} &\multicolumn{1}{c}{-} &\multicolumn{1}{c}{-} &\multicolumn{1}{c}{X} \\
Noninvasive &\multicolumn{1}{c}{-} &\multicolumn{1}{c}{X} &\multicolumn{1}{c}{X} &\multicolumn{1}{c}{-} &\multicolumn{1}{c}{-} &\multicolumn{1}{c}{X} &\multicolumn{1}{c}{X} &\multicolumn{1}{c}{X} &\multicolumn{1}{c}{-} &\multicolumn{1}{c}{X} &\multicolumn{1}{c}{X} \\
Accessible &\multicolumn{1}{c}{-} &\multicolumn{1}{c}{X} &\multicolumn{1}{c}{X} &\multicolumn{1}{c}{X} &\multicolumn{1}{c}{-} &\multicolumn{1}{c}{-} &\multicolumn{1}{c}{-} &\multicolumn{1}{c}{-} &\multicolumn{1}{c}{-} &\multicolumn{1}{c}{-} &\multicolumn{1}{c}{-} \\
Reproducibility &\multicolumn{1}{c}{X} &\multicolumn{1}{c}{-} &\multicolumn{1}{c}{-} &\multicolumn{1}{c}{-} &\multicolumn{1}{c}{X} &\multicolumn{1}{c}{-} &\multicolumn{1}{c}{-} &\multicolumn{1}{c}{-} &\multicolumn{1}{c}{X} &\multicolumn{1}{c}{-} &\multicolumn{1}{c}{-} \\
Robustness &\multicolumn{1}{c}{-} &\multicolumn{1}{c}{-} &\multicolumn{1}{c}{-} &\multicolumn{1}{c}{-} &\multicolumn{1}{c}{-} &\multicolumn{1}{c}{-} &\multicolumn{1}{c}{-} &\multicolumn{1}{c}{-} &\multicolumn{1}{c}{-} &\multicolumn{1}{c}{X} &\multicolumn{1}{c}{X} \\\cmidrule{1-12}
REASSURED & & & & & & & & & & & \\\cmidrule{1-12}
Real-time\par connectivity &\multicolumn{1}{c}{-} &\multicolumn{1}{c}{-} &\multicolumn{1}{c}{X} &\multicolumn{1}{c}{X} &\multicolumn{1}{c}{X} &\multicolumn{1}{c}{-} &\multicolumn{1}{c}{-} &\multicolumn{1}{c}{-} &\multicolumn{1}{c}{-} &\multicolumn{1}{c}{-} &\multicolumn{1}{c}{-} \\
Ease of specimen collection &\multicolumn{1}{c}{-} &\multicolumn{1}{c}{X} &\multicolumn{1}{c}{X} &\multicolumn{1}{c}{-} &\multicolumn{1}{c}{-} &\multicolumn{1}{c}{-} &\multicolumn{1}{c}{-} &\multicolumn{1}{c}{-} &\multicolumn{1}{c}{-} &\multicolumn{1}{c}{-} &\multicolumn{1}{c}{-} \\
Affordable &\multicolumn{1}{c}{-} &\multicolumn{1}{c}{X} &\multicolumn{1}{c}{X} &\multicolumn{1}{c}{X} &\multicolumn{1}{c}{-} &\multicolumn{1}{c}{X} &\multicolumn{1}{c}{X} &\multicolumn{1}{c}{X} &\multicolumn{1}{c}{-} &\multicolumn{1}{c}{-} &\multicolumn{1}{c}{-} \\
Sensitive &\multicolumn{1}{c}{X} &\multicolumn{1}{c}{-} &\multicolumn{1}{c}{-} &\multicolumn{1}{c}{X} &\multicolumn{1}{c}{X} &\multicolumn{1}{c}{-} &\multicolumn{1}{c}{-} &\multicolumn{1}{c}{-} &\multicolumn{1}{c}{X} &\multicolumn{1}{c}{-} &\multicolumn{1}{c}{-} \\
Specific &\multicolumn{1}{c}{X} &\multicolumn{1}{c}{-} &\multicolumn{1}{c}{-} &\multicolumn{1}{c}{X} &\multicolumn{1}{c}{X} &\multicolumn{1}{c}{-} &\multicolumn{1}{c}{-} &\multicolumn{1}{c}{-} &\multicolumn{1}{c}{X} &\multicolumn{1}{c}{-} &\multicolumn{1}{c}{-} \\
User-friendly &\multicolumn{1}{c}{-} &\multicolumn{1}{c}{X} &\multicolumn{1}{c}{X} &\multicolumn{1}{c}{-} &\multicolumn{1}{c}{-} &\multicolumn{1}{c}{-} &\multicolumn{1}{c}{-} &\multicolumn{1}{c}{-} &\multicolumn{1}{c}{-} &\multicolumn{1}{c}{X} &\multicolumn{1}{c}{X} \\
Rapid &\multicolumn{1}{c}{-} &\multicolumn{1}{c}{X} &\multicolumn{1}{c}{X} &\multicolumn{1}{c}{X} &\multicolumn{1}{c}{X} &\multicolumn{1}{c}{X} &\multicolumn{1}{c}{X} &\multicolumn{1}{c}{X} &\multicolumn{1}{c}{X} &\multicolumn{1}{c}{X} &\multicolumn{1}{c}{X} \\
Equipment-free/ Environmentally friendly &\multicolumn{1}{c}{-} &\multicolumn{1}{c}{X} &\multicolumn{1}{c}{X} &\multicolumn{1}{c}{-} &\multicolumn{1}{c}{-} &\multicolumn{1}{c}{-} &\multicolumn{1}{c}{-} &\multicolumn{1}{c}{-} &\multicolumn{1}{c}{-} &\multicolumn{1}{c}{-} &\multicolumn{1}{c}{-} \\
Deliverable &\multicolumn{1}{c}{-} &\multicolumn{1}{c}{X} &\multicolumn{1}{c}{X} &\multicolumn{1}{c}{X} &\multicolumn{1}{c}{-} &\multicolumn{1}{c}{-} &\multicolumn{1}{c}{-} &\multicolumn{1}{c}{-} &\multicolumn{1}{c}{-} &\multicolumn{1}{c}{-} &\multicolumn{1}{c}{-} \\
\bottomrule
\end{tabular}
\end{table}

% To (re)summarize the critical aspects, first, assessment must be quick and accurate so as to enable testing and treatment during the same visit. Second, the overall cost of testing (including transportation and storage costs) must be kept low. Third, the ability to actually deploy an assessment method is dependent on its cultural acceptability. For example, historically marginalized and exploited populations may be hesitant to work with state public health officials, or religious fasting may make the collection of some biosamples (e.g. urine) more complicated. Other requirements include real-time connectivity, equipment-free, environmentally friendly, and easy to use, which are largely satisfied by technological methods discussed herein. 

\subsection{Evaluating Emerging Methods}
Assay-based methods like lateral flow immunoassays offer high sensitivity and specificity due to their targeted biomarker reactions but face issues with batch variability, limited range sensitivity, and environmental instability. Their performance can vary across sample types and demographics. While assay-based methods are familiar to clinicians, they may burden patients due to reliance on blood or urine samples. 

Electrochemical methods are promising for PoC use due to speed and simplicity but can be affected by biofluid properties, electrode placement, and demographic differences. They often rely on less validated biosamples, limiting clinical adoption, and may require benchtop equipment, limiting practical deployment in limited resource settings. 

Spectroscopy-based methods, such as LC-MS/MS, are the clinical optimal reference standard due to their high reproducibility, but they require complex lab equipment and protocols, limiting field use. Emerging on-body spectroscopic approaches could improve accessibility and cultural acceptability, especially when integrated with smartphones. However, research on noninvasive, accessible spectroscopy for in-body micronutrient assessment remains inadequate.

Biofluid analytic methods, including those applying AI/ML to physiological or multi-modal sensor data, may offer the highest patient acceptability by avoiding additional testing. While potentially fast and accurate, their effectiveness depends heavily on training data quality. Therefore, these approaches face challenges in reproducibility, bias, and scalability, especially across diverse and/or underserved populations. Collecting equitable, high-quality data for training is resource-intensive, and it is likely that effective models will have to be highly specific to the communities they serve, as opposed to being fully generalizable to micronutrient malnutrition globally.

Although emerging methods for accessible and noninvasive micronutrient status assessment have promising capabilities within this set of requirements, few of these approaches are currently standardized and ready for reproducible deployment across diverse geographic settings and clinical populations. 
% This underscores the need for ongoing calibration, cross-site validation studies, and efforts to establish robust protocols.
% Therefore, we propose actionable opportunities, focused on each of our three identified limitations (
The actionable future opportunities we propose in Table \ref{tab:future_opportunities} aim to transition micronutrient status assessment from the laboratory to everyday use, enabling valuable micronutritional insights in a manner that is both easily accessible and noninvasive. 
% This approach could empower clinicians with reliable insights into nutritional status, enable researchers to conduct cost-effective, large-scale studies in resource limited settings, and help individuals manage and prevent micronutrient malnutrition proactively.
% This approach could empower clinicians to access reliable insights into nutritional status, researchers to run cost-effective, larger-scale studies to analyze more data, and individuals to manage micronutrient malnutrition effectively. These are necessary steps to not just address micronutrient malnutrition when it occurs, but to be proactive in its prevention. 
% Future work could further investigate specific details of integration into clinical practice and public health initiatives.

% TODO: Summary table
% \begin{table}
%     \centering
%     \begin{tabular}{cc}
%          Limitation/gap& Opportunity\\
%          Current non-clinical approaches to micronutrient assessment lack clinical relevance& \\
%          Holistic micronutrient assessment methods are not accessible& \\
%          PoC micronutrient assessments are invasive and specialized& \\
%     \end{tabular}
%     \caption{TODO: Caption}
%     \label{tab:future_opportunities}
% \end{table}

\begin{table}[ht]
    \caption{Limitations of Existing Micronutrient Status Assessment Methods and Opportunities to Address Them.}
    \label{tab:future_opportunities}
    \centering
    \small
    \begin{tabularx}{\textwidth}{p{2.5in}p{3.25in}}
        \toprule
        \textbf{Limitation} & \textbf{Opportunity} \\
        \midrule
        Limited clinical relevance & - Compare new approaches to the clinical optimal reference \\
        & - Evaluate assessment performance in routinely-assessed patients \\
        & - Adopt an interdisciplinary mindset when innovating \\
        & - Understand and integrate clinically relevant biofluid samples \\
        & - Measure clinically proven levels of circulating micronutrients \\
        \midrule
        Lack of holistic and comprehensive  approaches & - Employ precision nutrition by considering several types of data \\
        & - Utilize multi-modal solutions \\
        & - Gather many micronutrient statuses simultaneously \\
        & - Collect data from diverse populations and make data available \\
        \midrule
       Highly invasive and inaccessible & - Utilize commodity devices (smartphones, smartwatches) to collect data \\
       & - Make designs open-source \\
       & - Render insights actionable to non-experts \\
       & - Bypass the need for biofluid samples by using wearables \\
       & - Leverage less invasive biofluids such as urine \\
        \bottomrule
    \end{tabularx}
\end{table}

\subsection{Future Opportunities to Address Limitations of Micronutrient Assessment}
\subsubsection{Clinically Relevant Innovations}
We find that current innovative approaches to biofluid analysis for micronutrient status assessment lack clinical relevance (Sections \ref{subsec_assays} to \ref{subsec_analytic}). To overcome this limitation,
% and develop effective, meaningful assessment methods, 
future research could
% prioritize demonstrating clinical utility and 
shift focus toward clinical relevance as a primary goal.
% At a high level, 
This requires an interdisciplinary approach and recognition that technology is most effective when it complements clinical expertise, especially as micronutrition is inherently a clinical field where all solutions must be grounded in practical applications. 
% technological methods are best utilized as an aid to clinical expertise. 
% Micronutrition is fundamentally a clinical field, so all future solutions must contextualize their work within a clinical application.
Clinical, practical grounding can also be done by benchmarking new approaches against established optimal reference standards, detailed in Table \ref{tab: assessment}. Comparative validation is essential to demonstrate a novel method's relevance and its potential as a viable substitute for the clinical optimal reference standard. We note that appropriate use of ELISA tests are sufficient in most cases. 
% Such comparative validation is crucial to demonstrate the relevance and efficacy of a novel method, and to determine its potential as a viable substitute for the clinical optimal reference standard.
% Only a limited number of the reviewed methodologies conduct such comparative analyses with state-of-the-art clinical practices. 
% To aid in this endeavor, gold-standard methods for the assessment of each micronutrient are depicted in Tables \ref{tab:assess-vitamins-water-soluble}, \ref{tab:assess-vitamins-fat-soluble}, and \ref{tab:assess-minerals} in the Appendix. 
% We note that ELISA tests are sufficient in most cases, as long as the standard protocol is followed. 
% there is potential for studies to apply new assessment methods among patient populations that are routinely assessed by clinicians. Often, the clinical optimal reference standard is already being used to determine the status of these individuals. Thus, there is an opportunity to collaborate with medical professionals working in these populations, who may provide access to clinical test results for comparative analysis as well as valuable feedback. Selecting populations at risk for imbalance, such as candidates for bariatric surgery or individuals with diabetes, may be useful because they are frequently surveyed and assessed for nutritional status pre and post-treatment. 

We also find it is important for researchers to prioritize clinically relevant biofluids and ensure measurements reflect meaningful biomarker levels. For instance, reviewed papers claim to leverage sweat, yet sweat is an unreliable indicator of micronutrient status \cite{baker_physiological_2020, baker_physiology_2019}. Clinical relevance improves when methods align with the biomarkers used across the full spectrum of deficiency to excess. However, future research can also validate which biosamples and biomarkers may be suitable replacements for the optimal reference standard for specific micronutrients, and in specific usage contexts.
% This approach will further substantiate the relevance and efficacy of new methods.
% The development of new methods should focus on maintaining clinical relevance. Current methods often consist of stand-alone devices that measure a single micronutrient, primarily concentrating on the technology to achieve this \cite{}.  
% To maintain clinical relevance, it is crucial to consider clinical applications and adopt an interdisciplinary mindset. 
Since clinical validation can be challenging (Section \ref{clinical/biochemical analysis}),
studies could apply new assessment methods in patient populations already monitored by clinicians, offering opportunities for collaboration and access to clinical reference data for validation. Targeting at-risk groups, like bariatric surgery candidates or individuals with diabetes, is especially useful as they are routinely assessed for nutritional status.

% Researchers could also utilize clinically-relevant biofluids and consult clinical literature about the relevance of different biofluid samples (Section \ref{subsec_biofluids}), utilizing them where appropriate. For example, several reviewed papers claim to measure in-body micronutrient status from sweat, despite clinical evidence that sweat is an inaccurate representation of in-body status \cite{baker_physiological_2020, baker_physiology_2019}. Clinical relevance can be further improved by assuring novel methods can measure the same biomarkers used by clinicians across the spectrum of deficiency to excess.
% Future work should be grounded in clinically proven methodologies to ensure reliability and relevance.

% \subsection{Consider the individualized nature of micronutrient malnutrition using multi-modal solutions}

\subsubsection{Comprehensive Approaches through Individualized and Multi-Modal Solutions}
Micronutrition is complex and requires a holistic, individualized approach (Section \ref{intro}). 
Advancing assessment technologies calls for precision nutrition and multi-modal sensing that account for differences in diet, metabolism, and lifestyle \cite{harvard_school_of_public_health_precision_2020, srinivasan_precision_2017, kirk_precision_2021, zmora_harnessing_2021}. 
% To develop effective and applicable technologies, it is essential to embrace precision nutrition assessment and multi-modal sensing. 
% Precision nutrition leverages technology to account for individual differences in diet, metabolism, demographics, lifestyle, and more \cite{harvard_school_of_public_health_precision_2020, srinivasan_precision_2017, kirk_precision_2021, zmora_harnessing_2021}. 
% This allows nutritional diagnoses and interventions to be fine-tuned to a specific patient. Future micronutrient assessment methods could integrate elements of precision nutrition by considering these individual differences. 
% will be powerful tools in this endeavor, allowing for the consideration of the varied aspects of micronutrition during an assessment.
While current methods often analyze biofluids
% Multi-modal sensing is a promising opportunity to incorporate precision nutrition into new assessment methods. Current methods conduct biofluid analysis 
(Section \ref{biofluids section}) or physiology (Section \ref{phys section}) separately, combining them with clinical data, like treatment history and anthropomorphic measurements, can uncover patterns and offer a more complete view of micronutritional status. 
% . New technologies can combine these approaches with conventional health data, such as physical activity, demographics, and other clinical information.
% A multi-modal approach that leverages both biofluid and physiological analyses can reveal more relevant insights into nutritional status, alleviate over-reliance on subjective assessments, and ultimately create a comprehensive overview of an individual's health. 
% The combination of diverse data types also helps alleviate reliance on subjective assessments in current clinical methods, such as dietary reporting and physical exams \cite{reber_nutritional_2019}.
Multi-modal sensing also enables simultaneous analysis of multiple micronutrients, essential due to their complex interactions
% allows for the analysis of multiple micronutrients simultaneously, a functional necessity since micronutrient imbalances rarely manifest alone 
\cite{bailey_epidemiology_2015}. 
% Current biofluid analysis methods often focus on assessing a single micronutrient, but precision nutrition emphasizes the need to consider all aspects of an individual's health to gather the most relevant insights. 
% Micronutrients interact in complex ways 
(Tables \ref{tab: char-ws}, \ref{tab: char-fs}, \ref{tab: char-min}).
When single-device measurement is not feasible, future solutions could integrate multiple tools into a unified system.
% and there is opportunity for more methods that examine several micronutrients concurrently to understand their comprehensive effects on individual health. If it is not feasible for a single device to measure multiple micronutrients, future work could combine several devices or measurement modes reviewed here together in a unified solution.

We support calls for more micronutrition data \cite{brown_increasing_2021}, as access to individual nutrition profiles can accelerate progress toward precision assessment (Section \ref{subsec_analytic}). Innovative assessment devices and health sensors could be used in clinical studies to collect personalized micronutrient data over time. Valuable information includes demographics, clinical history, nutritional assessments, symptom images, biochemical results, and wearable data. Greater data availability would support deeper clinical investigations and aid early detection of imbalances. Disease etiology lies at the intersection of comprehensive clinical assessment, preventative care, and clinical relevance. Etiology stresses the multifactorial (and deeply individual, social, and cultural) nature of disease progression \cite{white_application_2020}. Comprehensive assessments, and data thereof, can contribute to the etiological understanding of micronutrient malnutrition and ultimately drive precise prevention strategies.

AI and ML tools can help integrate these factors, uncovering latent patterns across a vast quantity and variety of micronutrition data \cite{cote_artificial_2022}. We have discussed how micronutritional assessments are made more powerful when they are conducted and compared repeatedly over time. Accessible, multi-modal sensing would enable routine collection of health data relevant to micronutritional status, which can be combined and compared in real time, and \textit{en masse}, by ML models. However, the computational and environmental costs of AI and ML models, especially large-scale generative AI models, must also be considered.
It will also be essential to ensure demographic and cultural diversity to produce relevant and equitable models. Scalable and sustainable data collection through accessible PoC devices could be essential to achieving this.

\subsubsection{Accessible and Noninvasive Point-of-Care Devices}
This work highlights the critical need for accessible and noninvasive methods to assess micronutrient status. Such methods can generate valuable data though PoC assessments \cite{kirk_precision_2021} and therefore improve the overall understanding of micronutrition. 
% and be a byproduct of it, through a greater general understanding of micronutrition.
Current technologies fall short in accessibility, but potential solutions lie in leveraging commodity hardware, open-sourcing designs, prioritizing ease of use for non-experts, and utilizing less invasive biosamples.
% Our review of existing methods reveals that current technologies have yet to fully address these limitations. The solution to this issue involves leveraging commodity hardware, open-source designs, prioritizing ease-of-use for non-experts, and reducing invasiveness in assessment techniques.
% Often, micronutrient assessment devices that conduct biofluid analysis are over-specialized (Section \ref{subsec_biofluid limits}). While more specialized devices, sensors, or assays may be necessary given the complex nature of micronutrient assessment, the utility of these devices will increase when designed to be low-cost, open-source, simple to manufacture, and use off-the-shelf components. If future research follows these design criteria, multi-modal solutions with broad sensing capabilities will become more viable overall.
Smartphones and smartwatches, with built-in sensors, offer promising platforms for physiological insights. 
Their capabilities can be further expanded with features or accessories like electrochemical sensor chips or microfluidic pumps for biofluid analysis. As a result, multi-modal solutions with broad sensing and analysis capabilities will become more broadly available and low-cost.

Methods could be built considering some existing wearable devices.
% Although our review does not center on commercial devices, we offer a comparison of a subset of wearable health sensors, as emerging methods may build upon these platforms. 
The Empatica E4 and Polar H10 offer high-fidelity HRV and autonomic data, which is associated vitamin B12 and iron status (Section \ref{subsec_phys symptoms}) \cite{Schuurmans2020, giles2016validity}. Oura Ring, WHOOP, Fitbit, and Garmin devices track sleep, heart rate, and recovery metrics, which have been linked to nutrient-related fatigue and dysfunction \cite{henriksen2022polar, miller2020validation, benedetto2018assessment, garmin2020sleep}. Additionally, Dexcom continuous glucose monitors, while focused on glucose, demonstrate the clinical viability of wearable biochemical sensing and may inform future designs for micronutrient monitoring \cite{garg2022accuracy}. While these tools vary in accuracy, especially across sensor types and populations, they highlight growing opportunities for accessible physiological monitoring relevant to micronutrition.
Minimally invasive alternatives, such as urine testing (Section \ref{subsec_biofluids}) instead of serum or plasma, can enhance accessibility. Even more promising are wearable, on-body sensing devices which use spectroscopy and ML to monitor micronutrients continuously without the need for biosamples \cite{crocombe_portable_2018, watson_lumos_2022, monakhova_chemometrics-assisted_2010, monakhova_multicomponent_2017}. When care is taken to design innovative on-body, PoC assessments that are open-source and easy to use by laypeople (in addition to being accessible and noninvasive), comprehensive micronutrient status assessment can become viable for resource limited communities as well as individuals at home.

\section{Conclusion} \label{conc}
This article provides a comprehensive review of accessible and noninvasive methods for assessing in-body micronutrient status, focusing on biofluids (Section \ref{biofluids section}) and physiological (Section \ref{phys section}) approaches. We evlauate current techniques such as assays, electrochemistry, spectroscopy, optical sensors, and AI/ML for performance and clinical relevance. Key contributions include: (1) background on micronutrients for non-clinical audiences, (2) a synthesis of biofluid- and physiology-based methods, (3) future directions for noninvasive and accessible assessment, and (4) a unique focus on clinical applicability. Several summary tables provide an intuitive reference throughout.

% This article provides a comprehensive review of accessible and noninvasive in-body micronutrient status assessment methods. We take a critical look at a wide variety of research in the field, based on analyses of both biofluids (Section \ref{biofluids section}) and physiology (Section \ref{phys section}). These current methods, such as assays, electrochemistry, spectroscopy, optical sensors, and AI \& ML, are evaluated in the context of their performance and overall clinical relevance. Several tables are contributed throughout the paper and in the Appendix, intended to gather the most relevant information about micronutrients and their assessment together into an intuitive reference. Our major contributions to micronutrient status assessment as a research area include 
% (1) general background information on micronutrients for a non-clinical audience, (2) synthesis of existing technological and clinical micronutrient status assessment methods, split into biofluid and physiological-based techniques, (3) recommendations for future directions to develop accessible  and noninvasive assessment methods, and (4) a unique focus on clinical relevance throughout.

% both clinical and technological aspects, to motivate and support the development of novel accessible and noninvasive methods. 

% This review revealed xyz..
This review synthesizes micronutrient status assessment methods based on biofluid and physiological analyses. Biofluid approaches benefit from established biomarkers but often rely on blood samples and lack clinical validation. Physiological assessments face challenges like weak symptom associations, self-report bias, and limited micronutrient-specific research. From a clinical perspective, no current technology holistically integrates both types of analysis, despite this being standard in practice. Most research focuses on assessing a single micronutrient using either biofluid or physiological data alone.
This review outlines three key opportunities to advance micronutrient status assessment: improving clinical relevance, adopting holistic multi-modal approaches, and reducing invasiveness to enhance accessibility (Table \ref{tab:future_opportunities}). Addressing these gaps through innovative, non-invasive, and individualized PoC solutions could empower individuals to better manage micronutrient malnutrition.

\section*{Acknowledgments}
Special thanks to Sofia Luis for her assistance editing the revised version of this paper, and to the anonymous reviewers for their constructive feedback.

\bibliographystyle{ACM-Reference-Format}
\bibliography{references}

\newpage
\appendix 
\section{Tables} \label{appendix}
\begin{table}[!h]
\centering
\scriptsize
\caption{Alphabetized List of Abbreviations in Main Text} \label{tab:abbreviations}
\begin{tabular}{p{0.1\textwidth} p{0.30\textwidth} | p{0.1\textwidth} p{0.30\textwidth}}
\toprule
\textbf{Abbreviation} & \textbf{Definition} & \textbf{Abbreviation} & \textbf{Definition} \\
\midrule
AAS & atomic absorption spectrometry & LoD & limit of detection \\
AGP & acid glycoprotein & MILCA & mutual information least dependent component analysis \\
AI & artificial intelligence & ML & machine learning \\
ANN & artificial neural network & MS & mass spectroscopy \\
ANS & autonomic nervous system & NADH & nicotinamide adenine dinucleotide \\
AUC & area under the curve & NDNS & National Diet and Nutrition Surveys \\
BAC & blood alcohol content & NFC & near-field communication \\
BIA & bio-electrical impedance analysis & NFPE & Nutrition-Focused Physical Exam \\
BMI & body mass index & NHANES & National Health and Nutrition Examination Survey \\
BP & blood pressure & NIR & near infrared \\
Bphen & bathophenanthroline & PBS & phosphate buffered saline \\
BRI & body roundness index & PNS & parasympathetic nervous system \\
CBC & complete blood count & PoC & point of care \\
CoV & coefficient of variance & PPG & photoplethysmogram \\
CRP & C-reactive protein & PR & pulse rate \\
CUN-BAE & Clinical University of Navarra body adiposity estimator & PRV & pulse rate variability \\
CVD & cardiovascular disease & PWV & peak wavelength value \\
ECG & electrocardiogram & RBP & retinol binding protein \\
EDA & electrodermal activity & RDA & Recommended Dietary Allowance \\
EDR & estimated daily requirement & RF & radio frequency \\
ELISA & enzyme-linked immunosorbent assay & RMSE & root mean square error \\
GSH & glutathione & ROI & region of interest \\
HF & high-frequency & SNS & sympathetic nervous system \\
HMVA & human milk vitamin A & SPR & surface plasmon resonance \\
HPLC & high-performance liquid chromatography & SSR & spectral super-resolution \\
HPLC-IR & HPLC with IR detection & sTfR & soluble transferrin receptor \\
HPLC-UV & HPLC with UV detection & SWV & square wave voltammetry \\
HR & heart rate & TSH & thyroid-stimulating hormone \\
HRV & heart rate variability & UL & Upper Limit \\
ICA & independent component analysis & US & United States \\
ICP-MS & inductively coupled plasma mass spectrometry & UV & ultraviolet \\
IDA & iron-deficiency anemia & VAI & visceral adiposity index \\
IR & infrared & WC & waist circumference \\
LC & liquid chromatography & WHO & World Health Organization \\
LED & light emitting diode & WhtR & waist-to-height ratio \\
LF & low-frequency & & \\
\bottomrule
\end{tabular}
\end{table}

% \subsection{Characteristics of Micronutrients}
%Please add the following packages if necessary:
%\usepackage{booktabs, multirow} % for borders and merged ranges
%\usepackage{soul}% for underlines
%\usepackage[table]{xcolor} % for cell colors
%\usepackage{changepage,threeparttable} % for wide tables
%If the table is too wide, replace \begin{table}[!htp]...\end{table} with
%\begin{adjustwidth}{-2.5 cm}{-2.5 cm}\centering\begin{threeparttable}[!htb]...\end{threeparttable}\end{adjustwidth}

\tiny
% \begin{adjustwidth}{-2.5 cm}{-2.5 cm}\centering\begin{threeparttable}[!htb]
\begin{longtable}{p{0.45in}p{0.65in}p{0.8in}p{0.35in}p{0.6in}p{0.55in}p{0.7in}p{0.65in}}
\caption{Characteristics of Micronutrients: Water-Soluble Vitamins. Information from \cite{bailey_epidemiology_2015, collier_storage_2010, mueller_aspen_2017, berger_espen_2022, national_institutes_of_health_dietary_nodate}}\label{tab: char-ws}
% \begin{tabulary}{4.834in}{p{0.4in}LLp{0.5in}LLLL}
% \toprule
\\ \toprule
&\multicolumn{4}{c}{Overview} &\multicolumn{3}{c}{Interactions Impacting Status} \\\cmidrule(r){2-5}\cmidrule(l){6-8}
Micronutrient &Purpose &Storage &Risk of Excess &High Risk Populations &Micronutrients &Diseases (decrease) &Medications (decrease) \\ \midrule
\endfirsthead
\\ \toprule
&\multicolumn{4}{c}{Overview} &\multicolumn{3}{c}{Interactions Impacting Status} \\\cmidrule(r){2-5}\cmidrule(l){6-8}
Micronutrient &Purpose &Storage &Risk of Excess &High Risk Populations &Micronutrients &Diseases (decrease) &Medications (decrease) \\ \midrule
\endhead
\hline \multicolumn{8}{r}{{Continued on next page}} \\ \hline
\endfoot
\bottomrule
\endlastfoot
Vitamin B1 (thiamin); 3 forms (TMP, TTP, TPP) &Critical to energy metabolism and cell development, functionality &Small amounts in liver &Lack of evidence &Older adults &Absorption decreased by magnesium, folate deficiency &Alcoholism, Inflammatory bowel diseases, Obesity post bariatric surgery, chronic renal failure, critical illness, HIV/AIDS, diabetes &Furosemide, Fluorouracil \\
Vitamin B2 (riboflavin); 2 coenzyme derivatives (FMN and FAD) &Critical to energy metabolism, cell development and functionality, and metabolism of fats, drugs, and steroids (maintains homocysteine levels) &Small amounts in liver, heart, kidneys &Lack of evidence &Vegetarian athletes, pregnant and lactating people and their infants, people who are vegan and/or consume little milk, people with riboflavin transporter deficiency &Absorption decreased by copper, zinc, iron, manganese intake; deficiency associated with those of folate, pyridoxine, niacin &Alcoholism, Chronic intestinal failure &None \\
Vitamin B3 (niacin); 2 forms (NAD and NADP) &Critical to energy metabolism, NAD is needed in over 400 enzyme reactions &Some excess in red blood cells &Yes (in supplementation) &Those with undernutrition &Status decreased by inadequate riboflavin, pyridoxine, and/or iron intakes &Hartnup disease, carcinoid syndrome &Antidiabetes, isoniazid and pyrazinamide \\
Vitamin B5 (pantothenic acid) &Critical to energy metabolism, breaking down and making fats &Red blood cells and tissues &Lack of evidence &Those with a pantothenate kinase-associated neurodegeneration 2 mutation & & & \\
Vitamin B6 (pyridoxine); 3 forms (pyridoxine, pyridoxal, pyridoxamine) &Involved in a wide variety of enzyme reactions, protein metabolism, and cognitive development (maintaining homocysteine levels) &Majority bounded to Albumin &Yes (in supplementation) &Those with autoimmune disorders &Poor status associated with low concentrations of other B-complex vitamins &Alcoholism, Inflammatory bowel diseases, chronic renal failure, malabsorption (celiac, Crohn's, etc), homocystinuria &HIV therapy/treatment, therapies inhibiting vitamin activity, cycloserine, antiepileptics, theophyline \\
Vitamin B7 (biotin) &Critical to the metabolism of proteins, fats, and carbohydrates into energy &Most stored in liver &None &Those with biotinidase deficiency, chronic alcohol exposure, and pregnant and breastfeeding people & &Alcoholism, chronic intestinal failure &Anticonvulsants \\
Vitamin B9 (folate) &Used to create DNA and RNA, facilitate cell division, as well as to metabolize amino acids (conversion of homocysteine) &15-30 mg with 50\% in liver, rest in blood and body tissues &Yes (masks B12 deficiency) &Women of childbearing age, pregnancy, MTHFR genetic polymorphism &Absorption decreased by zinc deficiency, bioavailability increased by Vitamin C, excess can mask B12 deficiency &Alcoholism, chronic intestinal failure, Chronic (atrophic) gastritis, obesity post bariatric surgery, chronic renal failure &Methotrexate, antiepileptics, sulfasalazine \\
Vitamin B12 (cobalamin) \cite{adas_medical_knowledge_team_vitamin_2023}&Critical to CNS development and functionality, RBC formulation, DNA synthesis, conversion of homocysteine &80\% in liver; 1-5 mg (thousands times more than daily consumption); can last 2-5 years, 1-3 by some sources &None &Women, elderly, black people, those with low socioeconomic status, who have had gastrointestinal surgery, are vegetarian/vegan &Absorption decreased by excess vitamin C &Alcoholism, chronic intestinal failure, chronic (atrophic) gastritis, Liver diseases, obesity post bariatric surgery, critical illness &Gastric acid inhibitors, metformin \\
Vitamin C (ascorbic acid) &Required in synthesis of collagen and neurotransmitters, used in protein metabolism, and critical to immune function &High concentrations in cells and tissues, WBC, eyes, adrenal glands, pituitary gland, and brain; total content 300 mg (near acute deficiency) to 2g &Yes (mild nausea, diarrhea, cramps) &Smokers, those with low food variety, any disease causing oxidative stress &Shown to regenerate other antioxidants (ex vitamin E) &Alcoholism, chronic (atrophic) gastritis, obesity post bariatric surgery, critical illness &Chemotherapy/radiation, 3-hydroxy-3-methylglutaryl coenzyme A reductase inhibitors \\
\label{tab:char-vitamins-water-soluble}
\end{longtable}
% \end{tabulary}
% \end{threeparttable}\end{adjustwidth}
% \end{longtable}
% \end{center}

%Please add the following packages if necessary:
%\usepackage{booktabs, multirow} % for borders and merged ranges
%\usepackage{soul}% for underlines
%\usepackage[table]{xcolor} % for cell colors
%\usepackage{changepage,threeparttable} % for wide tables
%If the table is too wide, replace \begin{table}[!htp]...\end{table} with
%\begin{adjustwidth}{-2.5 cm}{-2.5 cm}\centering\begin{threeparttable}[!htb]...\end{threeparttable}\end{adjustwidth}
\tiny
% \begin{adjustwidth}{-2.5 cm}{-2.5 cm}\centering\begin{threeparttable}[!htb]
\begin{longtable}{p{0.45in}p{0.65in}p{0.8in}p{0.35in}p{0.6in}p{0.55in}p{0.7in}p{0.65in}}
\caption{Characteristics of Micronutrients: Fat-Soluble Vitamins. Information from \cite{bailey_epidemiology_2015, collier_storage_2010, mueller_aspen_2017, berger_espen_2022, national_institutes_of_health_dietary_nodate}}\label{tab: char-fs}
% \begin{tabulary}{4.834in}{p{0.4in}LLp{0.5in}LLLL}
% \toprule
\\ \toprule
&\multicolumn{4}{c}{Overview} &\multicolumn{3}{c}{Interactions Impacting Status} \\\cmidrule(r){2-5}\cmidrule(l){6-8}
Micronutrient &Purpose &Storage &Risk of Excess &High Risk Populations &Micronutrients &Diseases (decrease) &Medications (decrease) \\ \midrule
\endfirsthead
\\ \toprule
&\multicolumn{4}{c}{Overview} &\multicolumn{3}{c}{Interactions Impacting Status} \\\cmidrule(r){2-5}\cmidrule(l){6-8}
Micronutrient &Purpose &Storage &Risk of Excess &High Risk Populations &Micronutrients &Diseases (decrease) &Medications (decrease) \\ \midrule
\endhead
\hline \multicolumn{8}{r}{{Continued on next page}} \\ \hline
\endfoot
\bottomrule
\endlastfoot
Vitamin A &Critical for vision, cell growth, immune and reproductive functions &Most in liver (about 6 months), some in eyes &Yes &Infants, pregnant people in low/middle income/developing countries &Absorption decreased by zinc deficiency &Alcoholism, chronic intestinal failure, Inflammatory bowel diseases, liver diseases, obesity post bariatric surgery, cystic fibrosis &Orlistat, retinoids (results in toxicity) \\
Vitamin D; 2 forms: 25(OH)D (calcidiol) and 1,25(OH)D (calcitriol) \cite{ross_dietary_2011}&Bone growth and strength, absorption and control of calcium, reducing inflammation &Fatty tissue and liver &Yes &Breastfed infants, adults 20-39, those with kidney/liver dysfunction, with dark skin, limited sun exposure, conditions limiting fat absorption &Magnesium is critical to activation and binding, function is heavily interwoven with Calcium &Alcoholism, chronic intestinal failure, chronic (atrophic) gastritis, Inflammatory bowel diseases, liver diseases, obesity post bariatric surgery, chronic renal failure, critical illness &Orlistat, statins, steroids, thiazide diuretics \\
Vitamin E (alpha-tocotrienol form) &Function as antioxidants, aid in immune, cell signaling, metabolic processes &Liver (alpha-tocotrienol form) &Lack of evidence (UL of 1000 mg in adults) &Infants, those with fat malabsorption, dieting & &Alcoholism, chronic intestinal failure, inflammatory bowel diseases, liver diseases, obesity post bariatric surgery &Anticoagulant, antiplatlet, simvastatin, niacin, chemotherapy/radio treatment \\
Vitamin K &Involved in blood clotting and bone metabolism &Low blood and tissue stores, carried in lipoproteins &None (except for K3) &Newborns and those with fat malabsorption &Excretion stimulated by excess Vitamin D, absorption decreased by excess Vitamin E and A &Alcoholism, chronic intestinal failure, inflammatory bowel diseases, obesity post bariatric surgery, chronic renal failure, bleeding disorders &Antibiotics and anticoagulants, bile acid sequestrants, orlistat \\
\label{tab:char-vitamins-fat-soluble}
\end{longtable}
% \end{threeparttable}\end{adjustwidth}

%Please add the following packages if necessary:
%\usepackage{booktabs, multirow} % for borders and merged ranges
%\usepackage{soul}% for underlines
%\usepackage[table]{xcolor} % for cell colors
%\usepackage{changepage,threeparttable} % for wide tables
%If the table is too wide, replace \begin{table}[!htp]...\end{table} with
%\begin{adjustwidth}{-2.5 cm}{-2.5 cm}\centering\begin{threeparttable}[!htb]...\end{threeparttable}\end{adjustwidth}

\tiny
% \begin{adjustwidth}{-2.5 cm}{-2.5 cm}\centering\begin{threeparttable}[!htb]
\begin{longtable}{p{0.4in}p{0.75in}p{0.7in}p{0.4in}p{0.6in}p{0.6in}p{0.8in}p{0.5in}}
\caption{Characteristics of Micronutrients: Minerals. Information from \cite{bailey_epidemiology_2015, collier_storage_2010, mueller_aspen_2017, berger_espen_2022, national_institutes_of_health_dietary_nodate}}\label{tab: char-min}
% \begin{tabulary}{4.834in}{p{0.4in}LLp{0.5in}LLLL}
% \toprule
\\ \toprule
&\multicolumn{4}{c}{Overview} &\multicolumn{3}{c}{Interactions Impacting Status} \\\cmidrule(r){2-5}\cmidrule(l){6-8}
Micronutrient &Purpose &Storage &Risk of Excess &High Risk Populations &Micronutrients &Diseases (decrease) &Medications (decrease) \\ \midrule
\endfirsthead
\\ \toprule
&\multicolumn{4}{c}{Overview} &\multicolumn{3}{c}{Interactions Impacting Status} \\\cmidrule(r){2-5}\cmidrule(l){6-8}
Micronutrient &Purpose &Storage &Risk of Excess &High Risk Populations &Micronutrients &Diseases (decrease) &Medications (decrease) \\ \midrule
\endhead
\hline \multicolumn{8}{r}{{Continued on next page}} \\ \hline
\endfoot
\bottomrule
\endlastfoot
Iron \cite{gloria_iron_nodate,gerber_iron_2024}&Essential to oxygen transport through hemoglobin, energy metabolism, physical growth, neurological development, cell functioning, and hormone synthesis &60\% in blood hemoglobin, rest as ferritin in liver, spleen, bone marrow, muscles &Yes (especially those with hemochromatosis and elderly) &Infants, young children, teen females, pregnant people (especially if Mexican-American or Black), premenopausal, in food-insecure households, have increased menstrual bleeding &Absorption increased by Vitamin C intake, Absorption decreased by zinc, calcium, manganese intake and copper deficiency &Chronic intestinal failure, chronic (atrophic) gastritis, inflammatory bowel diseases, obesity post bariatric surgery, critical illness, cancer, heart failure &Levodopa, levothyroxine, proton pump inhibitors \\
Copper &Cofactor in energy production, iron absorption, neuropeptide activation, and synthesis of connective tissue and neurotransmitters &50-120mg total, 95\% carried by ceruloplasmin; 2-3 months in skeleton and muscle; tightly regulated, only ~1mg/d loss in bile &Yes &Pregnant people &Absorption decreased by high zinc &Chronic intestinal failure, obesity post bariatric surgery, chronic renal failure, critical illness, celiac disease, menkes disease & \\
Zinc \cite{pat_bass_symptoms_2023}&Physical growth and development, cellular metabolism, and immune functions &85\% in skeletal muscle and bone, ~0.1\% in plasma where 70\% of that is bound to albumin; 1.5g females, 2.5g males total &Yes &Children, teens, exclusively breastfed infants, pregnant people, vegetarian/vegan, have eating disorders, malabsorption, gastrointestinal disorders &Absorption decreased by high calcium/iron &Alcoholism, chronic intestinal failure, inflammatory bowel diseases, liver diseases, obesity post bariatric surgery, chronic renal failure, critical illness, sickle cell disease, HIV &Antibiotics, penicillamine, piuretics \\
Iodine &Thyroid gland function, protein synthesis, metabolism and enzyme activity &70-80\% in thyroid gland; 15-20 mg total &Yes &Infants, pregnant people, use uniodized salt, are in regions with iodine-deficient soils &Absorption decreased by iron intake and selenium deficiency & &Anti-thyroids, angiotensin-converting enzyme inhibitors, potassium-paring diuretics \\
Selenium &Reproduction, thyroid hormone metabolism, and DNA synthesis through selenoproteins; also acts as antioxidant &28-46\% in skeletal muscle, most in selenomethionine form &Yes &Kidney dialysis patients, those in selenium deficient regions & &Inflammatory bowel diseases, liver diseases, chronic renal failure, critical illness, obesity &Cisplatin \\
Magnesium &Regulates several chemical reactions, including blood glucose and blood pressure regulation, DNA, RNA, and protein synthesis, proper muscle and nerve functioning, bone development, and calcium and potassium ion transport &Approx 25g; 50-60\% in bone, <1\% in serum (tightly controlled), rest in soft tissue &Yes &Elderly &Absorption increased by Vitamin D &Alcoholism, gastrointestinal disease, bariatric surgery, T2D &Bisphosphonates, antibiotics, diuretics, proton pump inhibitors \\
\label{tab:char-minerals}
\end{longtable}
% \end{tabulary}
% \end{threeparttable}\end{adjustwidth}

% \subsection{Summary of Physiological Symptoms}
%Please add the following packages if necessary:
%\usepackage{booktabs, multirow} % for borders and merged ranges
%\usepackage{soul}% for underlines
%\usepackage[table]{xcolor} % for cell colors
%\usepackage{changepage,threeparttable} % for wide tables
%If the table is too wide, replace \begin{table}[!htp]...\end{table} with
%\begin{adjustwidth}{-2.5 cm}{-2.5 cm}\centering\begin{threeparttable}[!htb]...\end{threeparttable}\end{adjustwidth}
% \begin{adjustwidth}{-2.5 cm}{-2.5 cm}\centering\begin{threeparttable}[!htb]
\begin{center}
\tiny
\begin{longtable}{p{0.4in}p{0.75in}p{0.6in}p{1in}p{0.5in}p{0.6in}p{0.4in}p{0.5in}}
% \begin{tabulary}{5.934in}{LRRRRRRR}
\caption{Physiological Symptoms of Micronutrient Deficiencies: Water-Soluble Vitamins. Information from \cite{academy_of_nutrition_and_dietetics_micronutrient_2017, de_nails_2012, dibaise_hair_2019, radler_nutrient_2013, reber_nutritional_2019, mueller_aspen_2017, berger_espen_2022, national_institutes_of_health_dietary_nodate}}\label{tab: physio-ws}
\\ \toprule
Micronutrient &Eye &Nail &Oral &Disease &Autonomic &Misc &Timeframe \\\midrule
\endfirsthead
\\ \toprule
Micronutrient &Eye &Nail &Oral &Disease &Autonomic &Misc &Timeframe \\\midrule
\endhead
\hline \multicolumn{8}{r}{{Continued on next page}} \\ \hline
\endfoot
\bottomrule
\endlastfoot
Vitamin B1 &Disability in eye movement (ophthalmoplegia) & & &Cardiomyopathies/ heart failure, Sarcopenia & &Correlated with fatigue &Stores depleted within 20 days of insufficient intake \\
Vitamin B2 &Conjunctiva inflammation/grittiness (angular blepharitis), Redness/fissures in eyelid corners (Angular Palpebritis), conjunctiva redness/irritation, swollen/sticky eyelid, photophobia & &Bilateral cracks/redness at corners of lips/mouth (angular cheilosis), dry/swollen/ulcerated lips (cheilosis), redness in lips and tongue, swollen/inflamed/smooth tongue (glossitis), Atrophied papillae & & &Correlated with fatigue & \\
Vitamin B3 &Redness/fissures in eyelid corners (Angular Palpebritis), conjunctiva redness/irritation, swollen/sticky eyelid & &Bilateral cracks/redness at corners of lips/mouth (angular cheilosis), dry/swollen/ulcerated lips (cheilosis), redness in lips and tongue, swollen/inflamed/smooth tongue (glossitis), Atrophied papillae, inflamed gums (gingivitis) &Pellagra & &Correlated with fatigue &Biomarkers indicate insufficiency far before clinical symptoms appear \\
Vitamin B5 & & & & &Sleep issues, fall in diastolic bp and lability of systolic bp &Correlated with fatigue, numbness/burning in extremities & \\
Vitamin B6 &Conjunctiva inflammation/grittiness (angular blepharitis), conjuntiva pallor, Redness/fissures in eyelid corners (Angular Palpebritis), conjunctiva redness/irritation, swollen/sticky eyelid &Excessive thinness, hapalonychia &Bilateral cracks/redness at corners of lips/mouth (angular cheilosis), swollen/inflamed/smooth tongue (glossitis), dry/swollen/ulcerated lips (cheilosis), Atrophied papillae, redness in lips and tongue &Anemia, cardiomyopathies/heart failure &Supplementation improves blood pressure, reported to help regulate SNS &Correlated with fatigue &Borderline and mild status may not present symptoms for months or years; \citet{radler_nutrient_2013} say "deficiency often occurs within 2 months of inadequacy" \\
Vitamin B7 &Excessive dryness, excessive thinness, brittleness & &&Multiple sclerosis & &Correlated with fatigue & \\
Vitamin B9 &conjuntiva pallor &Central ridges &redness in lips and tongue, swollen/inflamed/smooth tongue (glossitis), inflamed gums (gingivitis), dry/swollen/ulcerated lips (cheilosis), Aphthous Stomatitis (canker sores), inflamed/burning mouth, Atrophied papillae &Anemia, diabetes mellitus & &Correlated with fatigue & \\
Vitamin B12 \cite{adas_medical_knowledge_team_vitamin_2023}&conjuntiva pallor &Pallor, clubbing (Koilonychia), transverse white lines (Muehrcke's lines), excessive dryness, darkness in nails, curved nail ends, central ridges, longitudinal melanonychia &Bilateral cracks/redness at corners of lips/mouth (angular cheilosis), swollen/inflamed/smooth tongue (glossitis), dry/swollen/ulcerated lips (cheilosis), pallor, Aphthous Stomatitis (canker sores), inflamed/burning mouth, Atrophied papillae, redness in lips and tongue, bleeding gums, tooth loss, tooth cavities &Anemia, Osteoporosis, sarcopenia &Deficiency lowers HRV measurements, levels negatively correlated with sleep duration, levels positively correlated with sleep movement and self-assessed quality, night sweats, oxidative stress &Correlated with fatigue &Clinical symptoms can take years (typically 2-5) to appear because of storage levels, glossitis may present initially \\
Vitamin C & &Splinter hemorrhage, excessive thinness, hapalonychia &Intraoral mucosa and tongue inflammation, inflamed gums (gingivitis), thrush, tooth loss, tooth cavities &Scurvy &Supplementation improves blood pressure, helps regulate SNS &Correlated with fatigue &Deficiency can occur after 3-6 months of poor intake, signs of scurvy appear within 1 month of <10mg/day intake \\
\label{tab:phys-vitamins-water-soluble}
\end{longtable}
\end{center}
% \end{tabulary}
% \end{threeparttable}\end{adjustwidth}

\begin{center}
\tiny
% \begin{longtable}{p{0.5in}p{1in}p{0.85in}p{1.25in}p{0.6in}p{0.85in}p{0.5in}p{0.75in}}
\begin{longtable}{p{0.4in}p{0.6in}p{0.5in}p{0.4in}p{0.95in}p{0.9in}p{0.4in}p{0.6in}}
\caption{Physiological Symptoms of Micronutrient Deficiencies: Fat-Soluble Vitamins. Information from \cite{academy_of_nutrition_and_dietetics_micronutrient_2017, de_nails_2012, dibaise_hair_2019, radler_nutrient_2013, reber_nutritional_2019, mueller_aspen_2017, berger_espen_2022, national_institutes_of_health_dietary_nodate}}\label{tab: physio-fs}
\\ \toprule
Micronutrient &Eye &Nail &Oral &Disease &Autonomic &Misc &Timeframe \\\midrule
\endfirsthead
\\ \toprule
Micronutrient &Eye &Nail &Oral &Disease &Autonomic &Misc &Timeframe \\\midrule
\endhead
\hline \multicolumn{8}{r}{{Continued on next page}} \\ \hline
\endfoot
\bottomrule 
\endlastfoot
Vitamin A &Bitot's spots, yellowish lumps around eyes (xanthelasma), cornea softening (kerotomalcia), night blindness &excessive dryness, excessive thinness, leukonychia, hapalonychia & &Obesity (beta-carotene), measles &Depletion led to increased norepinephrine and epinephrine in heart and spleen of rats &Heavily associated with antioxidants and immune processes; &Plasma retinol lowers only after storage in liver and eyes are nearly depleted, then Xerophthalmia (progressive eye dryness leading to night blindness) develops after that \\
Vitamin D \cite{ross_dietary_2011}& &Beau's lines, longitudinal melanonychia, excessive thinness, hapalonychia &inflamed gums (gingivitis) &Cancer cachexia, cardiomyopathies/heart failure, Chronic obstructive pulmonary disease, osteoporosis, sarcopenia, critical to formation of hypocalcemia, depression &Deficiency lowers HRV measurements, calcidiol deficiency lowers resting sympathovagal balance, calcitriol deficiency to worse reactions to stress, supplementation improves blood pressure & & \\
Vitamin E & & & &Obesity & & & \\
Vitamin K & & & &Osteoporosis & &Impaired clotting and bleeding & \\
& & & & & & & \\
\label{tab:phys-vitamins-fat-soluble}
\end{longtable}
\end{center}

\begin{center}
\tiny
% \begin{longtable}{p{0.5in}p{1in}p{0.85in}p{1in}p{0.6in}p{0.85in}p{0.75in}p{0.75in}}
\begin{longtable}{p{0.35in}p{0.5in}p{0.6in}p{1in}p{0.6in}p{0.6in}p{0.6in}p{0.5in}}
\caption{Physiological Symptoms of Micronutrient Deficiencies: Minerals. Information from \cite{academy_of_nutrition_and_dietetics_micronutrient_2017, de_nails_2012, dibaise_hair_2019, radler_nutrient_2013, reber_nutritional_2019, mueller_aspen_2017, berger_espen_2022, national_institutes_of_health_dietary_nodate}}\label{tab: physio-min}
\\ \toprule
Micronutrient &Eye &Nail &Oral &Disease &Autonomic &Misc &Timeframe \\\midrule
\endfirsthead
\\ \toprule
Micronutrient &Eye &Nail &Oral &Disease &Autonomic &Misc &Timeframe \\\midrule
\endhead
\hline \multicolumn{8}{r}{{Continued on next page}} \\ \hline
\endfoot
\bottomrule
\endlastfoot
Iron \cite{gloria_iron_nodate,gerber_iron_2024}&conjuntiva pallor, Redness/fissures in eyelid corners (Angular Palpebritis), conjunctiva redness/irritation, swollen/sticky eyelid, blue-tinted sclera &Pallor, clubbing (Koilonychia), transverse white lines (Muehrcke's lines), brittleness, excessive dryness, excessive thinness, darkness in nails, curved nail ends, central ridges, Onycholysis, onychorrhexis &Bilateral cracks/redness at corners of lips/mouth (angular cheilosis), pallor, swollen/inflamed/smooth tongue (glossitis), Atrophied papillae, dry/swollen/ulcerated lips (cheilosis), thrush, inflamed/burning mouth, redness in lips and tongue &Anemia, cardiomyopathies/heart failure, osteoporosis &Disrupts optimal function of endocrine and immune systems; positively correlated with sleep quality (disputed); IDA affects temp regulation and HRV (HRV disputed); low levels associated with higher HR &Status has relation to energy levels and fatigue according to some sources; critical to oxygen binding; weakness; impaired cognitive function &Multiple phases: depletion of stores (mild deficiency, can take several months), iron-deficiency erythropoiesis (erythrocyte production), then iron deficiency anemia (IDA) \\
Copper &Conjuntiva pallor & & &Anemia, chronic obstructive pulmonary disease, fatty liver disease, osteoporosis &negatively correlated with sleep quality, reported to help regulate sns &Abnormal lipid metabolism &Some weeks to develop and not readily recognized, Usually manifests in acute conditions \\
Zinc \cite{pat_bass_symptoms_2023}&Conjunctiva inflammation/grittinesss (angular blepharitis) &Beau's lines, onychorrhexis, leukonychia, brittleness &Changes in taste (inconsistently observed), dryness (Xerostomia), inflamed gums (gingivitis) &Alcoholic hepatitis, cancer cachexia, chronic obstructive pulmonary disease, obesity, osteoporosis, sarcopenia, increased pneumonia risk &Deficiency linked to increased blood pressure, positively correlated with sleep quality, reported to help regulate SNS, critical to ANS functionality according to some sources &Light evidence of relationship between low dietary zinc and un-ideal metabolic response, correlated with fatigue &Symptoms after "several months of low levels" \\
Iodine & &clubbing (Koilonychia) & &Goiter, hypothyroidism & &Critical to metabolic function &Hypothyroidism occurs when intake falls below 10-20 $\mu$g/d, goiter appears fairly quickly \\
Selenium & &excessive dryness, excessive thinness, pallor & &Cardiomyopathies/ heart failure, chronic obstructive pulmonary disease, obesity &Intake reduces hypertrophy and oxidative stress, negatively effects blood pressure &Effects on metabolism & \\
Calcium & &Beau's lines, transverse leukonychia, brittleness, excessive dryness, excessive thinness, onychomadesis, onychorrhexis, hapalonychia & &Osteoporosis, rickets, osteomalacia, congestive heart failure, seizures & &Hypocalcemia can be asymptomatic or have a wide range of symptoms; most common are numbness, tingling, muscle spasms & \\
Magnesium & &Excessive dryness, excessive thinness, brittleness &inflamed/burning mouth &Cardiovascular disease, hypertension, metabolic syndrome, type 2 diabetes, depression, hypocalcemia, hypokalemia, seizures &Abnormal heart rhythms observed &Overt signs of clinical deficiency are not routinely recognized; correlated with fatigue, nausea, numbness, tingling, muscle spasms & \\
\label{tab:phys-minerals}
\end{longtable}
\end{center}

% \subsection{Gold-Standard Biochemical Methods of Micronutrient Status Assessment}
% \input{Tables/assessment}

% \input{Tables/test_cost}

\end{document}